\documentclass[11pt]{iopart}
\usepackage{color}
\usepackage{nicefrac}
\usepackage{textcomp}
\usepackage{iopams}
\usepackage{subfig}
\usepackage{graphicx}

\newcommand{\halfcolumn}{0.35\columnwidth}
\newcommand{\mass}{}

\begin{document}

\title[]{A numerical study of the Navier-Stokes transport coefficients for \textcolor{black}{2D} granular hydrodynamics}
\date{\today}

\author{Lidia Almaz\'an$^1$, Jos\'e A. Carrillo$^2$, Clara Salue\~na$^3$, Vicente Garz\'o$^4$ and Thorsten P\"oschel$^1$}
\address{$^1$ Institute for Multiscale Simulation, Universit\"at Erlangen-N\"urnberg,
D-91052 Erlangen, Germany}
\address{$^2$ Department of Mathematics, Imperial College London, London
SW7 2AZ, UK}
\address{$^3$ Departament d'Enginyeria Mec\`anica, Universitat Rovira i Virgili,
43007 Tarragona, Spain}
\address{$^4$ Departamento de F\'isica, Universidad de Extremadura, E-06071 Badajoz, Spain}

\ead{lidia.t.almazan@cbi.uni-erlangen.de}

\begin{abstract}
A numerical study is presented to analyze the thermal mechanisms
of unsteady, supersonic granular flow, by means of hydrodynamic
simulations of the Navier-Stokes granular equations. For this
purpose a paradigmatic problem in granular dynamics such as the
Faraday instability is selected. Two different approaches for the
Navier-Stokes transport coefficients for granular materials are
considered, namely the traditional Jenkins-Richman theory for
moderately dense quasi-elastic grains, and the improved
Garz\'o-Dufty-Lutsko theory for arbitrary inelasticity, which we
also present here. Both solutions are compared with event-driven
simulations of the same system under the same conditions, by
analyzing the density, the temperature and the velocity field.
Important differences are found between the two approaches leading
to interesting implications. In particular, the heat transfer
mechanism coupled to the density gradient which is a distinctive
feature of inelastic granular gases, is responsible for a major
discrepancy in the temperature field and hence in the diffusion
mechanisms.
\end{abstract}

\submitto{\NJP}
\maketitle

\section{Introduction}

The hydrodynamics of granular materials is far from being well understood. The
first difficulty comes from the kinetic theory level, where the
far-from-equilibrium nature of the problem leads to both conceptual and
technical limitations. Many contributions, starting in the '80 of the last
century \cite{Jenkins-Richman,Jenkins-Richman2}, have helped to develop a
well established hydrodynamic theory of granular gases, including mixtures and
polydisperse materials. However the application to other types of granular
materials is still uncertain.

In academy as well as in industry, one would like to have a good theory for a
variety of granular flow problems under different conditions. In the process of
going from theory to real applications, one must resort to good choices of
transport coefficients to ensure the appropriate modeling of the system.
The Navier-Stokes transport coefficients have been obtained for dilute and
semi-dilute granular gases for selected problems within the framework of kinetic
theory. However, their validity cannot be guaranteed beyond the conditions for
which they were derived and as we enter the realm of moderately dense materials,
where basic assumptions like molecular chaos are not fulfilled. On the other
hand, a purely empirical approach, like the one used for regular liquids and
where one measures the transport coefficients, to use them later in the
Navier-Stokes equations, does not apply for granular hydrodynamics. The reason
is that the properties of the flow depend strongly and nonlinearly on conditions
like the preparation of the system, flow rate, and phenomena like dilatancy;
plus the fact that, in laboratory measurements, effects due to the surface
properties of particles, wall roughness, the coupling with the
interstitial fluid, etc, are generally important. From the theoretical point of
view, the treatment of granular materials by means of the available
statistical-mechanics techniques faces inherent difficulties brought out by the
dissipative character of real grain interactions, which is responsible for
microscopic irreversibility, lack of scale separation, mesoscopic nature of
the flow, and strong nonlinearities in the governing equations.

One of the first attempts to determine the Navier-Stokes transport
coefficients from the revised Enskog theory was carried out by
Jenkins and Richman (JR) \cite{Jenkins-Richman,Jenkins-Richman2}.
However, the technical difficulties of the analysis entailed
approximations that limited their accuracy. In particular, given
that their analysis is restricted to nearly elastic systems, the
inelasticity of collisions only influences the energy balance
equation by a sink term, and so the expressions of the
Navier-Stokes transport coefficients are the same as those
obtained for elastic collisions. The JR approach has been
numerically validated by molecular dynamics (MD) simulations in
Ref.\ \cite{BizonShattuckSwiftSwinney:1999} and in experiments
such as granular flow past an obstacle \cite{RBSS02} and
vertically oscillated granular layers
\cite{BMSS02,Bougieetal04,CarrilloPoeschelSaluena:2008,B10,BD11}.
The choice of vibrated granular material as a test case for
hydrodynamic theories comes from being one of the simplest
experiments in which all different regimes of the granular flow
are present while leading to interesting standing-wave pattern
formation and dynamics
\cite{MeloUmbanhowarSwinney:1994,MeloUmbanhowarSwinney:1995},
clustering \cite{HM03,BSSP04} and phase transitions
\cite{MeersonPoeschelBromberg:2003,MeersonEtAl:2004,EshuisEtAl:2010}.

One of the problems which has been repeatedly revisited as an interesting
example of granular collective behavior is the granular Faraday instability.
Similarly as in that described by Faraday for regular liquids
\cite{Faraday:1831}, a vibrated granular layer develops characteristic patterns
of stripes, hexagons and squares, for certain intervals of frequencies and
amplitudes of oscillation \cite{MeloUmbanhowarSwinney:1995}. Experimentally and
by means of particle simulations, the analogy of vibrated granular materials and
regular fluids has been clearly shown. In two dimensions, at a certain stage
of the motion the Faraday waves appear as shown in Fig.
\ref{fig:balls}, while the time evolution of the periodic pattern, with twice
the periodicity of that of the driving oscillation, can be seen in Fig.
 \ref{fig:pattern}.

However, in many applications the dynamics of granular flow is
supersonic, in a regime where the typical velocity of the flow is
often many times or even orders of magnitude larger than the
thermal velocity. To clarify concepts, the latter is proportional
to the square root of the fluctuational part of the kinetic energy
(which gives origin to the so-called granular temperature). More
precisely, the vibrational regimes which are often imposed in real
applications in order to mobilize granular material lead to an
interplay of alternating diffusion and inertial regimes, giving
rise to a rich although extremely complex dynamics which we will
analyze in detail here.

Beyond the weak dissipation limit, however, it is expected that the functional
form of the Navier-Stokes transport coefficients for a granular gas differ from
their corresponding elastic counterparts. Thus, in subsequent works Garz\'o and
Dufty, and Lutsko (GDL) \cite{GD99,Lutsko05}, based on the application of the
Chapman-Enskog method \cite{CC70} to the Enskog equation, do not impose any
constraints at the level of collisional dissipation and take into account the
(complete) nonlinear dependence of the Navier-Stokes transport coefficients on
the coefficient of restitution $\alpha$. In particular, and in contrast to the
JR results \cite{Jenkins-Richman,Jenkins-Richman2}, the heat flux has a
contribution proportional to the density gradient which defines a new transport
coefficient $\mu$, which is not present in the elastic case. On the other hand,
as for ordinary fluids \cite{CC70}, the Navier-Stokes transport coefficients are
given in terms of the solutions of a set of coupled linear integral equations
that are approximately solved by considering the leading terms in a Sonine
polynomial expansion. In spite of this approximation, the corresponding forms
for the transport coefficients compare well with computer simulations
\cite{BR04,BR04-1,GSM07}, even for quite strong inelasticity.

In a previous paper \cite{CarrilloPoeschelSaluena:2008} we studied
computationally the Faraday instability  in
vibrated granular disks, comparing the output from particle and
Navier-Stokes hydrodynamic simulations in detail: the onset of the
instability, the characteristic wavelength, and the pattern itself
by studying the density, temperature and velocity fields. This
served to validate a Navier-Stokes code for granular material
based on a WENO (Weighted Essentially Non-Oscillatory) approach
\cite{Shu98} which is capable of capturing the features of the
highly supersonic flow generated by the impact of a piston. For
this purpose we used the JR expressions
\cite{Jenkins-Richman,Jenkins-Richman2} for the Navier-Stokes
transport coefficients, valid for \emph{elastic} hard spheres at
moderate densities. The conclusion of the study was that the JR
results showed qualitative and quantitative agreement with those
from event-driven MD simulations, in a range of parameters which
covered the entire bifurcation diagram of the Faraday instability
at the coefficient of restitution $\alpha=0.75$.

As already mentioned, the JR approach however fails describing the heat flux
accurately, since the transport coefficient $\mu$ coupled to the density
gradient vanishes in the latter approach. The presence of this new term in the
heat flux is crucial to explain for instance the dependence of the granular
temperature with height in MD simulations in dilute vibrated systems with
gradients only in the vertical direction \cite{Soto,BRM01,BRMG02}.
Apart from that, a value of the coefficient of restitution of 0.75 justifies
the use of the correct forms of the Navier-Stokes transport coefficients
proposed in the GDL approach \cite{GD99,Lutsko05} which include the effect of
dissipation on momentum and heat transport.

In the present paper, we follow a similar approach to Ref.\
\cite{CarrilloPoeschelSaluena:2008}, in order to study numerically
the thermal mechanisms that an oscillating boundary imposes on
granular material under gravity. That is, we will use the
expressions of the Navier-Stokes transport coefficients derived in
Refs.\ \cite{GD99,Lutsko05} to compare the performance of the
granular Navier-Stokes hydrodynamics with respect to particle
simulations. We will also analyze the differences between the
results provided by the JR approach
\cite{Jenkins-Richman,Jenkins-Richman2} and those from the current
approximation \cite{GD99,Lutsko05} to the Navier-Stokes transport
coefficients.

The outline of the paper is as follows: In Section 2 we will review the
Navier-Stokes theory, introducing the GDL kinetic coefficients for dilute and
moderately dense 2D granular gases, opposed to the JR kinetic coefficients
which are only valid for vanishing inelasticity. We will also explain briefly
how to treat numerically the Navier-Stokes equations, while section 3 will be
devoted to the results obtained with JR and GDL and their comparison with
MD simulations. These will lead to interesting implications which will be
discussed in more detail in the conclusions section.

%

\section{Navier-Stokes hydrodynamic theory of granular gases}

We consider a granular fluid composed of smooth inelastic hard disks of mass $m$
and diameter $\sigma$. Collisions are characterized by a (constant) coefficient
of normal restitution $0<\alpha\leq 1$. In a kinetic theory description, the
relevant information on the system is contained in the one-particle velocity
distribution function. At \emph{moderate} densities and assuming molecular
chaos, the velocity distribution function obeys the (inelastic) Enskog kinetic
equation \cite{GS95,BDS97}. Starting from this kinetic theory, one can easily
obtain the (macroscopic) Navier-Stokes hydrodynamic equations for the number
density $n({\vec{r}}, t)$, the flow velocity $\vec u({\vec{r}}, t)$, and
the local temperature $T({\vec{r}}, t)$ \cite{BP04}. In the case of
two-dimensional granular gases, the balance equations read
\begin{equation}
\label{density}
\frac{\partial n}{\partial t}+ \vec\nabla\cdot(n \vec u) = 0\,,
\end{equation}
\begin{equation}
\label{velocity} \rho\left(\frac{\partial \vec u}{\partial t} + \vec
u\cdot\vec\nabla\vec u \right) = -\vec\nabla\cdot \hat P +n\vec
F\,,
\end{equation}
and
\begin{equation}\label{temperature}
n\left(\frac{\partial T}{\partial t} +\vec u\cdot\vec\nabla T\right) = -\nabla\cdot\vec q-
\hat P:\vec \nabla \vec u-\zeta nT\,.
\end{equation}
In the above equations, $\rho=mn$ is the mass density, $\vec F$ is the external
force acting on the system, $\hat P$ is the pressure tensor, $\vec q$ is the
heat flux, and $\zeta$ is the cooling rate due to the energy dissipated in
collisions. It is worthwhile to note that the macroscopic equations given in
Eqs.\ \eref{density}-\eref{temperature} differ from their counterparts for
elastic fluids only via the appearance of the cooling rate $\zeta$ on the
right-hand side of Eq.\ \eref{temperature}. On the other hand, the
corresponding transport coefficients defining the momentum and heat fluxes must
depend in general on the coefficient of restitution $\alpha$.

As it happens for elastic fluids, the usefulness of the balance equations
\eref{density}-\eref{temperature} is limited unless the fluxes and the cooling
rate are specified in terms of the hydrodynamic fields and their spatial
gradients. To first order in the spatial gradients, the Navier-Stokes
constitutive equations provide a link between the exact balance equations and a
closed set of equations for the hydrodynamic fields. The constitutive
relation of the pressure tensor $P_{ij}$ is
\begin{equation}
\label{vic1} P_{ij}=p\delta _{ij}-\eta \left( \partial_{j}u_{i
}+\partial_{i}u_{j} -\delta_{ij} \vec\nabla \cdot \vec
u\right)-\gamma \delta_{ij} \vec\nabla \cdot \vec u,
\end{equation}
where $p$ is the hydrostatic pressure, $\eta$ is the shear viscosity, and
$\gamma$ is the bulk viscosity. The constitutive equation for the heat flux is
\begin{equation}
\label{vic2}
\vec q= -\kappa \vec\nabla T - \mu \vec\nabla n,
\end{equation}
where $\kappa$ is the coefficient of thermal conductivity, and $\mu$ is a new
coefficient which does not have an analogue for a gas of elastic particles.
Finally, to first order in gradients, the cooling rate $\zeta$ can be written as
\cite{GS95}
\begin{equation}
\label{eq:zeta_LG}
\zeta= \zeta_0 + \zeta_1 \nabla\cdot\vec{u} \, .
\end{equation}

It is important to remark that the derivation of the Navier-Stokes order
transport coefficients does not limit \emph{in principle} their application to
weak inelasticity. The Navier-Stokes hydrodynamic equations themselves may or
may not be limited with respect to inelasticity, depending on the particular
states considered. In particular, the derivation of these equations by means of
the Chapman-Enskog method assumes that the spatial variations of the
hydrodynamic fields $n$, $\vec u$, and $T$ are small on the scale of the mean
free path. In the case of ordinary fluids, the strength of the gradients can be
controlled by the initial or boundary conditions. However, the problem is
more complicated for granular fluids since in some cases (e.g., steady states
such as the simple shear flow \cite{G03,SGD04}) there is an intrinsic relation
between dissipation and some hydrodynamic gradient and so, the two cannot be
chosen independently. Consequently, there are examples for which the
Navier-Stokes approximation is never valid or is restricted to the quasielastic
limit. On the other hand, the transport coefficients characterizing the
Navier-Stokes order hydrodynamic equations are well-defined functions of
$\alpha$, regardless of the applicability of those equations.

As said in the Introduction, the evaluation of the explicit forms of the
hydrostatic pressure $p$, the Navier-Stokes transport coefficients $\eta$,
$\gamma$, $\kappa$, and $\mu$ and the coefficients $\zeta_0$ and $\zeta_1$
requires to solve the corresponding Enskog equation. However, due to the
mathematical complexity of this kinetic equation, only approximate results for
the above coefficients can be obtained. Here, we consider two independent
approaches for hard disks proposed by Jenkins and Richman
\cite{Jenkins-Richman2} and Garz\'o and Dufty \cite{GD99} and Lutsko
\cite{Lutsko05}. Let us consider each method separately.

\subsection{Jenkins-Richman (JR) results}

The results derived by Jenkins and Richman
\cite{Jenkins-Richman,Jenkins-Richman2} are obtained by solving the Enskog
equation for spheres \cite{Jenkins-Richman} and disks \cite{Jenkins-Richman2} by
means of Grad's method \cite{G49}. The idea behind Grad's moment method is to
expand the velocity distribution function in a complete set of orthogonal
polynomials (generalized Hermite polynomials), the coefficients being the
corresponding velocity moments. Next, the expansion is truncated after a certain
order $k$. When this truncated expansion is substituted into the hierarchy of
moment equations up to order $k$ one gets a closed set of coupled equations. In
the case of a two-dimensional system, the eight retained moments are the
hydrodynamic fields ($n$, $\vec u$, and $T$) plus the irreversible momentum and
heat fluxes ($P_{ij}-p \delta_{ij}$ and $\vec q\,$).

Although the application of Grad's method to the Enskog equation is not
restricted to nearly elastic particles, the results derived by Jenkins and
Richman \cite{Jenkins-Richman2} neglect the cooling effects on temperature due
to the cooling rate in the expressions of the transport coefficients [see for
instance, Eqs. (70), (89), (98), (99), and (100) of Ref.\
\cite{Jenkins-Richman2} when the disks are smooth]. Given that this assumption
can only be considered as acceptable for nearly elastic systems, the authors of
Ref. \cite{Jenkins-Richman2} conclude that their theory only holds in the
quasielastic limit ($\alpha \to 1$).

The explicit forms of the hydrostatic pressure, the Navier-Stokes transport
coefficients and the cooling rate in the JR theory are given by \begin{equation}
\label{hydrostatic}
p_\textrm{\small{JR}}=\frac{4}{\pi \sigma^2}\phi T[1+(1+\alpha)G(\phi)],
\end{equation}
\begin{equation}
\label{eq:shearviscosity_JR} \eta_\textrm{\small{JR}}= \frac{\phi}{2\sigma}
\sqrt{\frac{m T}{\pi}}
\left[\frac{1}{G(\phi)}+2+\left(1+\frac{8}{\pi}\right)G(\phi)\right],
\end{equation}
\begin{equation}
\label{eq:bulkviscosity:JR} \gamma_\textrm{\small{JR}}=\frac{8}{\pi\sigma}
\phi G(\phi)\sqrt{\frac{m T}{\pi}}\,,
\end{equation}
\begin{equation}
\label{eq:thermalconductivity_JR}
\eqalign{
\kappa_\textrm{\small{JR}}&=\frac{2\phi}{\sigma} \sqrt{\frac{T}{\pi m}}\left[
\frac{1}{G(\phi)}+3+\left(\frac94+\frac{4}{\pi}\right)G(\phi)\right]\,,\cr
\mu_\textrm{\small{JR}}&=0,}
\end{equation}
\begin{equation}
\label{eq:zeta_JR}
\eqalign{
\zeta_\textrm{0,\small{JR}}&=\frac{4}{\sigma}(1-\alpha^2)\sqrt{\frac{T}{\pi m}}G(\phi),\cr
\zeta_\textrm{1,\small{JR}}&=0.}
\end{equation}
In the above equations, $\phi=n\pi \sigma^2/4$ is the (dimensionless) volume
fraction occupied by the granular disks, also called packing fraction,
$G(\phi)=\phi \chi(\phi)$, and $\chi(\phi)$ is the pair correlation function.

Because of the assumption of near elastic particles in the JR theory, Eqs.\
\eref{hydrostatic}--\eref{eq:zeta_JR} show clearly that the coefficient of
restitution $\alpha$ only enters in the equation of state \eref{hydrostatic}
and in the expression \eref{eq:zeta_JR} for the zeroth-order cooling rate
$\zeta_0$. At this level of approximation, the expressions of the Navier-Stokes
transport coefficients $\eta_\textrm{\small{JR}}$, $\gamma_\textrm{\small{JR}}$, and
$\kappa_\textrm{\small{JR}}$ are the same as those given by the Enskog equation for
elastic disks \cite{G70}.

In order to get the dependence of the transport coefficients and the cooling
rate in both JR and GDL approaches, one has to chose an approximate form for the
pair correlation function $\chi(\phi)$. In this paper, we have chosen the forms
proposed by Torquato
\cite{T95},
\begin{equation}\label{eq:chi}
 \chi(\phi) = \left\{ \begin{array}{cl}
                       \displaystyle\frac{1-\frac{7}{16}\phi}{(1-\phi)^2 }
                      & \textrm{for } 0\leq \phi < \phi_f, \\[5mm]
               \displaystyle\frac{1-\frac{7}{16}\phi_f}{(1-\phi_f)^2 } \frac{\phi_c-\phi_f}{\phi_c - \phi}
                        & \textrm{for }  \phi_f \leq \phi \leq \phi_c,
                        \end{array} \right.
\end{equation}
which go through the freezing point $\phi_f=0.69$ and approach the random close
packing fraction, $\phi_c=0.82$ with reasonable accuracy.

\subsection{Garz\'o-Dufty-Lutsko (GDL) results}

The dependence of the Navier-Stokes transport coefficients on the coefficient of
restitution was first obtained by Garz\'o and Dufty \cite{GD99} for hard spheres
($d=3$) by solving the Enskog equation from the Chapman-Enskog method
\cite{CC70}. These results were then extended to an arbitrary number of
dimensions by Lutsko \cite{Lutsko05}. Here, we refer to the above theories as
the GDL theory. The Chapman-Enskog method \cite{CC70} is a procedure to
construct an approximate perturbative solution to the Enskog equation in powers
of the spatial gradients. As said in the Introduction, the GDL theory considers
situations where the spatial gradients are sufficiently small and
\emph{independent} of the coefficient of restitution $\alpha$. As a consequence,
the corresponding forms of the Navier-Stokes transport coefficients are not
limited \emph{a priori} to weak inelasticity since they incorporate the complete
nonlinear dependence on $\alpha$. This is the main difference with respect to
the JR approach.

On the other hand, as for elastic collisions \cite{CC70}, the Navier-Stokes
transport coefficients in the Chapman-Enskog method cannot be \emph{exactly}
determined since they are defined in terms of the solutions of a set of coupled
linear integral equations. It is useful to represent these solutions as an
expansion in a complete set of polynomials (Sonine polynomials) and generate
approximations by truncating the expansion. In practice the leading terms in
these expansions provides an accurate description over the full range of
dissipation and density since in general they yield good agreement with Monte
Carlo simulations, except for the heat flux transport coefficients at high
dissipation \cite{BR04,BR04-1}. Motivated by this disagreement, a modified
version of the first Sonine approximation has been recently proposed
\cite{GSM07,Garzo2011}. The modified Sonine approximation replaces the Gaussian
weight function (used in the standard Sonine method) by the homogeneous cooling
state distribution. This new method significantly improves the
$\alpha$-dependence of $\kappa$ and $\mu$ since partially eliminates the
discrepancies between simulation and theory for quite strong dissipation (see
for instance, Figs. 1-3 of Ref.\ \cite{GSM07}).

The results obtained in the GDL approach for the equation of
state and the Navier-Stokes transport coefficients for hard disks ($d=2$) are
\begin{equation}
\label{hydrostaticLG}
p_{\textnormal{\small{GDL}}}=p_\textrm{\small{JR}}=\frac{4}{\pi \sigma^2}\phi T[1+(1+\alpha)G(\phi)],
\end{equation}
\begin{equation}\label{eq:bulkviscosity:LG}
\gamma_\textrm{\small{GDL}}=\frac{4}{\pi\sigma} \phi G(\phi)\sqrt{\frac{m
T}{\pi}}(1+\alpha)\left( 1- \frac{c}{32} \right),
\end{equation}
\begin{equation}
\label{eq:shearviscosity_LG}
\fl
\eta_\textrm{\small{GDL}}= \frac{\sqrt{mT/\pi}}{2\sigma}
\frac{\left[1-\frac{1}{4}(1+\alpha)(1-3\alpha)G(\phi)\right]\left[1+\frac{1}{2}
G(\phi)(1+\alpha)\right]}{\nu_\eta^*-\frac{1}{2}\zeta_0^*}+\frac{1}{2}\gamma_\textrm{\small{GDL}},
\end{equation}
\begin{equation}
\label{eq:thermalconductivity_LG}
\fl
\kappa_\textrm{\small{GDL}}=\frac{2}{\sigma} \sqrt{\frac{T}{\pi m}}\left
\{ \left[ 1+\frac34 G(\phi)(1+\alpha)\right]\kappa_k^* +
\frac{2}{\pi}\phi G(\phi)(1+\alpha)\left( 1+\frac{7c}{32} \right)
\right \},
\end{equation}
\begin{equation}
\label{muLG} \mu_\textrm{\small{GDL}} =  \frac{T\sigma}{\phi} \sqrt{\frac{\pi T}{m}}
\,  \left [ 1+\frac34 G(\phi)(1+\alpha) \right ] \, \mu_k^*
\end{equation}
where the (reduced) kinetic contributions $\kappa_k^*$ and $\mu_k^*$ are
\begin{equation}
\label{eq:conduc1} \kappa_k^* = \frac{1+c+\frac38 G(\phi)(1+\alpha)^2\left[2\alpha-1+\frac{c}{2}(1+\alpha)\right]}
{2(\nu_{\kappa}^*-2\zeta_0^*)},
\end{equation}
\begin{equation}
\label{eq:mu}\fl \mu_k^* =  \frac{\zeta_0^*\kappa_k^*(1+\phi\partial_\phi \ln \chi)
+ \frac{c}{4} + \frac38 G(\phi)(1+\alpha)(1+\frac{1}{2}\phi\partial_\phi \ln \chi)
\left[ \alpha(\alpha-1)+\frac{c}{12} (14-3\alpha + 3\alpha^2) \right]}
{2\nu_{\kappa}^*-3\zeta_0^*}.
\end{equation}
In Eqs.\ \eref{eq:shearviscosity_LG}--\eref{eq:mu} we have introduced the quantities \cite{Garzo2011}
\begin{equation}
\zeta_0^* = \frac12   \chi(\phi)(1-\alpha^2) \left( 1+
\frac{3c}{32} \right),
\end{equation}
\begin{equation}
\label{nueta}
\nu_{\eta}^* =  \frac18 \chi(\phi)(7-3\alpha)(1+\alpha)\left( 1+
\frac{7c}{32} \right),
\end{equation}
\begin{equation}
\label{nukappa}
\nu_{\kappa}^* = \frac14
\chi(\phi)(1+\alpha)\left[ 1+\frac{15}{4}(1-\alpha) +
\frac{365-273\alpha}{128} \, c\right],
\end{equation}
where
\begin{equation}
\label{eq:Sonine_c}
c(\alpha)= \frac{32(1-\alpha)(1-2\alpha^2)}{57 -25 \alpha +30 \alpha^2 (1-\alpha) }
\end{equation}
is the fourth cumulant coefficient measuring the deviation of the homogeneous
reference state from its Gaussian form. Also taking into account Eq.\
\eref{eq:chi}, we obtain the expression
to be used in Eq.\ \eref{eq:mu}.

\begin{figure}
 \centerline{\includegraphics[width=0.45\columnwidth]{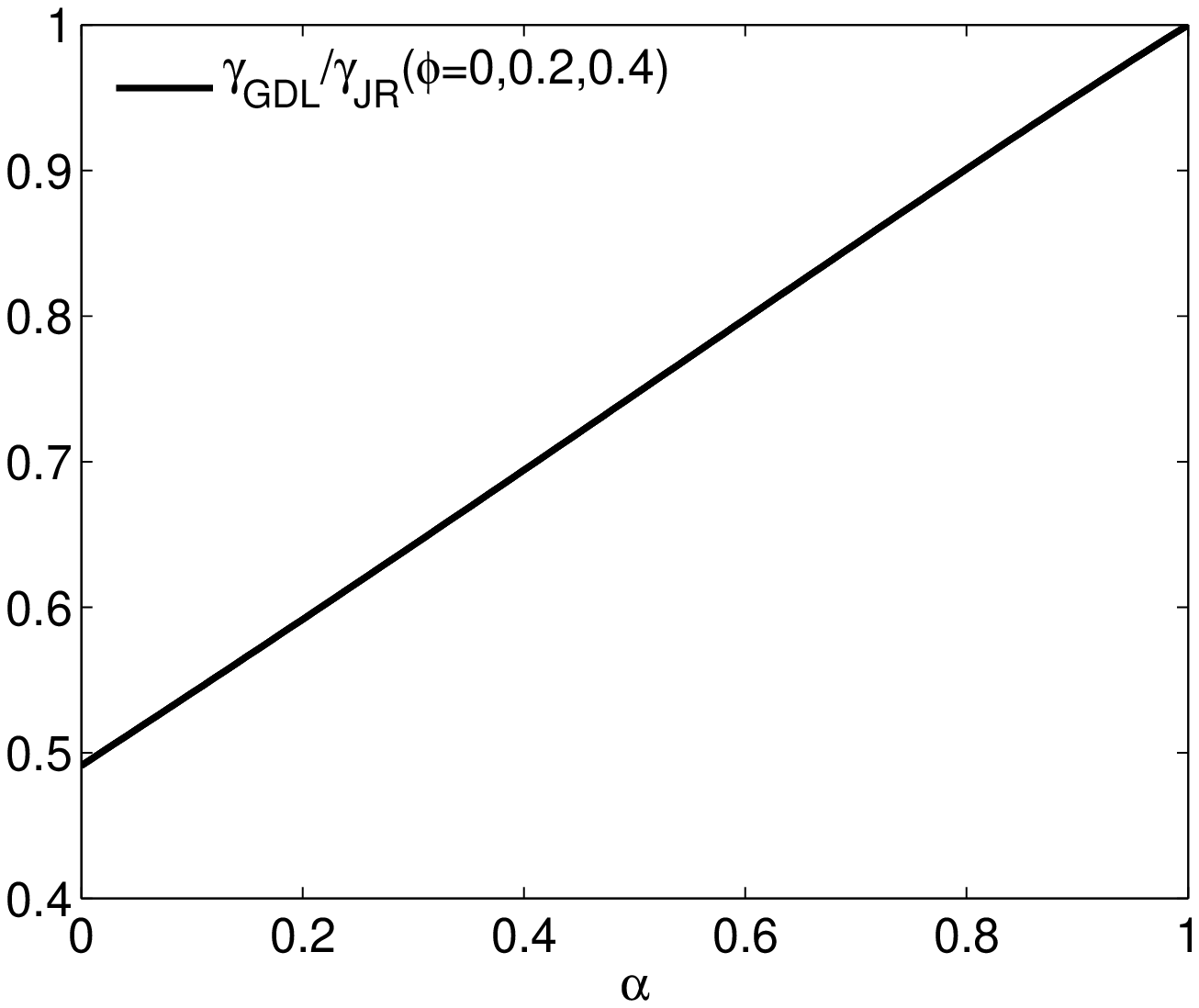}\includegraphics[width=0.45\columnwidth]{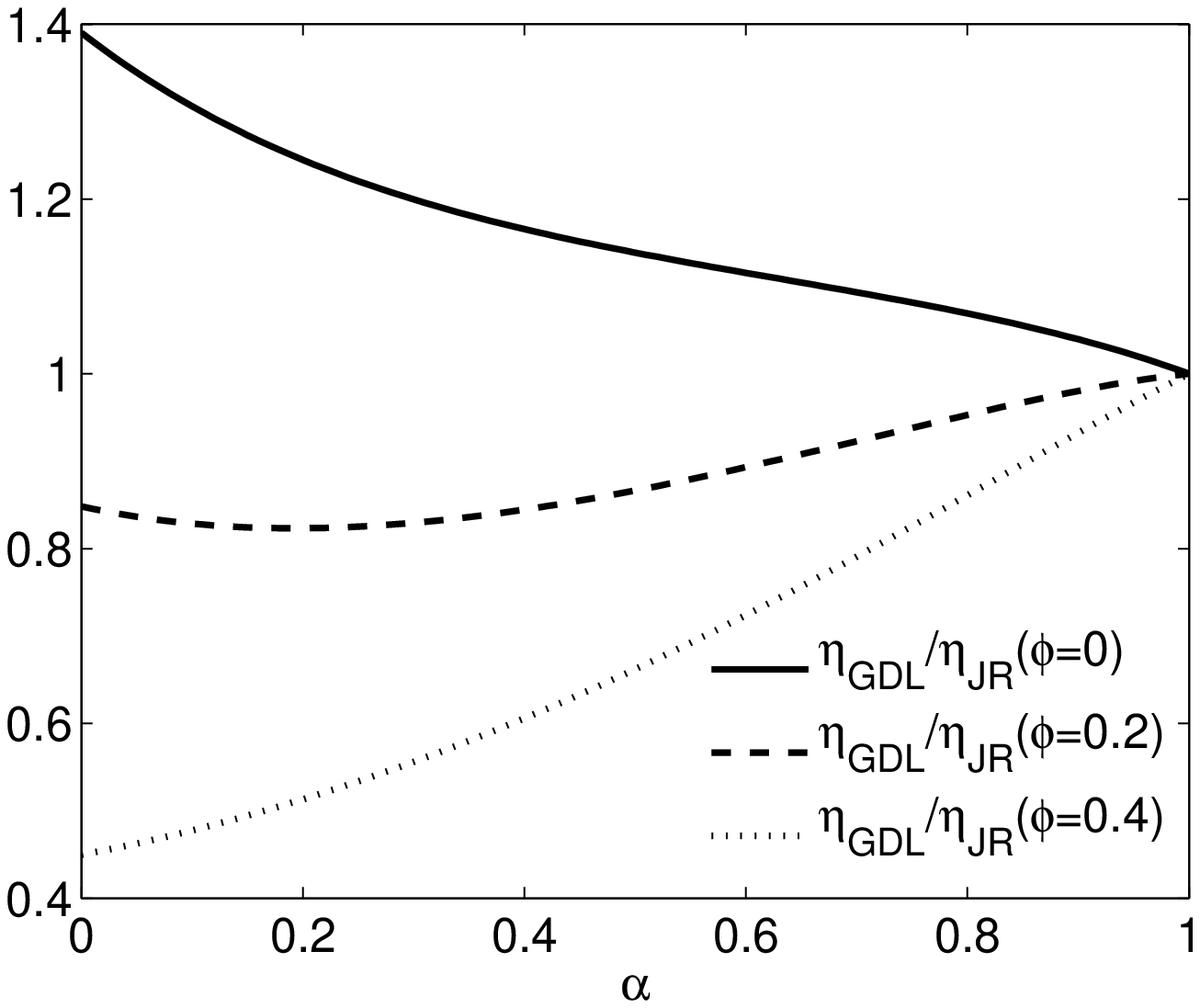}}
 \centerline{\includegraphics[width=0.45\columnwidth]{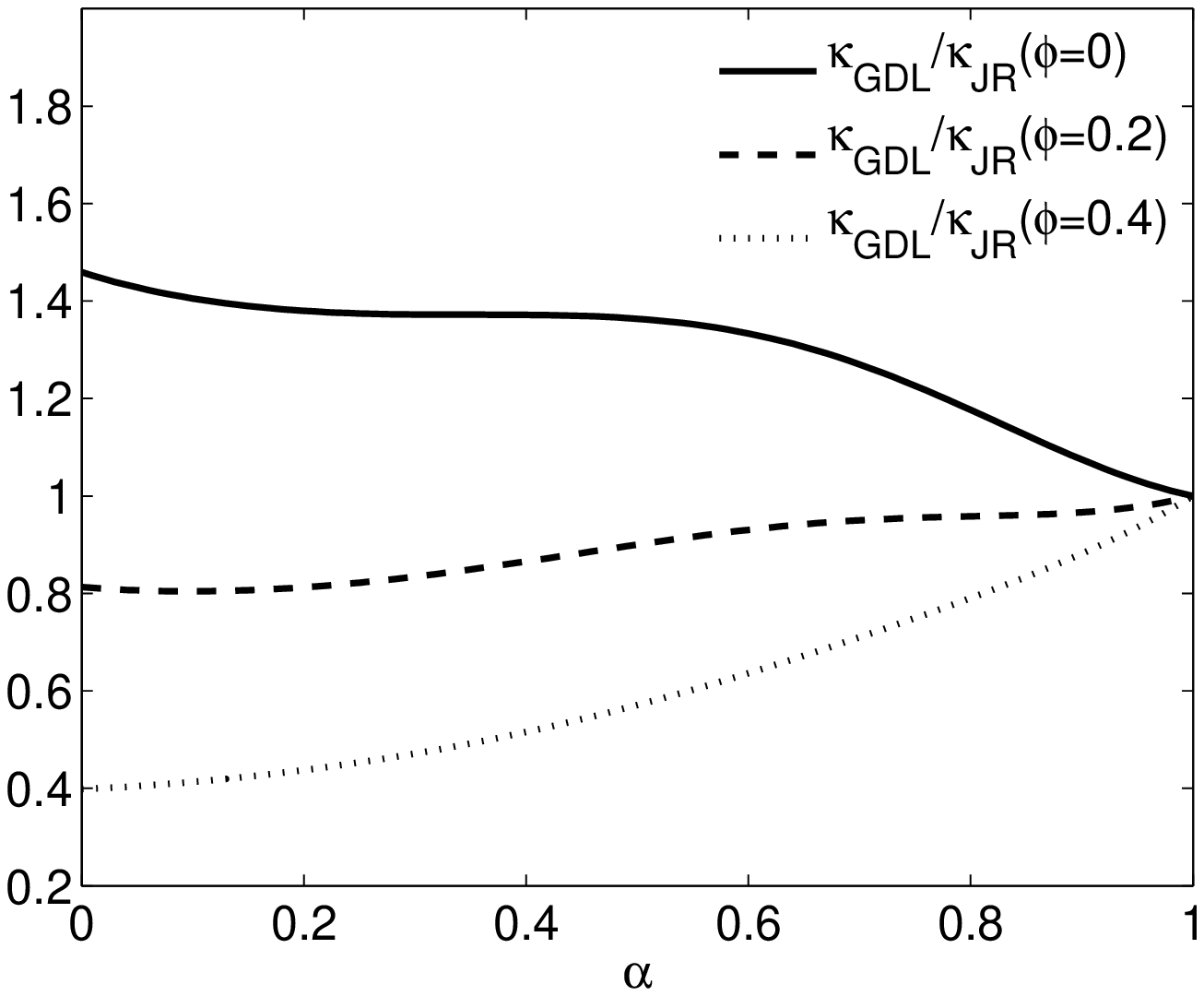}\includegraphics[width=0.45\columnwidth]{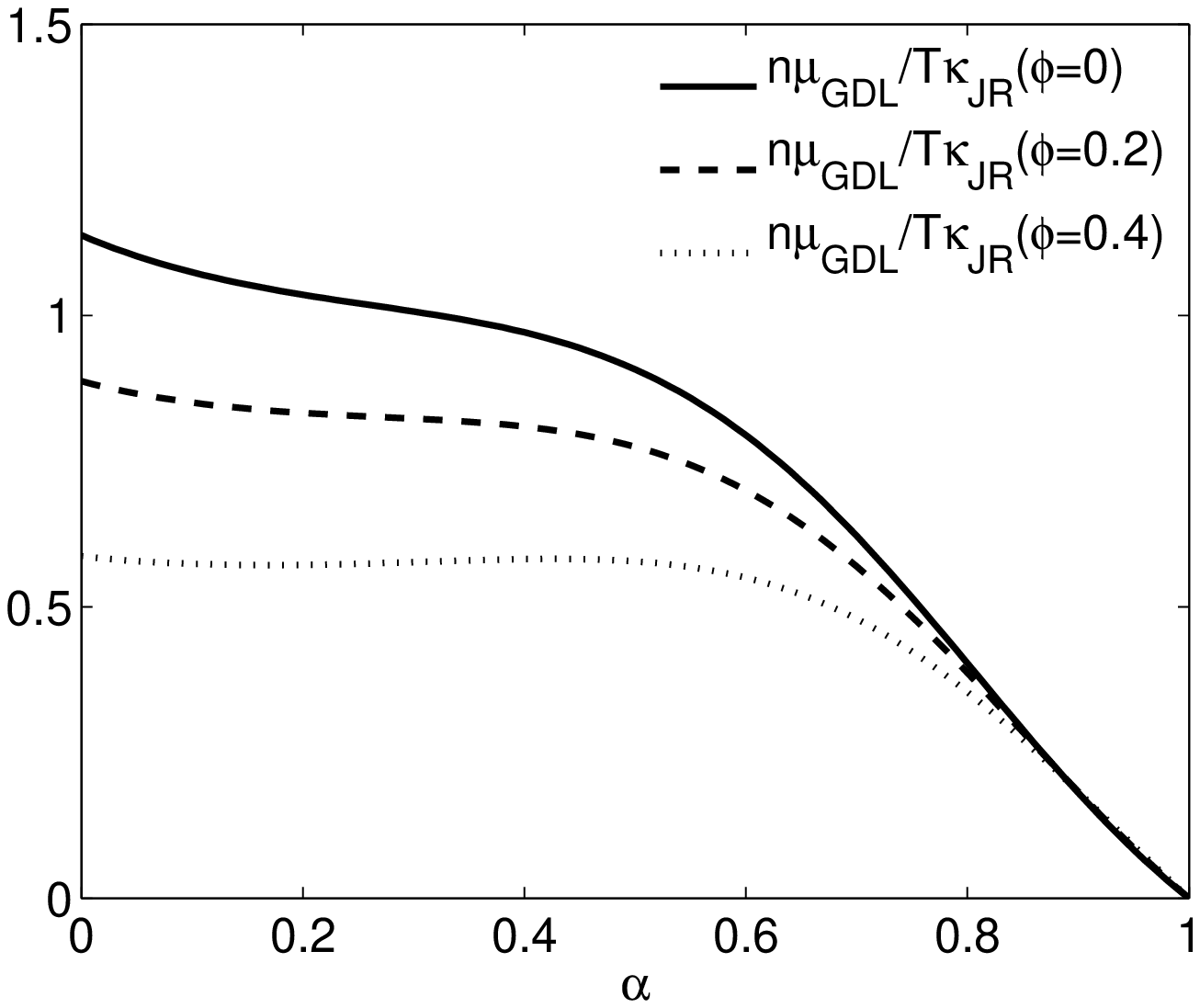}}
\caption{Bulk viscosity ratio $\gamma_\textrm{\small{GDL}}/\gamma_\textrm{\small{JR}}$
(top left), shear viscosity ratio $\eta_\textrm{\small{GDL}}/\eta_\textrm{\small{JR}}$
(top right), thermal conductivity ratio
$\kappa_\textrm{\small{GDL}}/\kappa_\textrm{\small{JR}}$ (bottom left), and
$n\mu_\textrm{\small{GDL}}/T\kappa_\textrm{\small{JR}}$ ratio (bottom right) as a
function of the restitution coefficient $\alpha$ for three
different values of the packing fraction $\phi$: $\phi=0$ (solid
line), $\phi=0.2$ (dashed line), and $\phi=0.4$ (dotted line).}
\label{fig:ratiotransportcoef}
\end{figure}

It is quite apparent that, except the equation of state \eref{hydrostaticLG},
the expressions for the Navier-Stokes transport coefficients of the GDL results
clearly differ from those obtained in the JR approach. In fact, Eqs.\
\eref{eq:bulkviscosity:LG}, \eref{eq:shearviscosity_LG},
 \eref{eq:thermalconductivity_LG},
and \eref{muLG} of the GDL theory reduce to Eqs.\
\eref{eq:shearviscosity_JR}, \eref{eq:bulkviscosity:JR}, and
\eref{eq:thermalconductivity_JR}, respectively, in the elastic
limit ($\alpha=1$, and so $\zeta_0^*=c=0$). Note that the expressions derived by
Lutsko \cite{Lutsko05} neglect in the expressions \eref{nueta} and
\eref{nukappa} of $\nu_{\eta}^*$ and $ \nu_{\kappa}^*$, respectively, the
factors of $c$ coming from the non-Gaussian corrections to the reference
state. These extra factors will be accounted for in our numerical results since
their effect on transport becomes non negligible at small values of $\alpha$. In
Fig. \ref{fig:ratiotransportcoef} we show the ratio between the bulk viscosity,
shear viscosity, and thermal conductivity given by the GDL and JR approaches as
a function of the coefficient of restitution $\alpha$ for different packing
fractions $\phi$. Note that the bulk viscosity ratio does not depend on $\phi$.
We also observe the order of magnitude of the new term in the heat flux due to
the density gradient in the GDL theory with respect to the heat flux of the JR
theory. The quantitative percentage of deviation of the transport coefficients
with the GDL theory from the JR theory is quite significant for $\alpha=0.8$ and
the different packing fractions $\phi$ used. We emphasize how the GDL-term
related to the density gradient in the heat flux becomes very important for
$\alpha\leq 0.8$.

\begin{figure}
\centerline{\includegraphics[width=0.6\columnwidth]{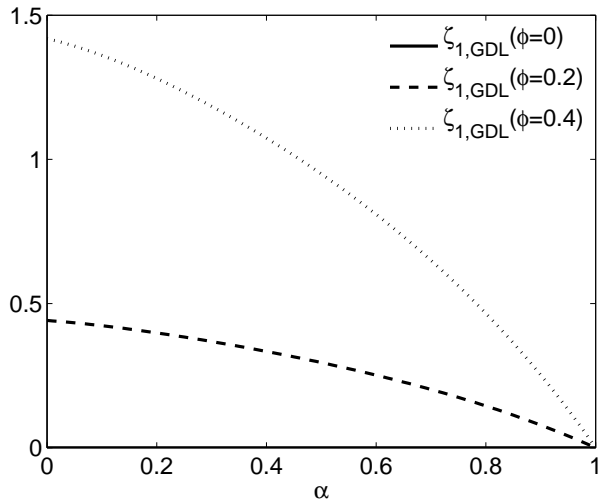}}
\caption{First order correction of the cooling coefficient for GDL
theory as a function of the coefficient of restitution $\alpha$
for three different values of the packing fraction $\phi$:
$\phi=0$ (solid line), $\phi=0.2$ (dashed line), and $\phi=0.4$
(dotted line). }
 \label{fig:cooling}
\end{figure}

Finally, the contributions to the cooling rate are given by
\begin{equation}
\label{zeta0LG}
\zeta_\textrm{0,\small{GDL}}=\frac{4}{\sigma}(1-\alpha^2)\sqrt{\frac{T}{\pi
\mass}}G(\phi)\left( 1+ \frac{3c}{32} \right),
\end{equation}
\begin{equation}
\label{zeta1LG}
\zeta_\textrm{1,\small{GDL}}= \frac32 G(\phi) (1-\alpha^2)
\left[\frac{3}{32} \, \frac{\frac18\omega^* -c(1+\alpha)(\frac13
-\alpha)}{\nu_{\zeta}^*-\frac34 (1-\alpha^2)} -1 \right],
\end{equation}
where
\begin{equation}
\label{n1}
\nu_{\zeta}^* =-\frac{1+\alpha}{192}(30\alpha^3-30\alpha^2+153\alpha-185),
\end{equation}
\begin{equation}
\label{n2} \omega^*= (1+\alpha)\left[ (1-\alpha^2)(5\alpha-1) -
\frac{c}{12} (15\alpha^3-3\alpha^2+69\alpha-41)\right].
\end{equation}
Equation \eref{zeta0LG} agrees with its corresponding counterpart in the JR
theory, Eq.\ \eref{eq:zeta_JR}, when one neglects the non-Gaussian corrections
to the reference state ($c=0$). Note that $\zeta_1$ vanishes in limits of
elastic gases ($\alpha=1$, arbitrary volume fraction $\phi$) and of dilute
inelastic gases ($\phi=0$, arbitrary values of the coefficient of restitution
$\alpha$). In Fig. \ref{fig:cooling}, we plot the $\alpha$-dependence of
$\zeta_\textrm{1,\small{GDL}}$. We observe that the first-order contribution to the total
cooling rate appears to be more significant as the gas becomes denser.

\subsection{Numerical scheme for the hydrodynamic granular equations}

The compressible Navier-Stokes-like equations for granular materials
\eref{density}, \eref{velocity}, and \eref{temperature} are solved in
conservation form for the convective terms, that is, we numerically solve the
system for the density, the momentum and the total energy: $(n, n\vec u, W)$
where the total energy density $W$ is given by
\begin{equation}\label{totalenergy}
W=n T + \frac12 n |\vec u|^2\,.
\end{equation}
This system can be rewritten as a system of nonlinear conservation laws with
sources as in Ref.\ \cite{CarrilloPoeschelSaluena:2008}. Local eigenvalues and
both local left- and right-eigenvectors of the Jacobian matrices of the fluxes
are explicitly computable (see Appendix of Ref.\
\cite{CarrilloPoeschelSaluena:2008}). We only mention here that the
characteristic speeds of the waves in the hyperbolic part of the equation can be
written in terms of the speed of sound, given by
\begin{equation}
c_s^2 = \frac{\partial p}{\partial n} + \frac{p}{n^2}
\frac{\partial p}{\partial \epsilon} ,
\label{soundspeed}
\end{equation}
for a general equation of state where $p=p(n,\epsilon)$ with the enthalpy
$\epsilon=T$ for a two dimensional system. We refer to Ref.\
\cite{CarrilloPoeschelSaluena:2008} for the full details of the numerical scheme
that here is applied to both the GDL and the JR Navier-Stokes hydrodynamic
equations since they share the same structure. Let us just briefly mention that
Navier-Stokes terms are treated by simple centered high-order explicit in time
finite difference approximations and considered as sources for the method of
lines in the time approximation. Meanwhile the Euler (convective) terms are
solved in local coordinates by a fifth-order explicit in time finite difference
characteristic-wise WENO method in a uniform grid following Refs.\
\cite{JS96,Shu98}. Thus, typical wave speeds and vectors, eigenvalues and
eigenvectors of the purely hyperbolic part, are correctly resolved.

\section{Results}

We have applied the traditional MD approach to compare the results
obtained from the Navier-Stokes hydrodynamic equations with the
GDL kinetic coefficients, showing also those results provided by
the previously used JR model as a reference. In all simulations,
the frequency of the piston motion is $f =$ 3.75 Hz and the
amplitude is $A=$ 5.6 particle diameters. The system size is tuned
to fit three pattern wavelengths in the (horizontal) $x$-direction
($125\,\sigma$), which is periodic. In the (vertical)
$y$-direction, the hydrodynamic simulations are constrained into a
box of finite height of 60 diameters, whereas the MD system is not
limited (particles reach the height of 60 diameters very rarely).
The particles are 783 disks of diameter $\sigma = 1$ cm and mass
$m=1$ mg, and $g=9.81$ m/s$^2$ is the acceleration of gravity. The
coefficient of restitution is $\alpha=0.80$, however a similar
behavior is found regardless the value of the coefficient of
restitution between $\alpha$ = 0.60 and 0.80. At $\alpha$ = 0.85
and beyond instead, the pattern does not form in our system. Since
JR and GDL will differ less and less, the differences will shrink
at high values of $\alpha$ anyway. The interesting region is found
at intermediate values of $\alpha$, whereas the use of JR is
clearly wrong at very low values of the coefficient of
restitution. Therefore we will show the results for $\alpha=0.80$
as a representative case of what one will observe under the
conditions of the Faraday instability.

The top and bottom walls in both hydrodynamic simulations are adiabatic and
impenetrable. More precisely, the normal velocity is zero at the walls, the
energy flux is zero, and the tangential velocity remains unchanged. The
simulation is carried over in the comoving frame of the wall, and thus the force
per unit mass of the simulated system is $\vec{F}=-g(1+A \sin(2\pi f
t))\vec{j}$, with $\vec{j}=(0,1)$.

\begin{figure}[]
\includegraphics[width=0.14\textwidth]{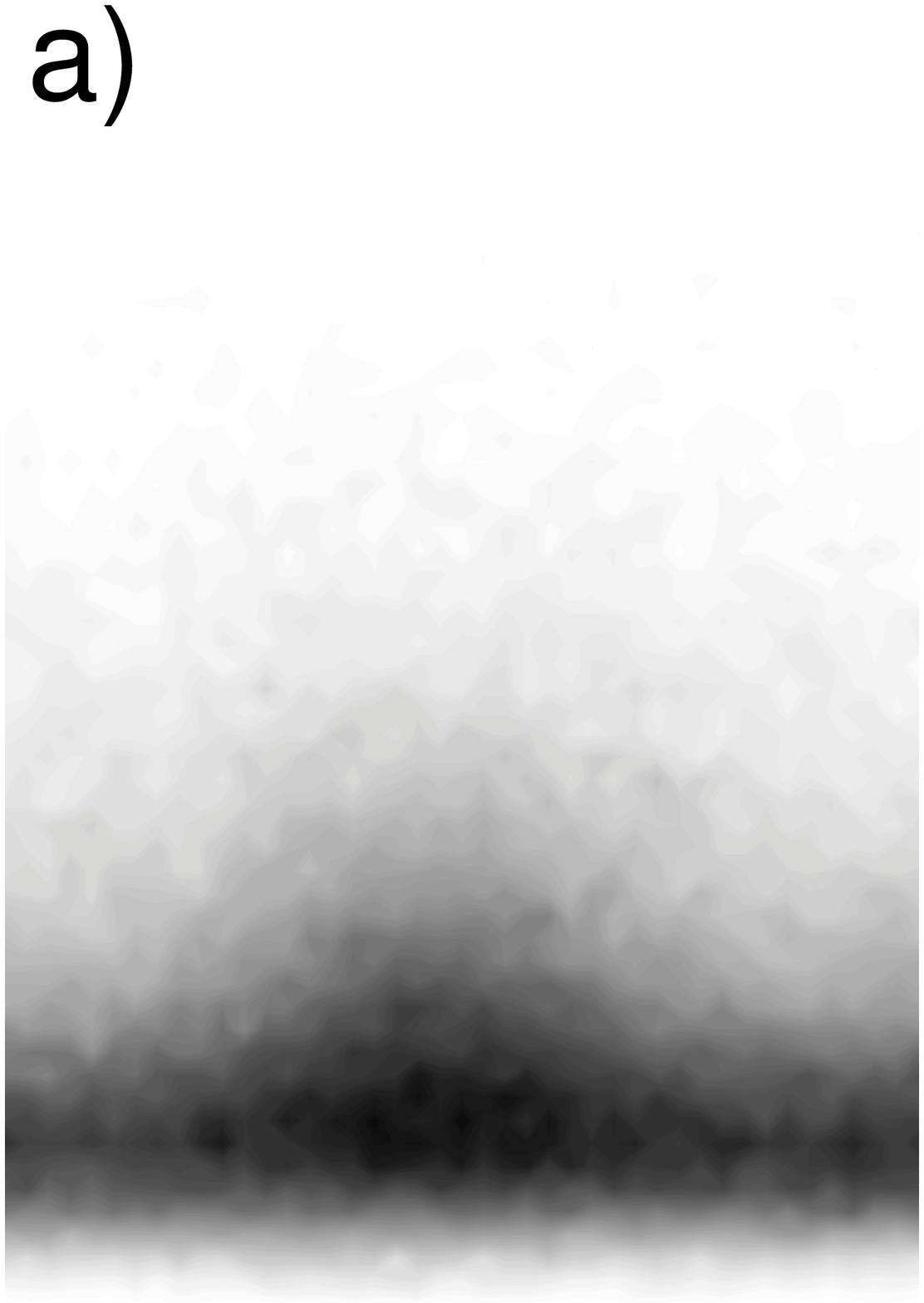} \hspace{-0.4cm}
\includegraphics[width=0.14\textwidth]{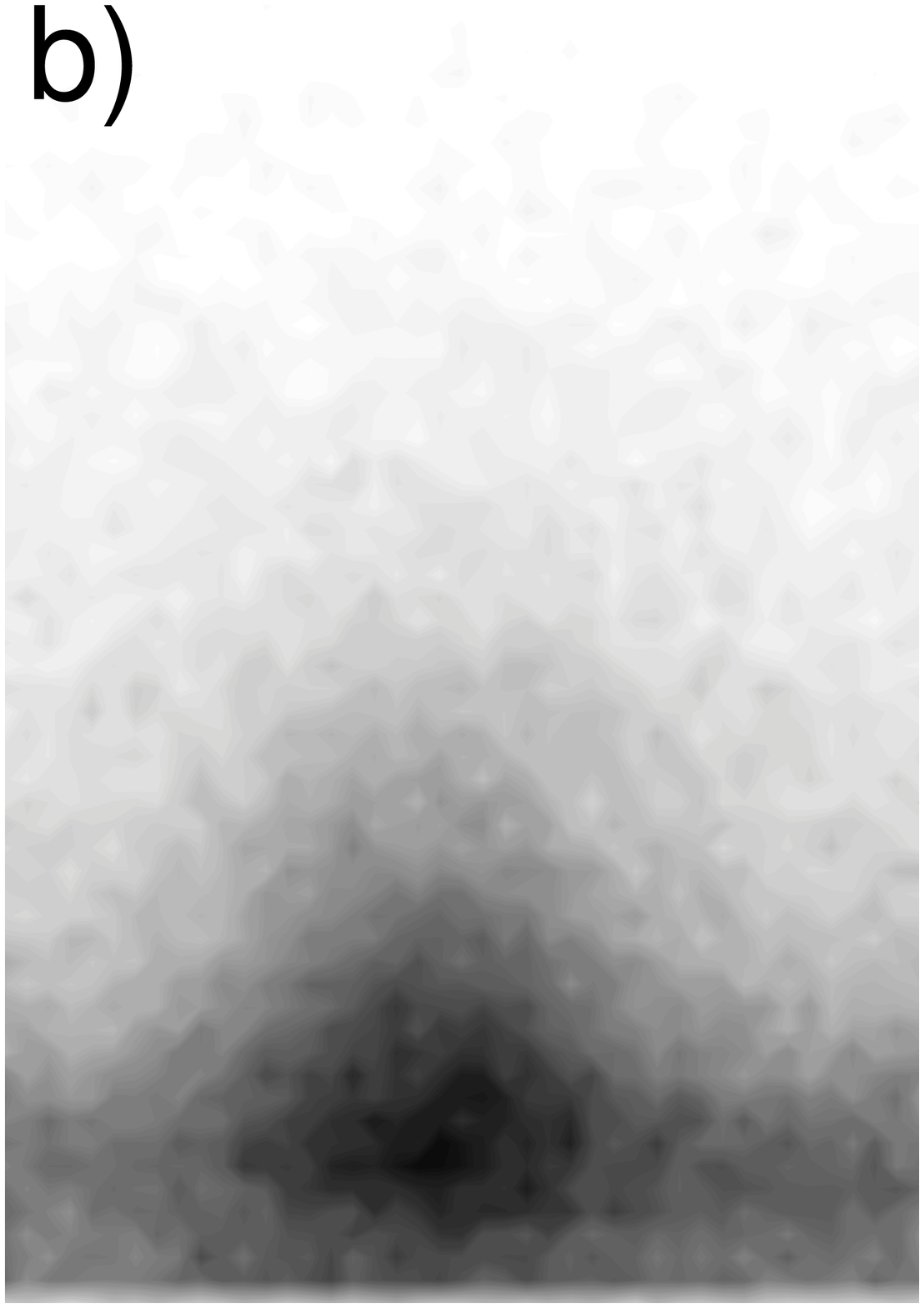}\hspace{-0.4cm}
\includegraphics[width=0.14\textwidth]{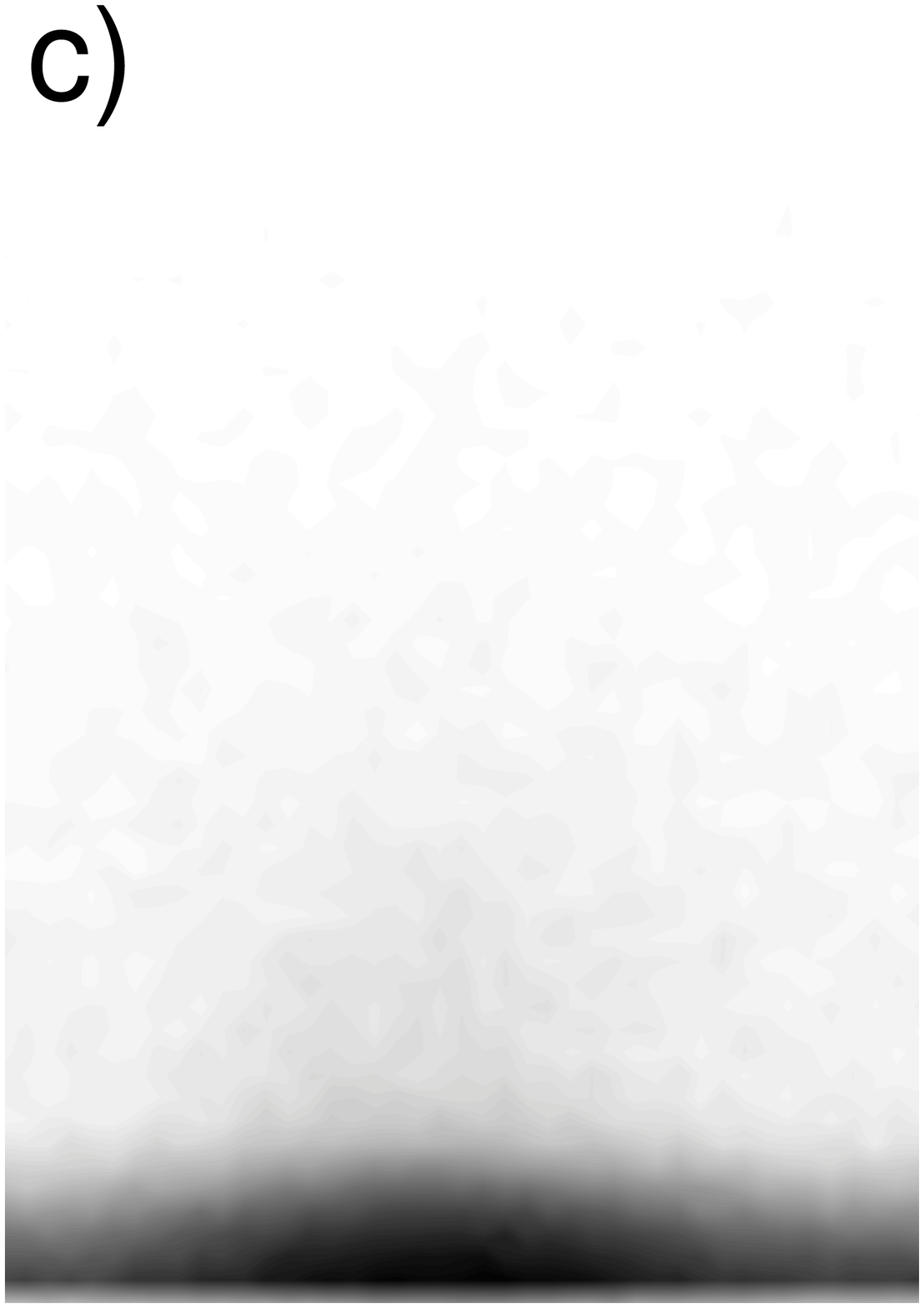}\hspace{-0.4cm}
\includegraphics[width=0.14\textwidth]{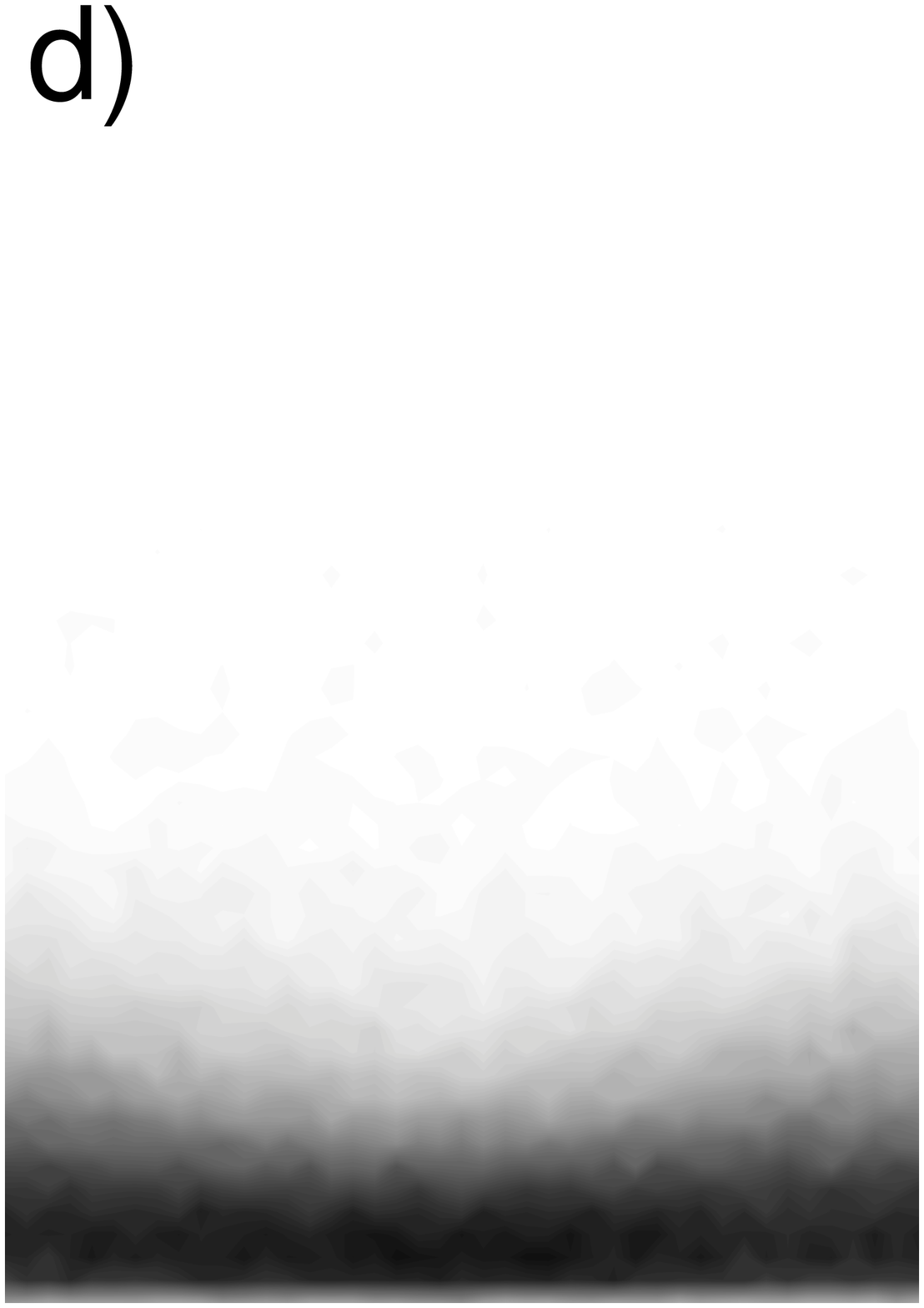}\hspace{-0.4cm}
\includegraphics[width=0.14\textwidth]{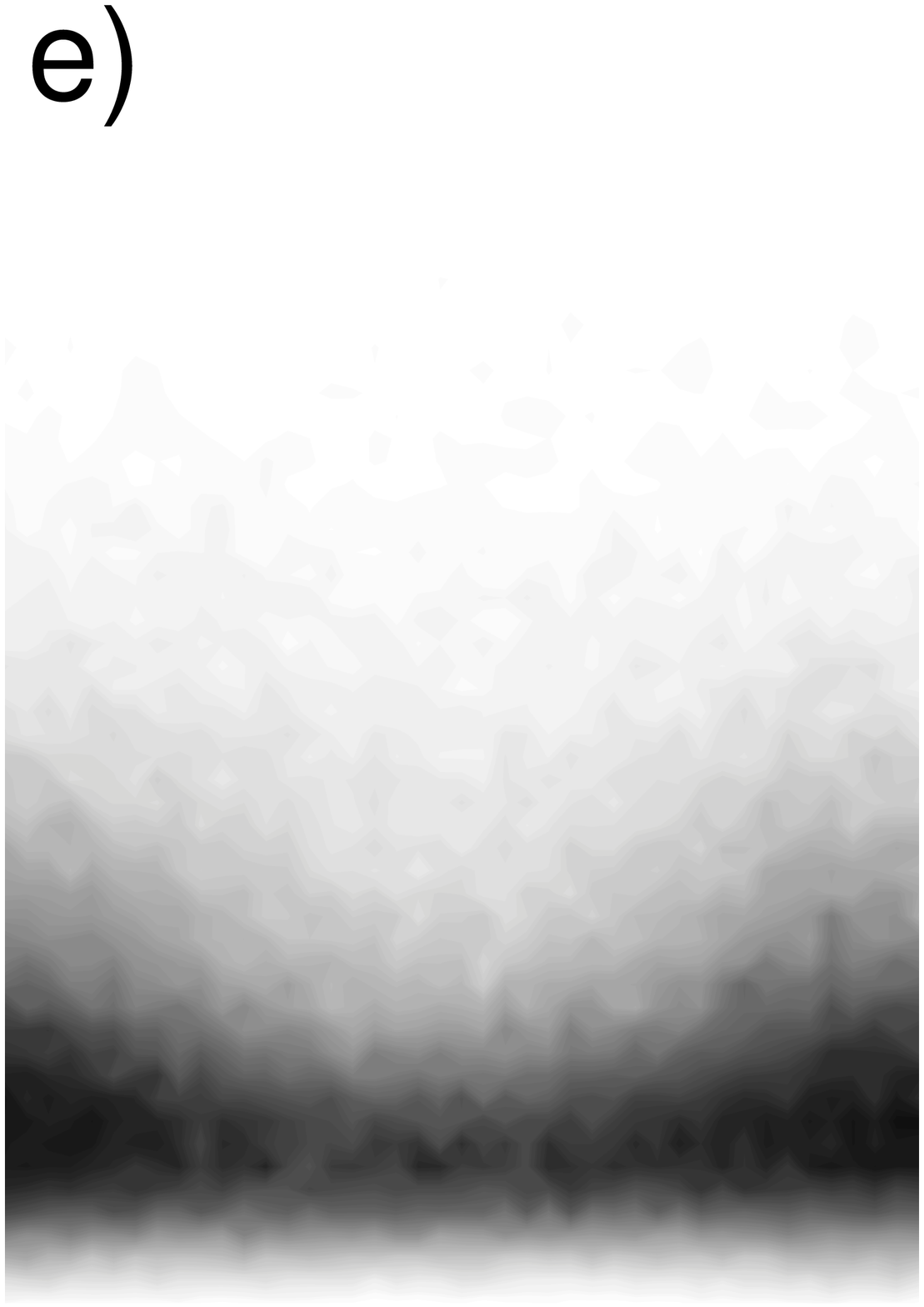}\hspace{-0.4cm}
\includegraphics[width=0.14\textwidth]{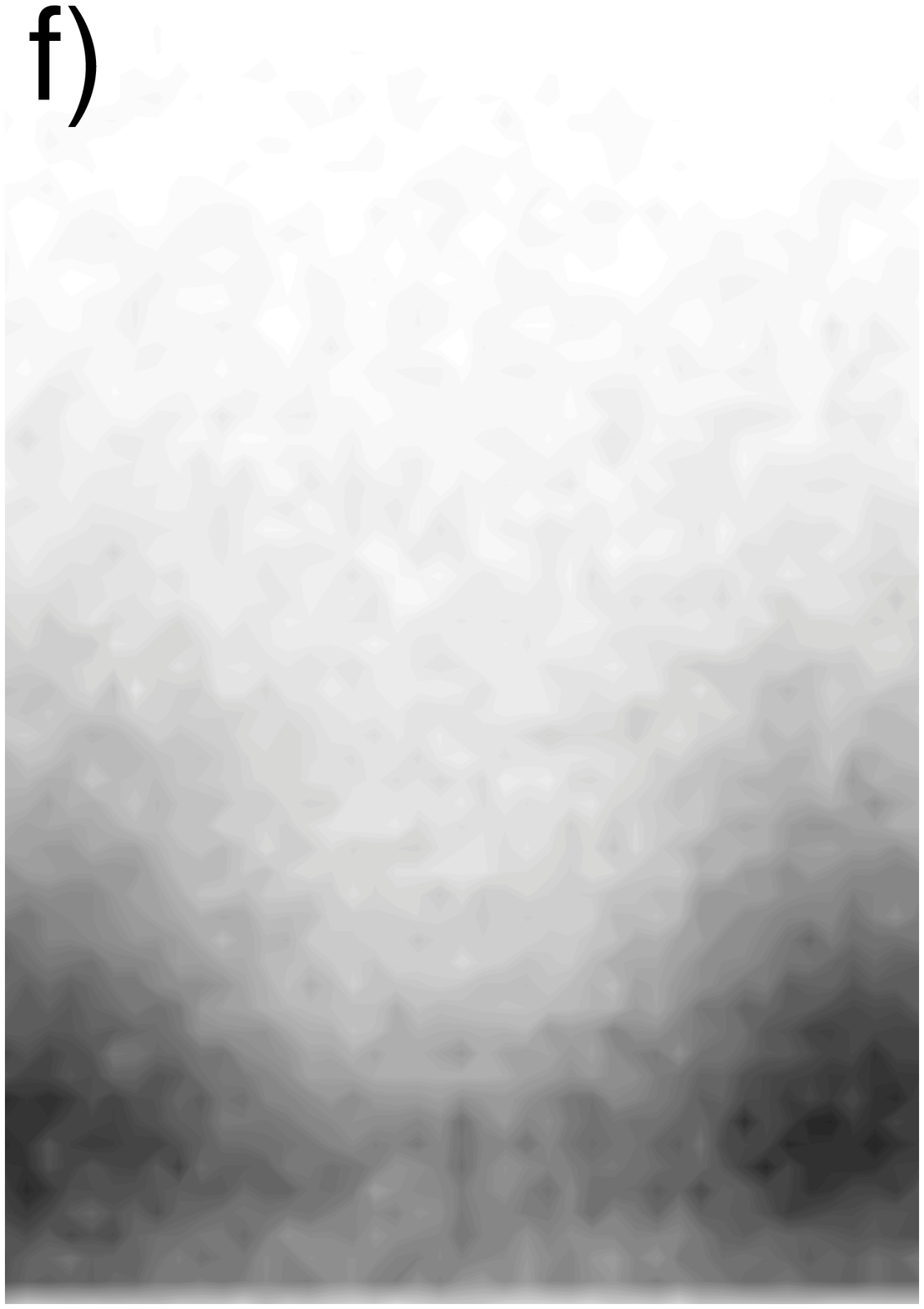}\hspace{-0.4cm}
\includegraphics[width=0.14\textwidth]{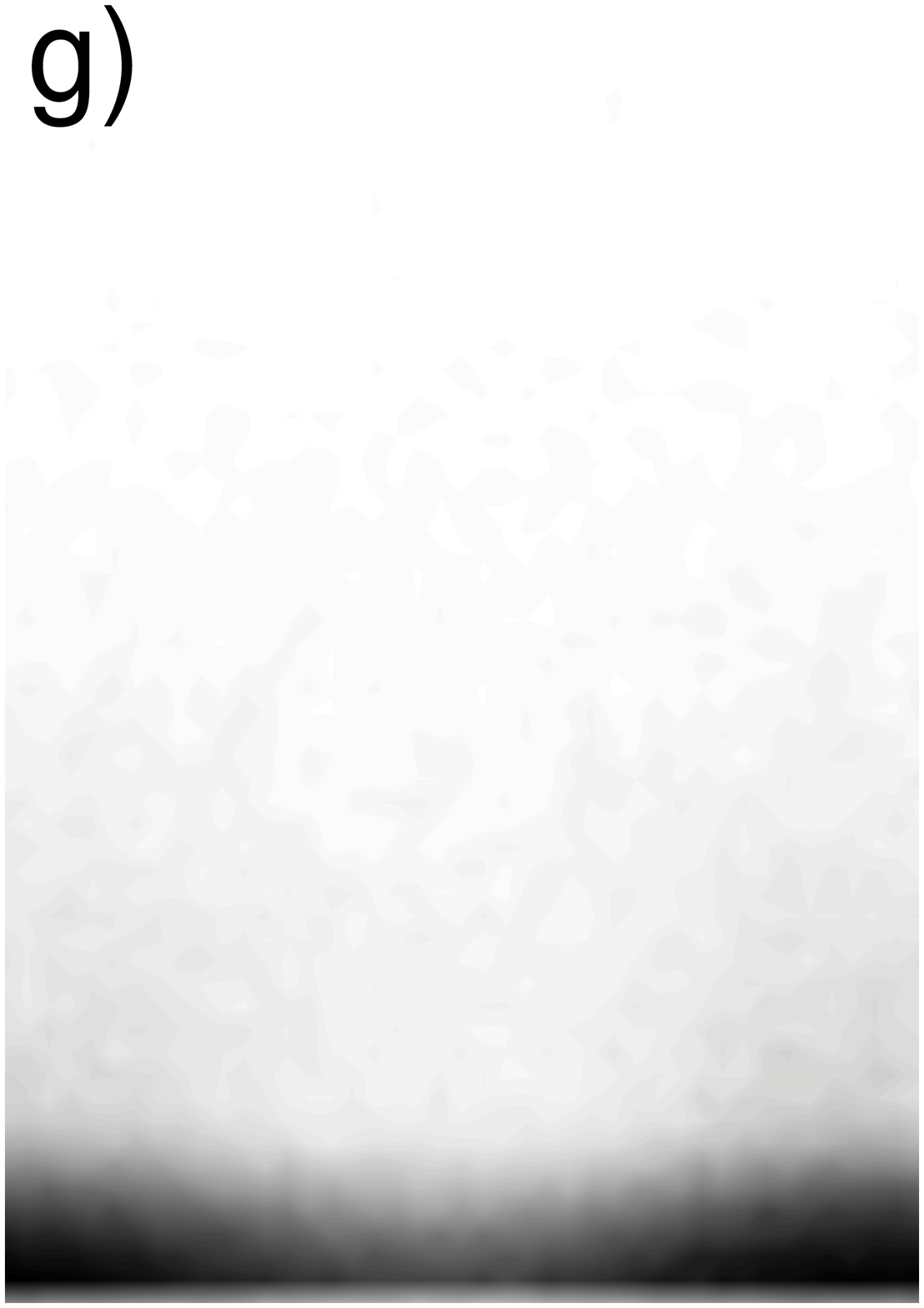}\hspace{-0.4cm}
\includegraphics[width=0.14\textwidth]{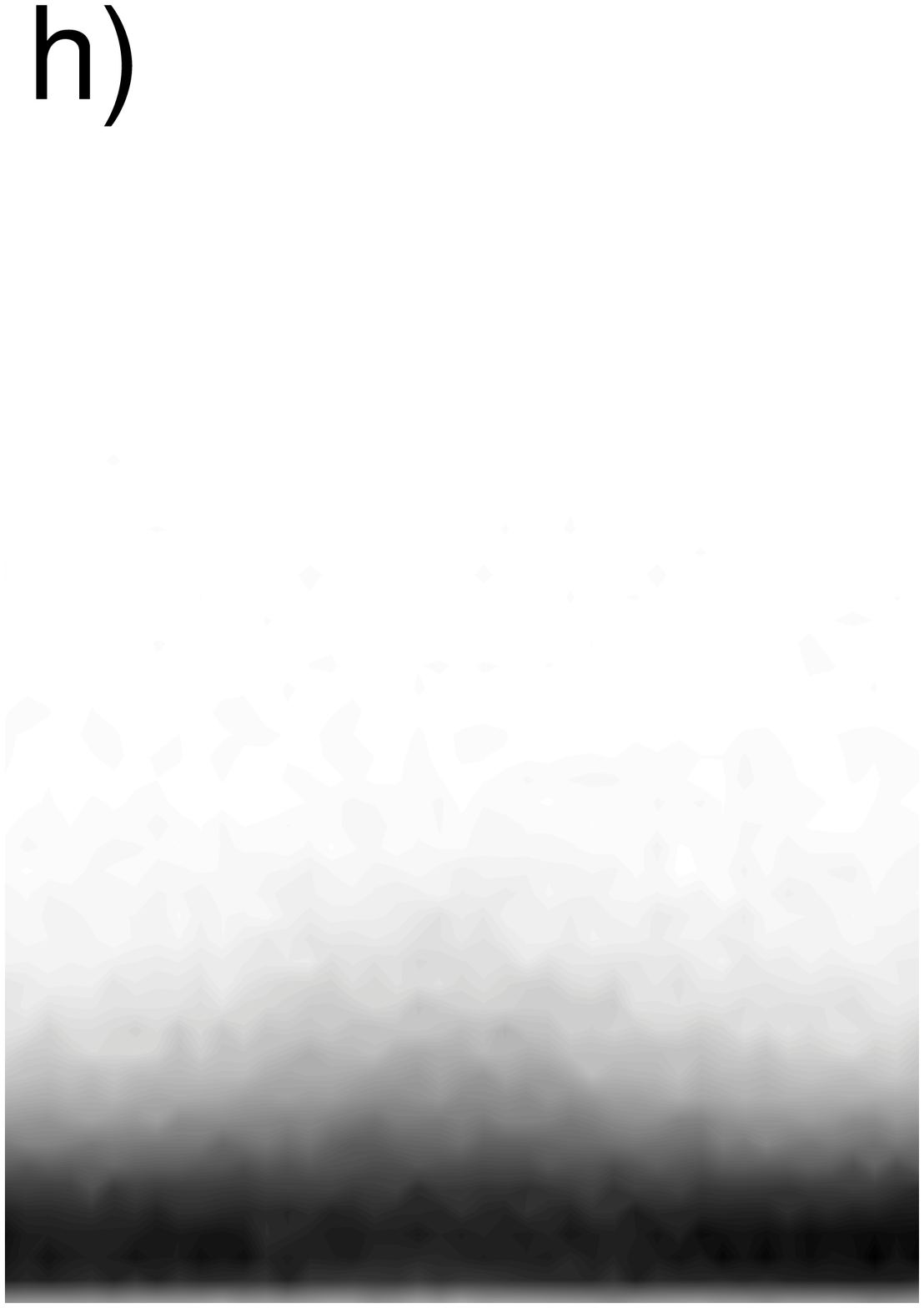}
\caption{Density field obtained by phase- and space-averaging particle
positions from the MD simulation. One single wavelength is
shown along eight equidistant phases (a) to (h). }
\label{fig:pattern}
\end{figure}

We refer the reader to Ref.\ \cite{CarrilloPoeschelSaluena:2008} regarding the
details of the averaging procedure applied to the MD sequence, here consisting
of 1,000 cycles, which leads to the averaged (2D) MD hydrodynamic fields for the
density (packing fraction) (Fig. \ref{fig:pattern}), linear momentum and thermal
energy. From the latter, the temperature field is also obtained. These are
compared to the corresponding ones generated by the two hydrodynamic
simulations.

\begin{figure}
 \begin{center}
 \subfloat[$t=0$]{\includegraphics[width=\halfcolumn]{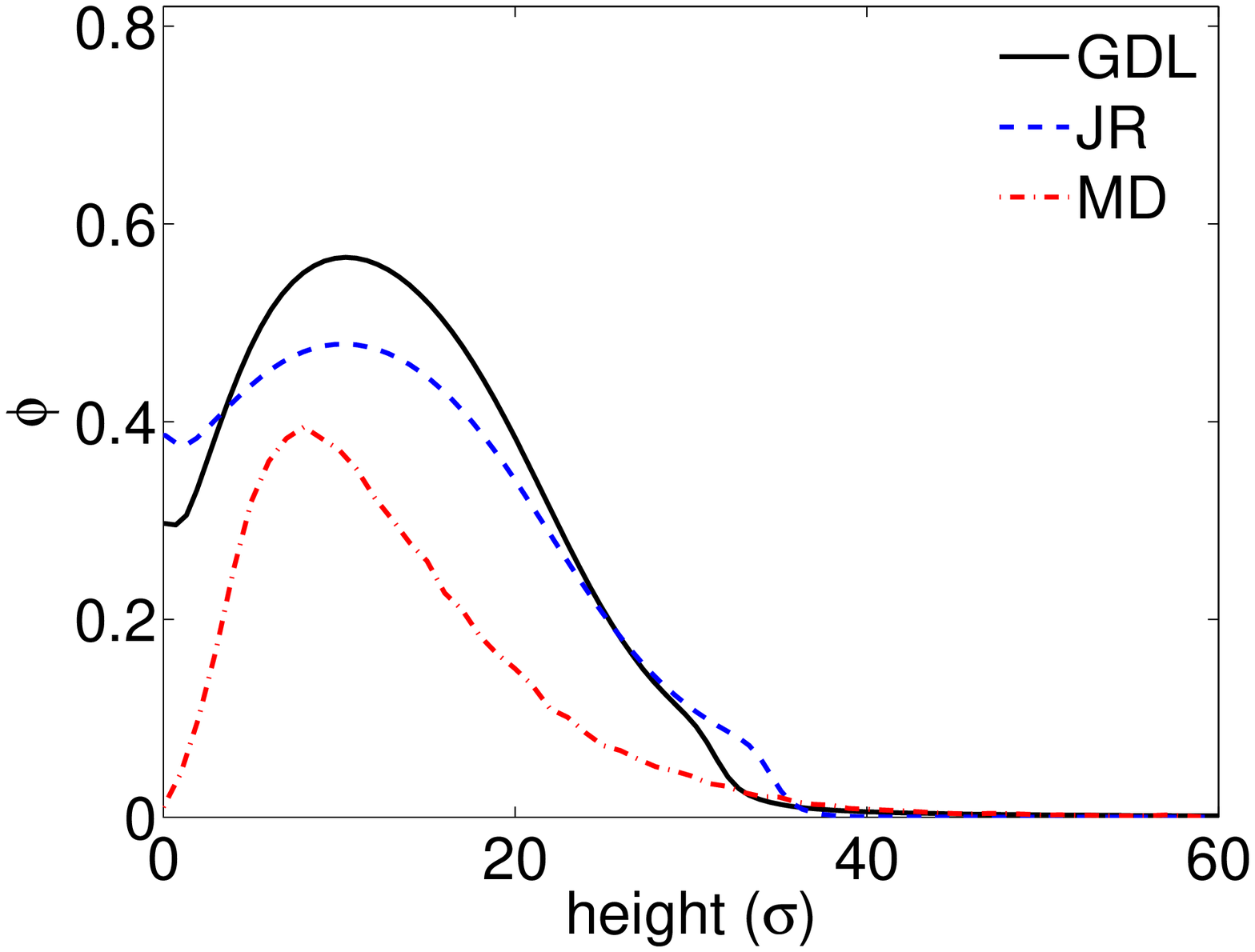}}
 \subfloat[$t=\nicefrac{1}{4} f^{-1}$]{\includegraphics[width=\halfcolumn]{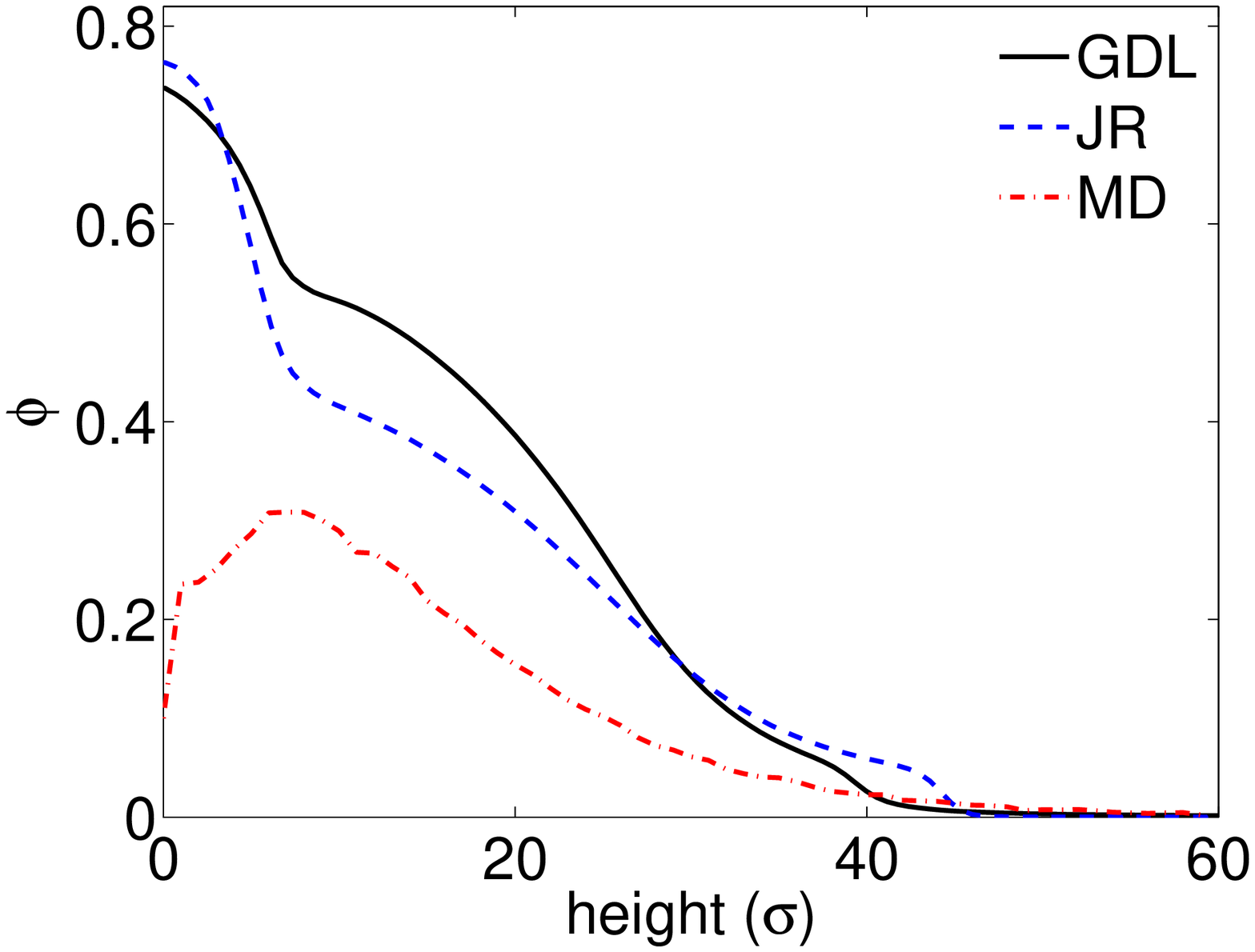}}

 \subfloat[$t=\nicefrac{2}{4} f^{-1}$]{\includegraphics[width=\halfcolumn]{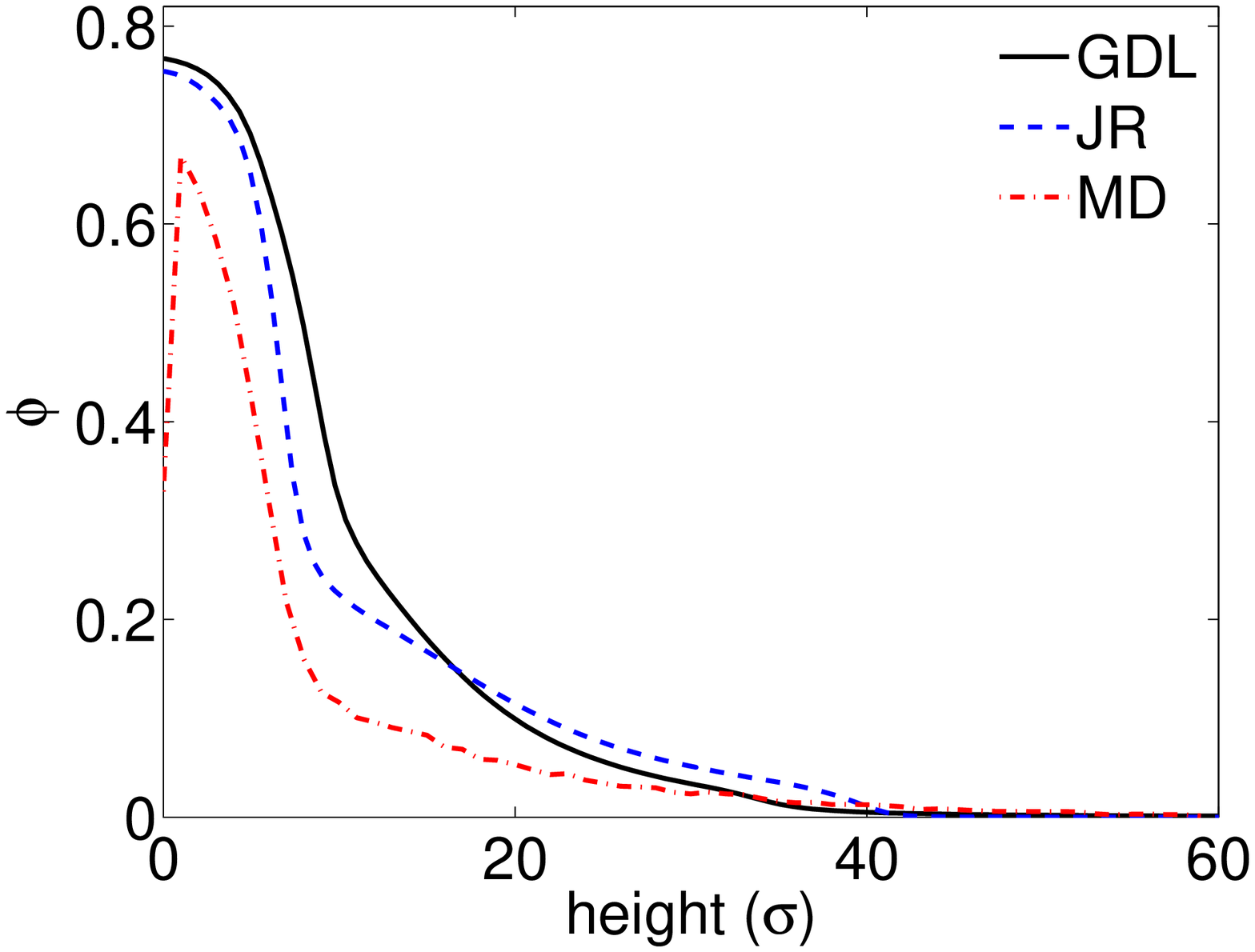}}
 \subfloat[$t=\nicefrac{3}{4} f^{-1}$]{\includegraphics[width=\halfcolumn]{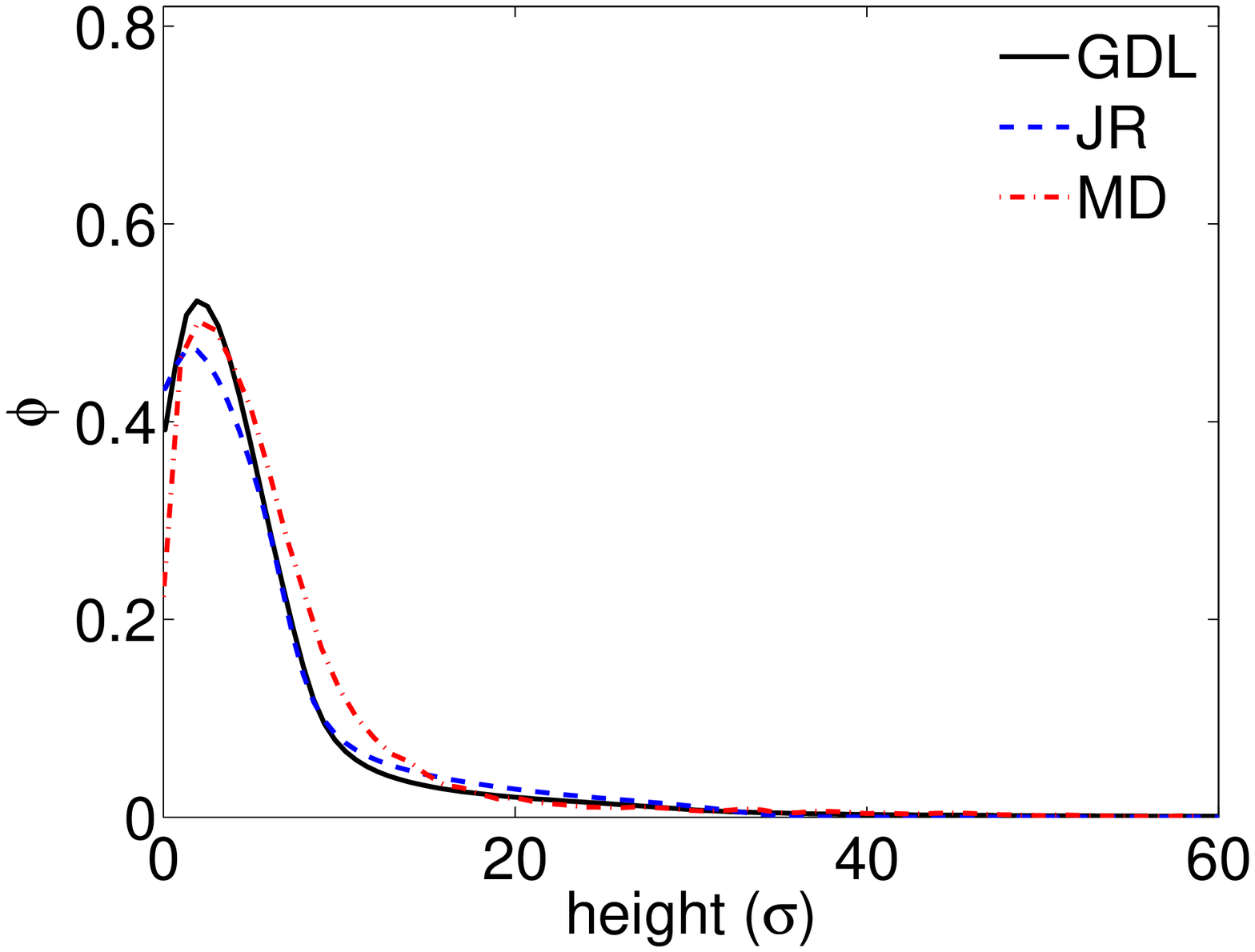}}\\

 \subfloat[$t=\nicefrac{4}{4} f^{-1}$]{\includegraphics[width=\halfcolumn]{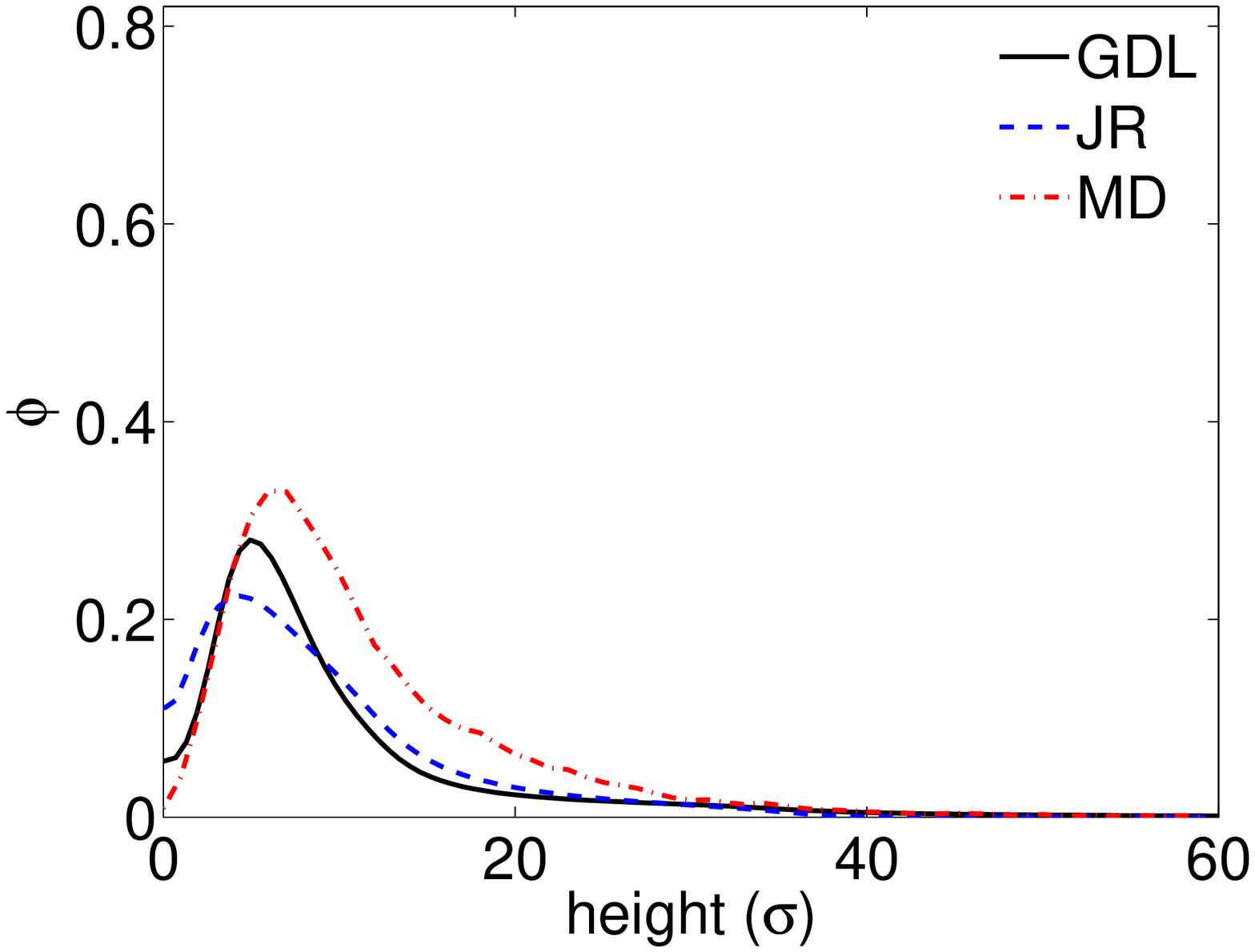}}
 \subfloat[$t=\nicefrac{5}{4} f^{-1}$]{\includegraphics[width=\halfcolumn]{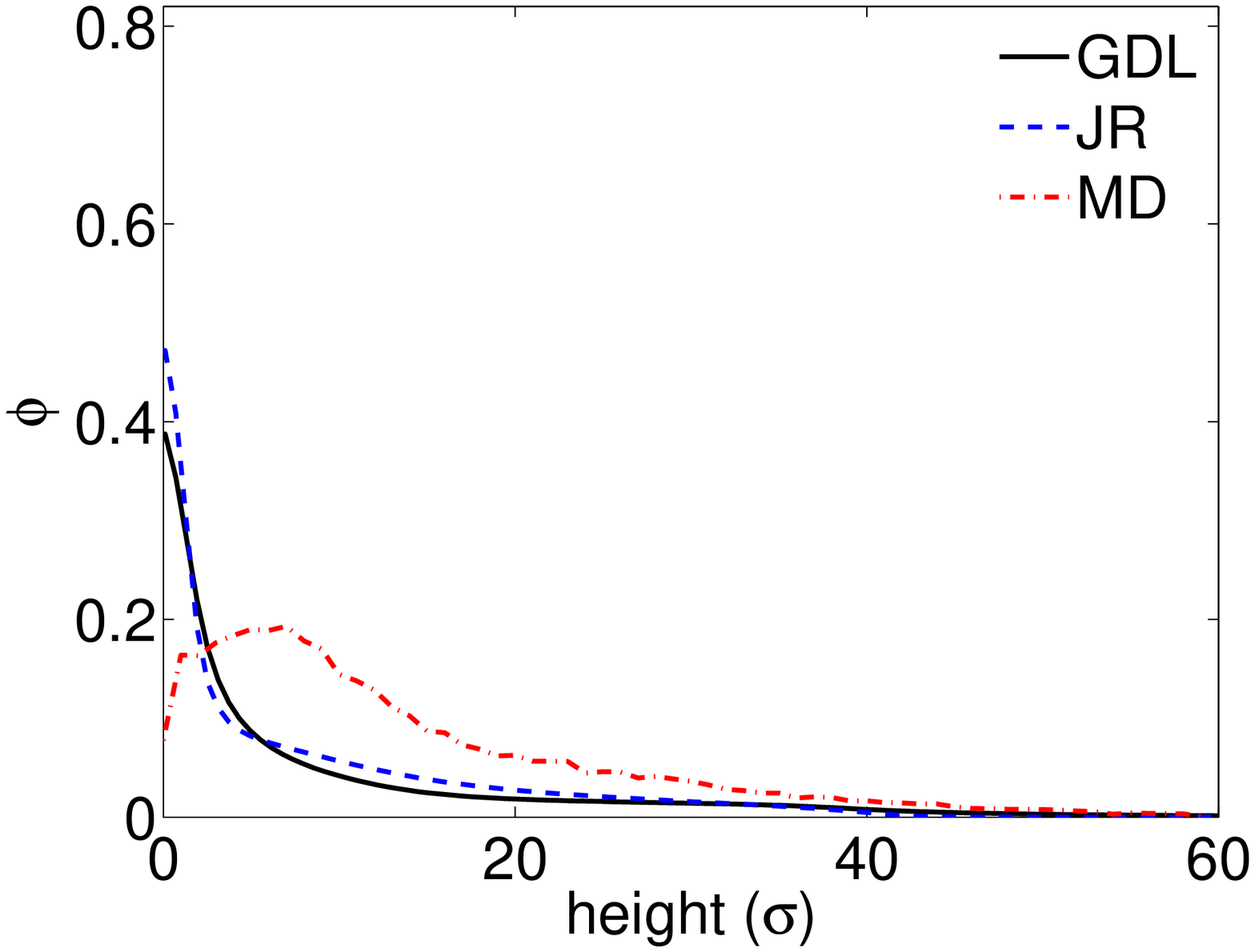}}

 \subfloat[$t=\nicefrac{6}{4} f^{-1}$]{\includegraphics[width=\halfcolumn]{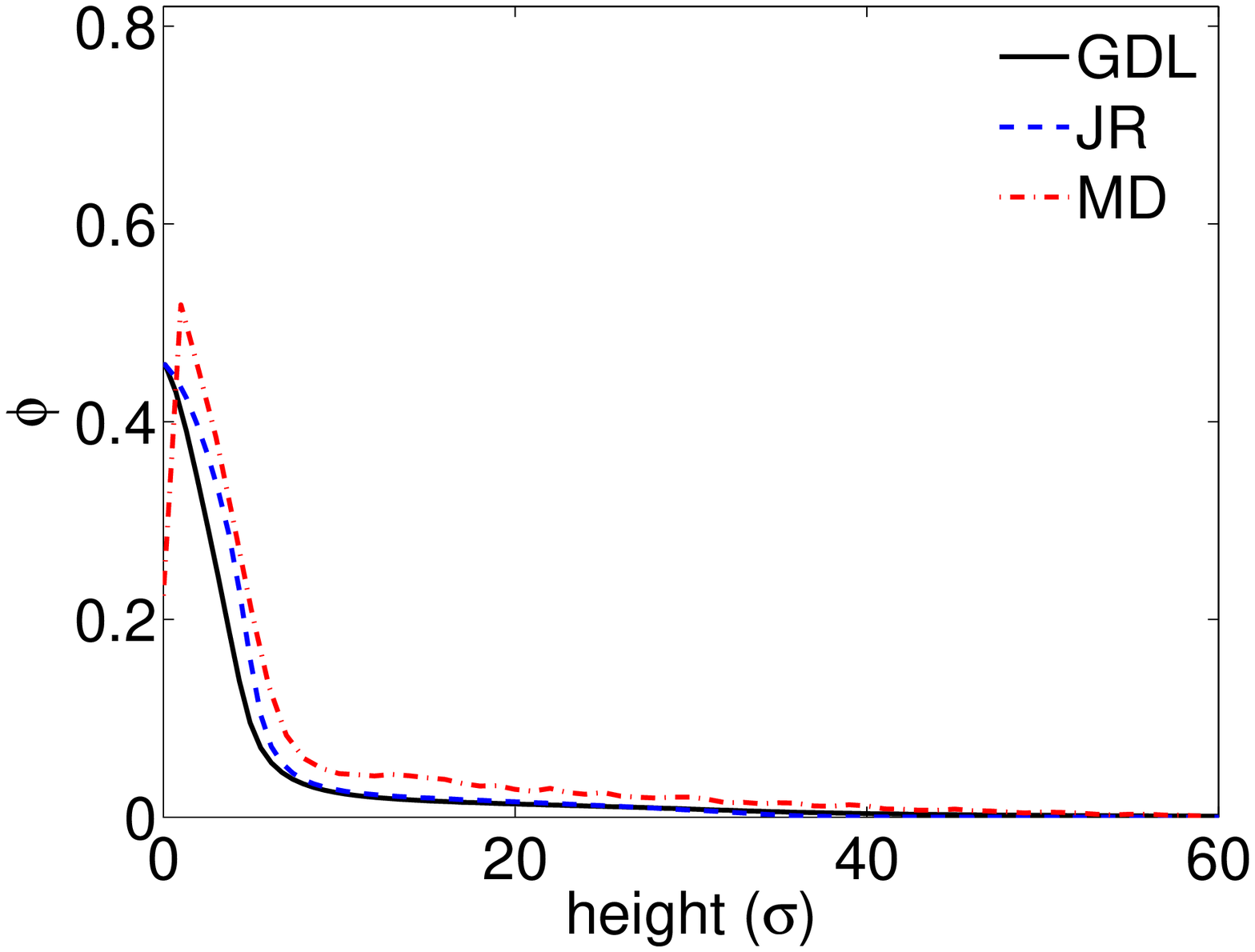}}
 \subfloat[$t=\nicefrac{7}{4} f^{-1}$]{\includegraphics[width=\halfcolumn]{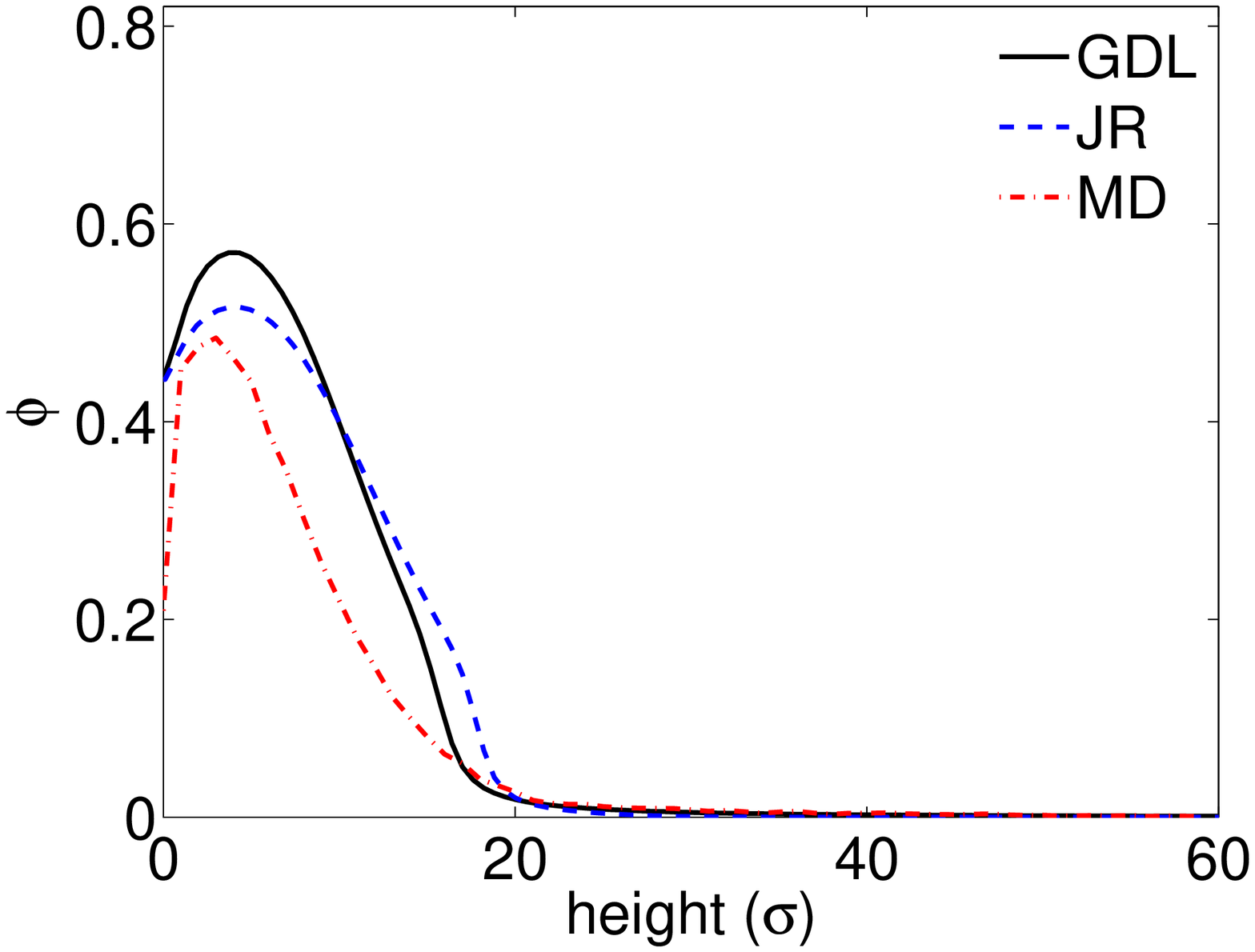}}
\end{center}
\caption{(color online) The  profiles of the packing fraction
($\phi$) as a function of height (in units of $\sigma$) at
selected times over two oscillation periods. For time evolution of
the profiles see \cite{online}.} \label{fig:xnu}
\end{figure}

\begin{figure}
 \begin{center}
 \subfloat[$t=0$]{\includegraphics[width=\halfcolumn]{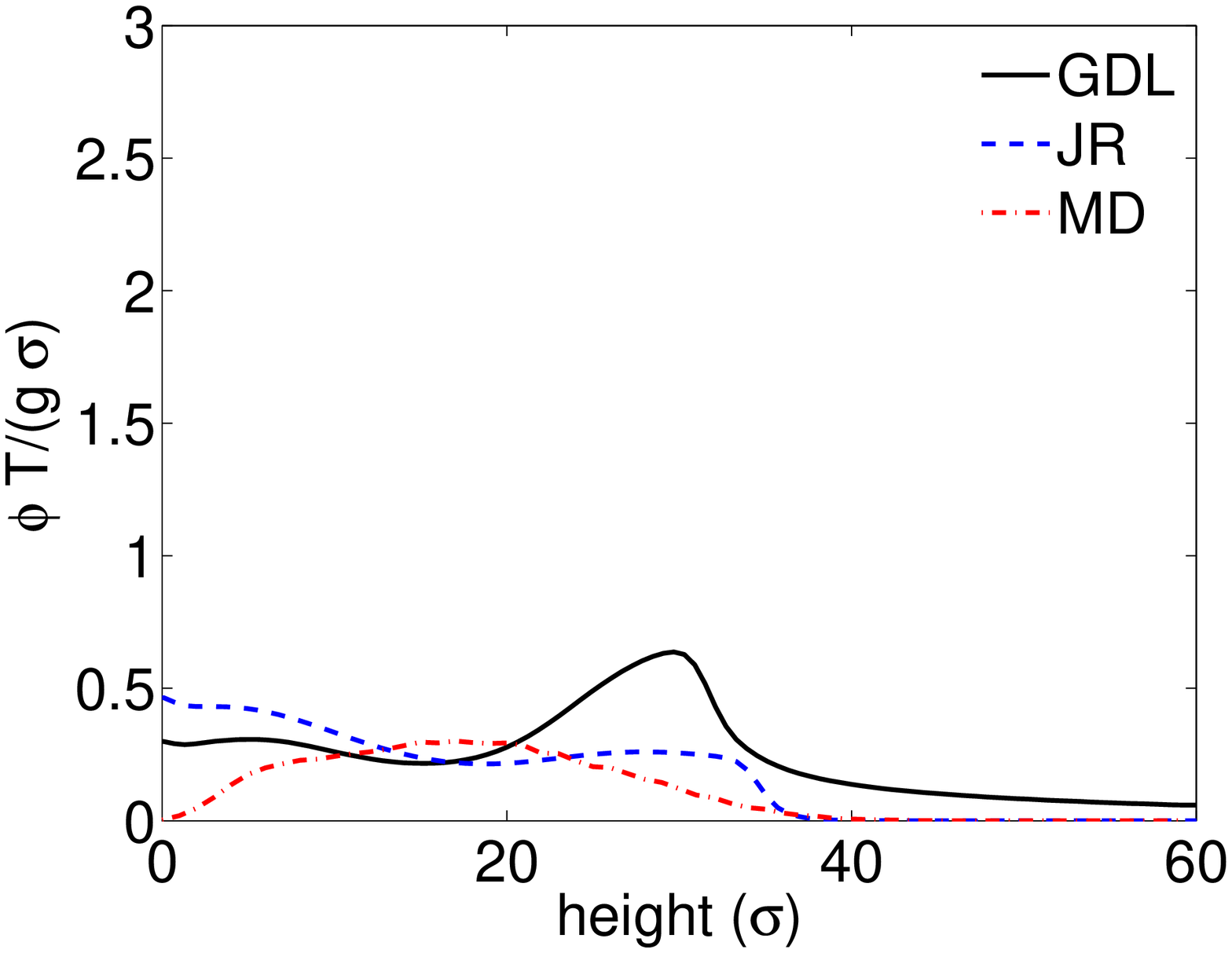}}
 \subfloat[$t=\nicefrac{1}{4} f^{-1}$]{\includegraphics[width=\halfcolumn]{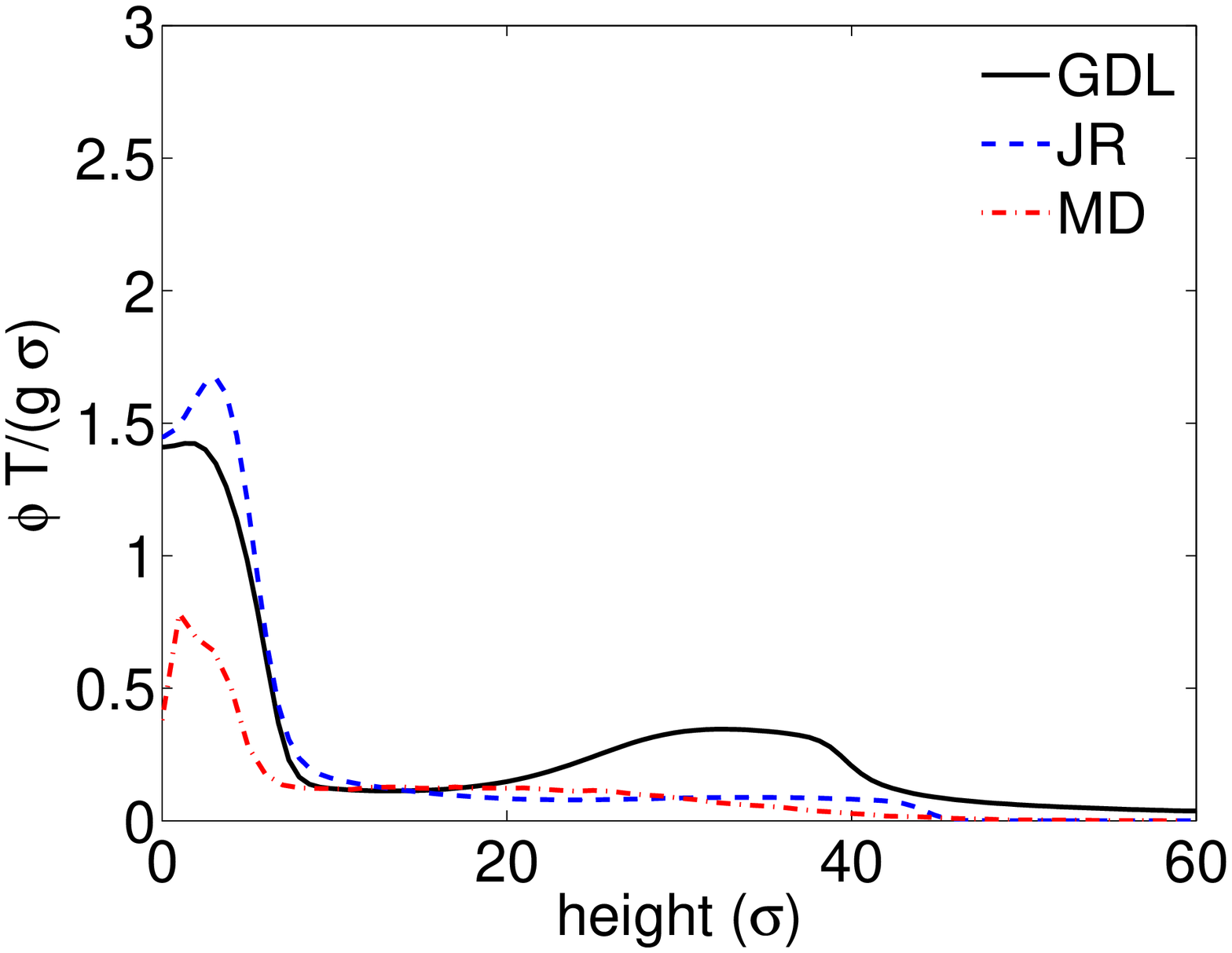}}

 \subfloat[$t=\nicefrac{2}{4} f^{-1}$]{\includegraphics[width=\halfcolumn]{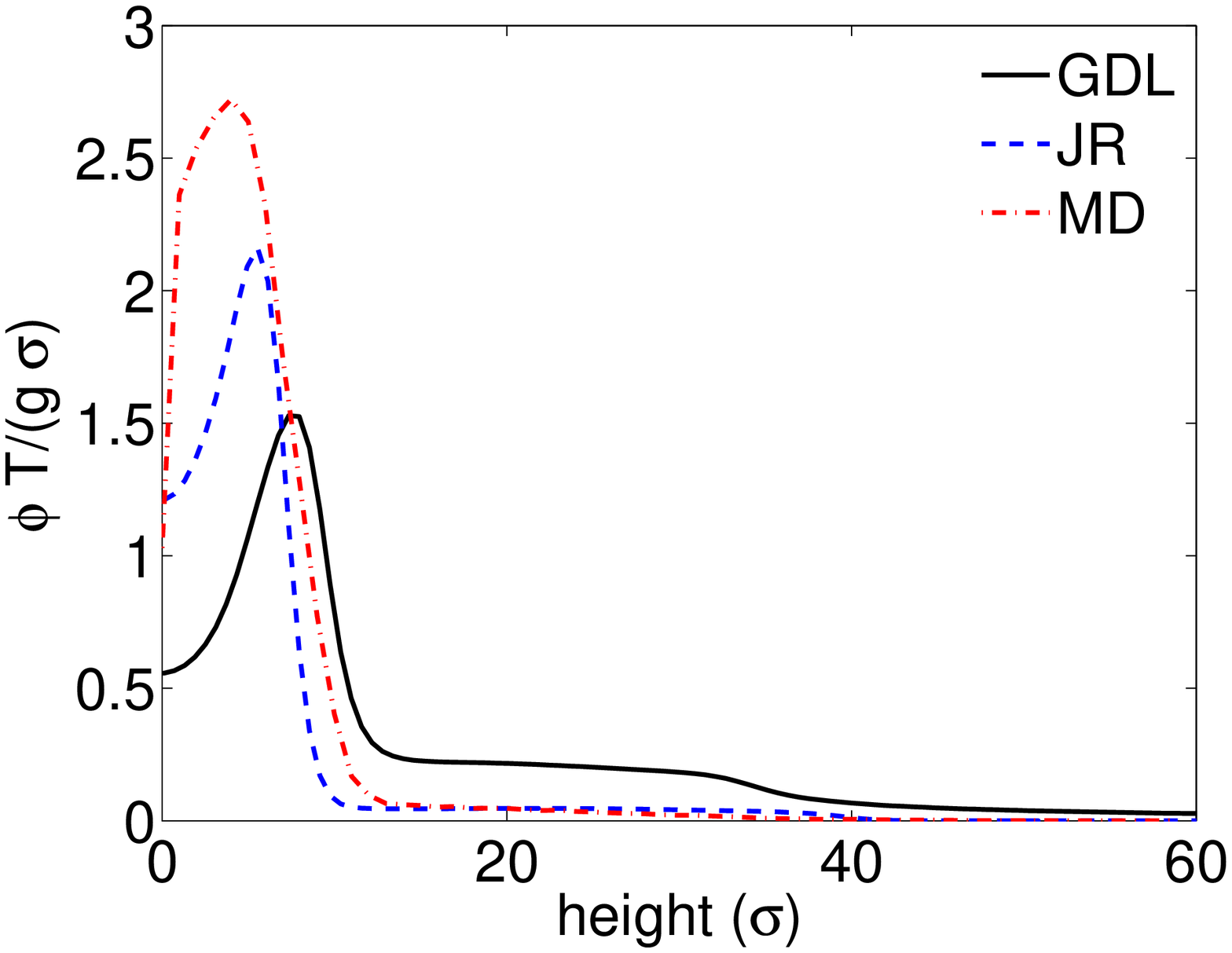}}
 \subfloat[$t=\nicefrac{3}{4} f^{-1}$]{\includegraphics[width=\halfcolumn]{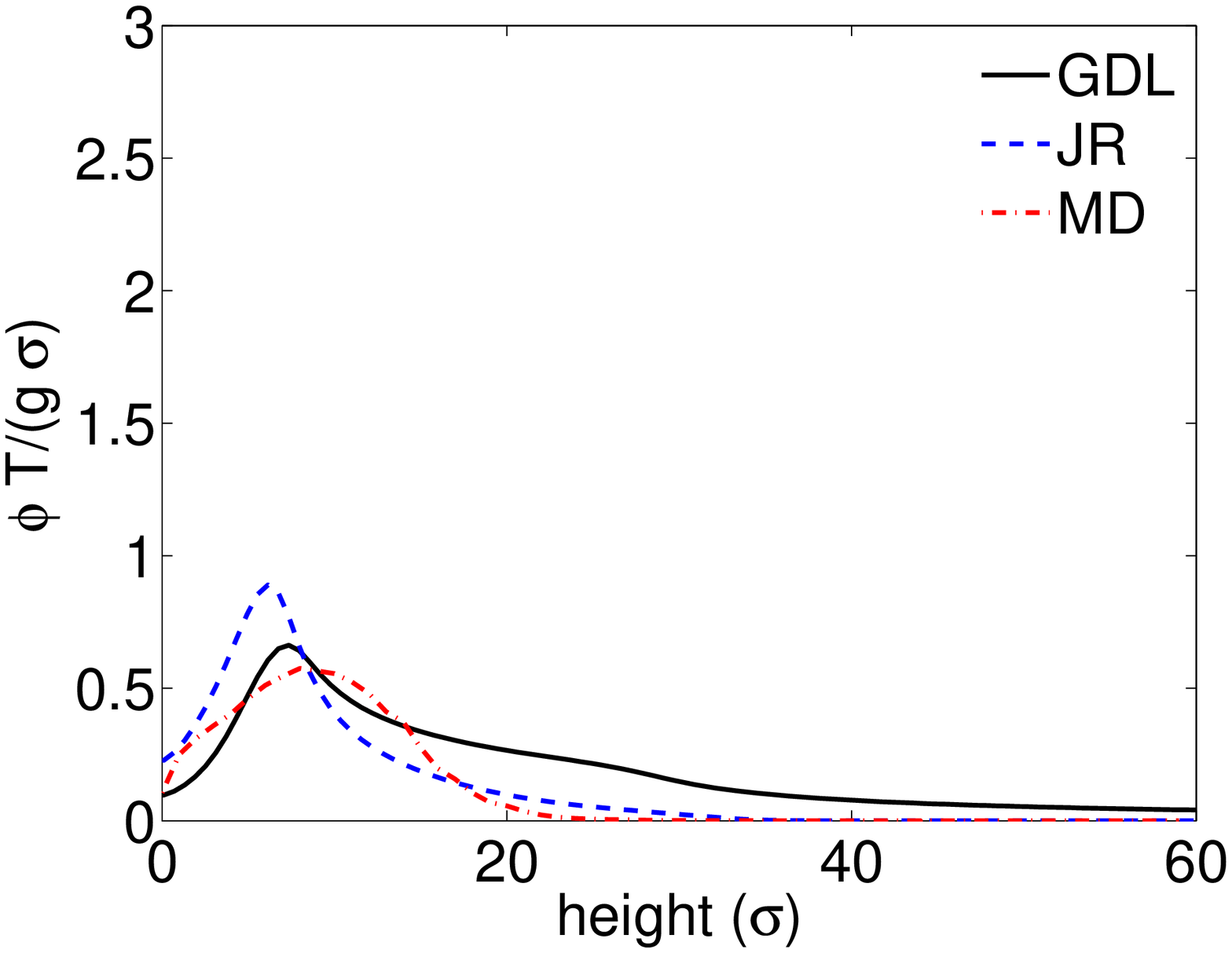}}

 \subfloat[$t=\nicefrac{4}{4} f^{-1}$]{\includegraphics[width=\halfcolumn]{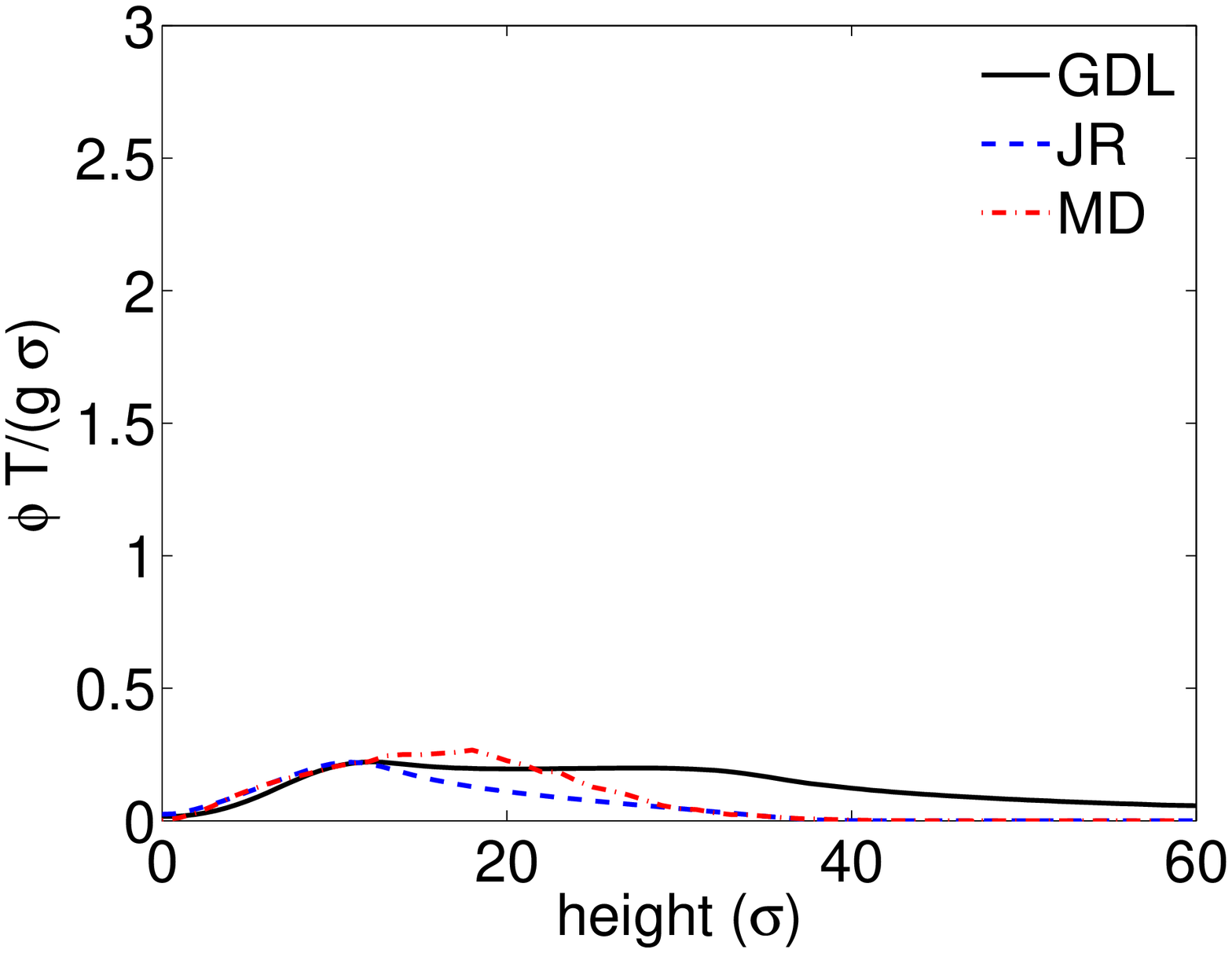}}
 \subfloat[$t=\nicefrac{5}{4} f^{-1}$]{\includegraphics[width=\halfcolumn]{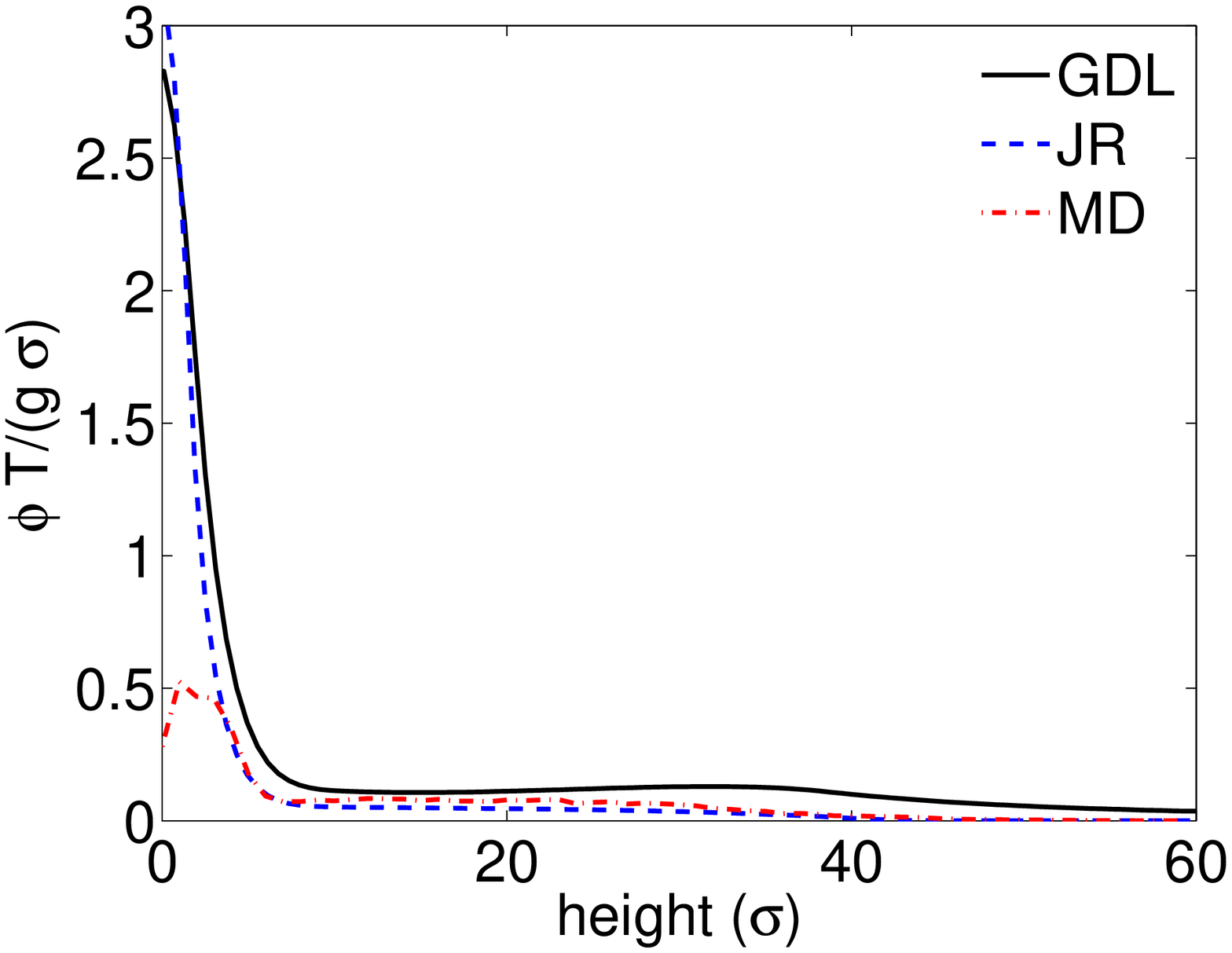}}

 \subfloat[$t=\nicefrac{6}{4} f^{-1}$]{\includegraphics[width=\halfcolumn]{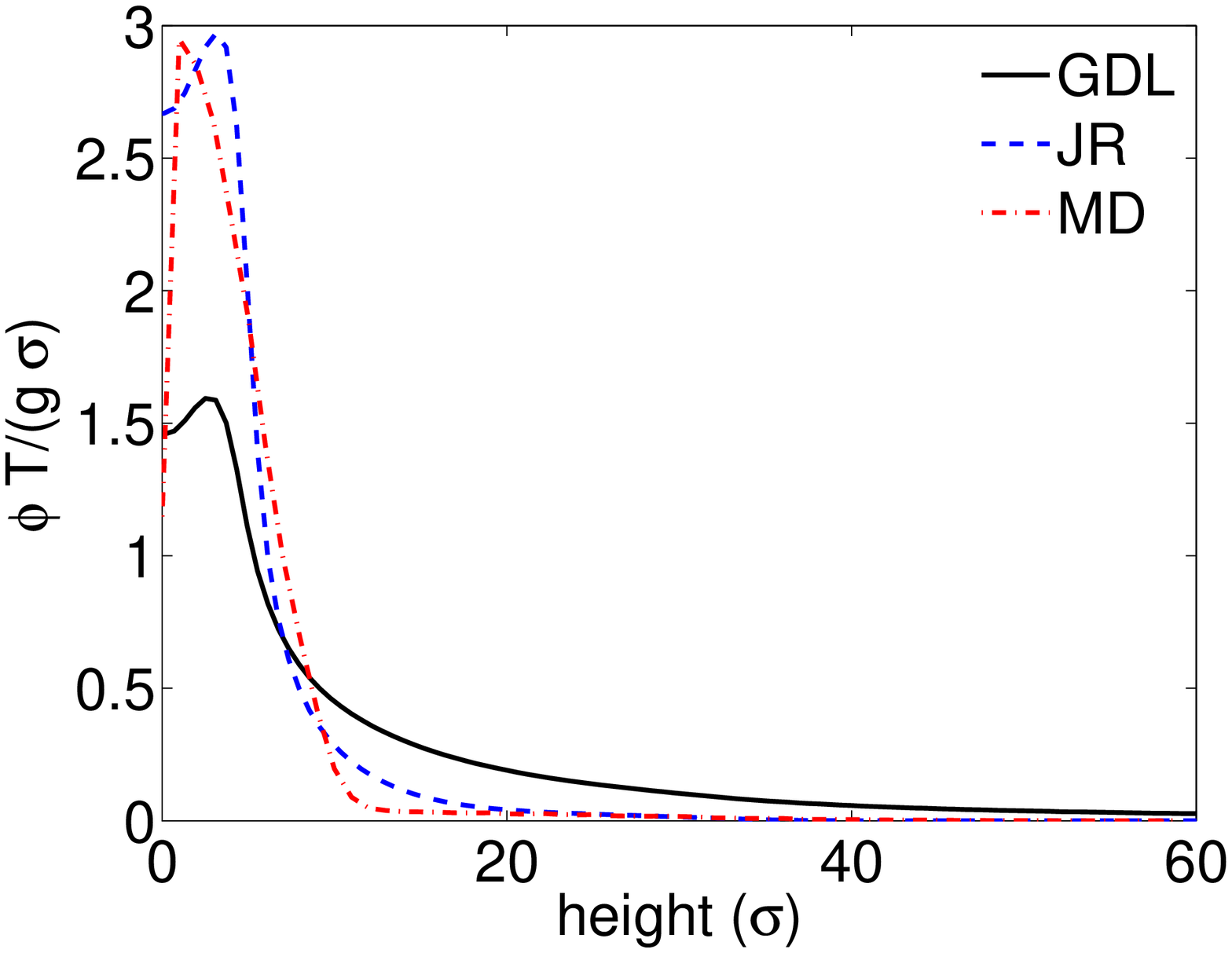}}
 \subfloat[$t=\nicefrac{7}{4} f^{-1}$]{\includegraphics[width=\halfcolumn]{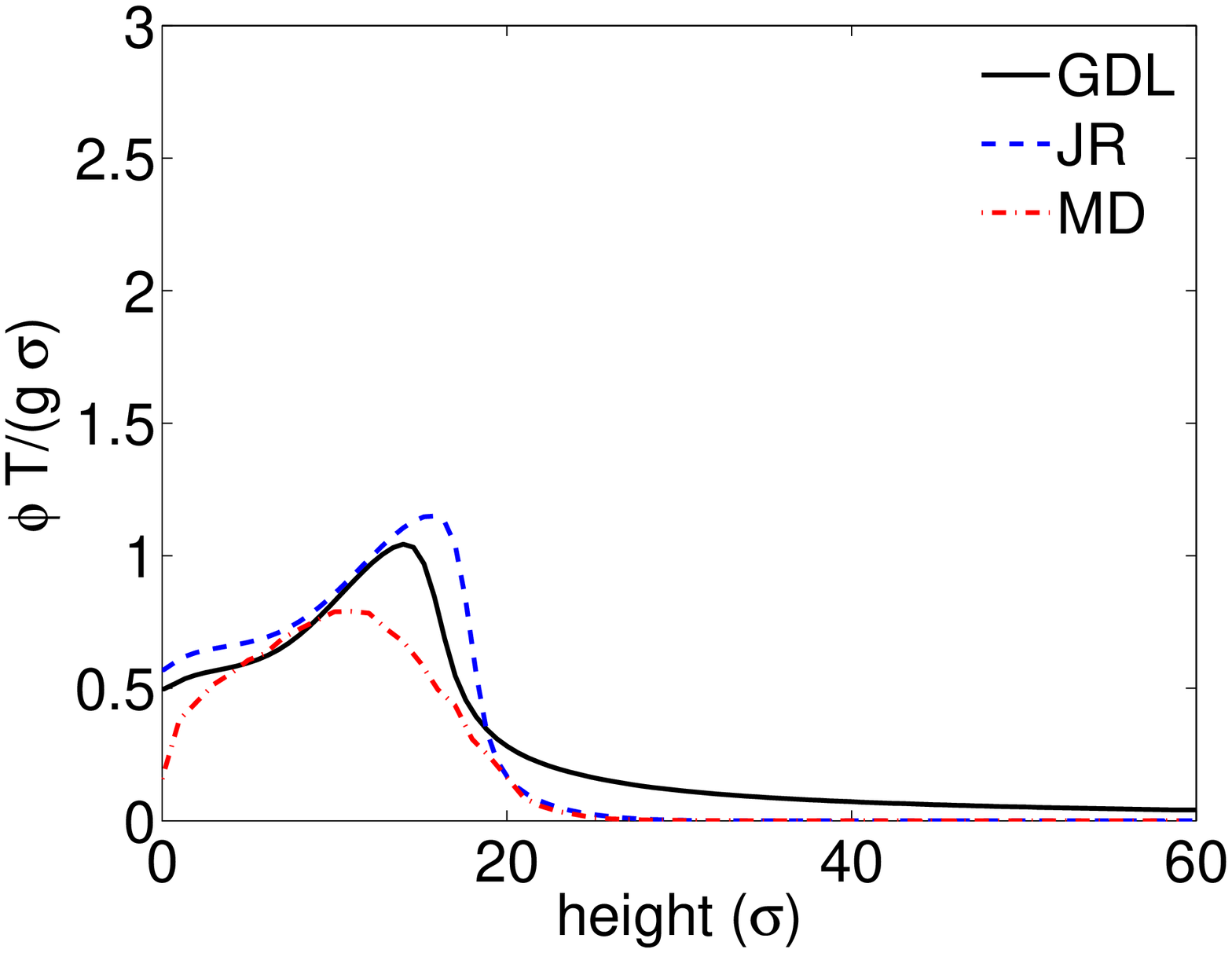}}

\end{center}
\caption{(color online) Scaled internal energy ($\phi T/(\sigma g)$) as a
function of height (in units of $\sigma$) at selected times over two oscillation
periods. For time evolution of the profiles see \cite{online}.}
\label{fig:energy}
\end{figure}

\begin{figure}
 \begin{center}
 \subfloat[$t=0$]{\includegraphics[width=\halfcolumn]{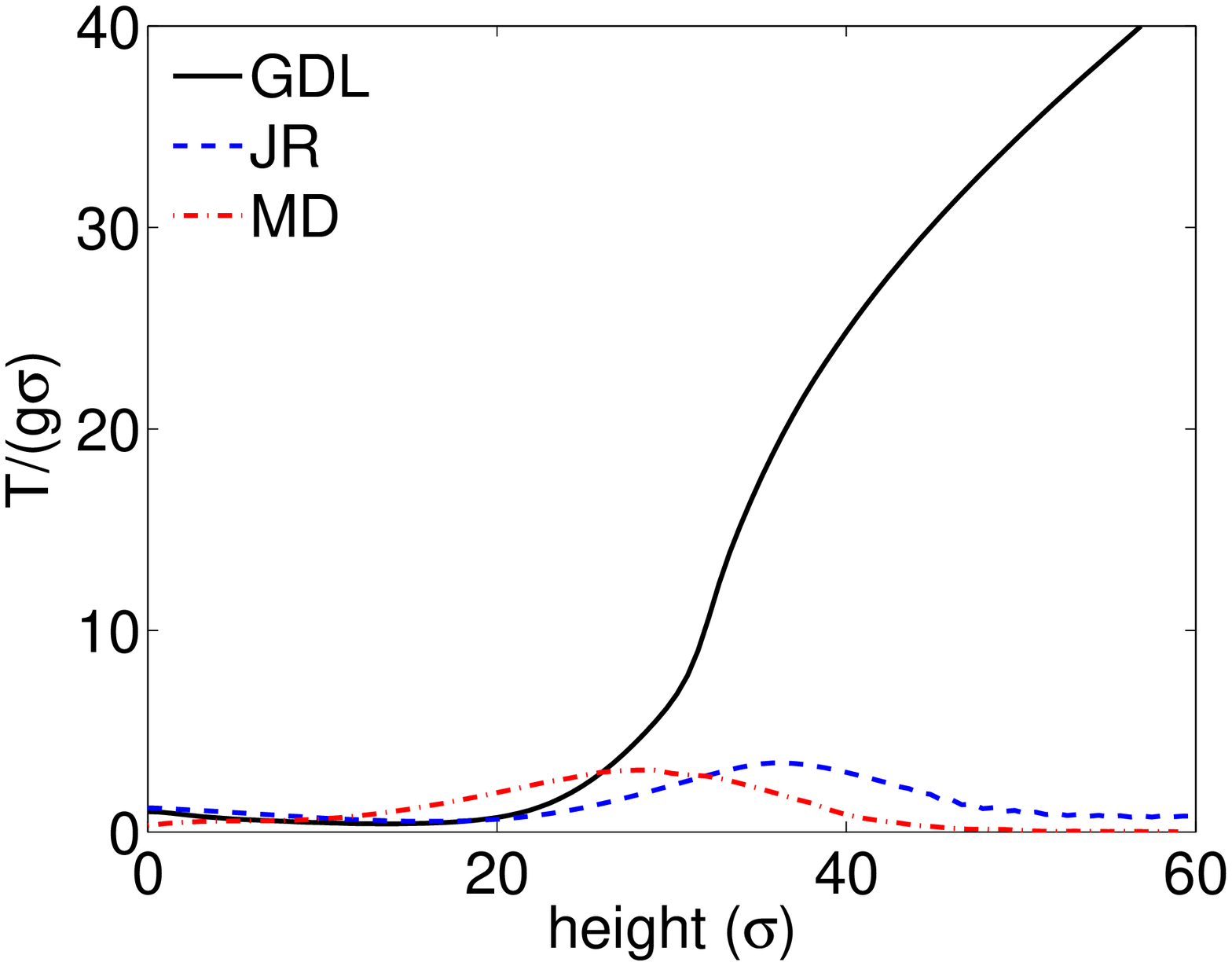}}
 \subfloat[$t=\nicefrac{1}{4} f^{-1}$]{\includegraphics[width=\halfcolumn]{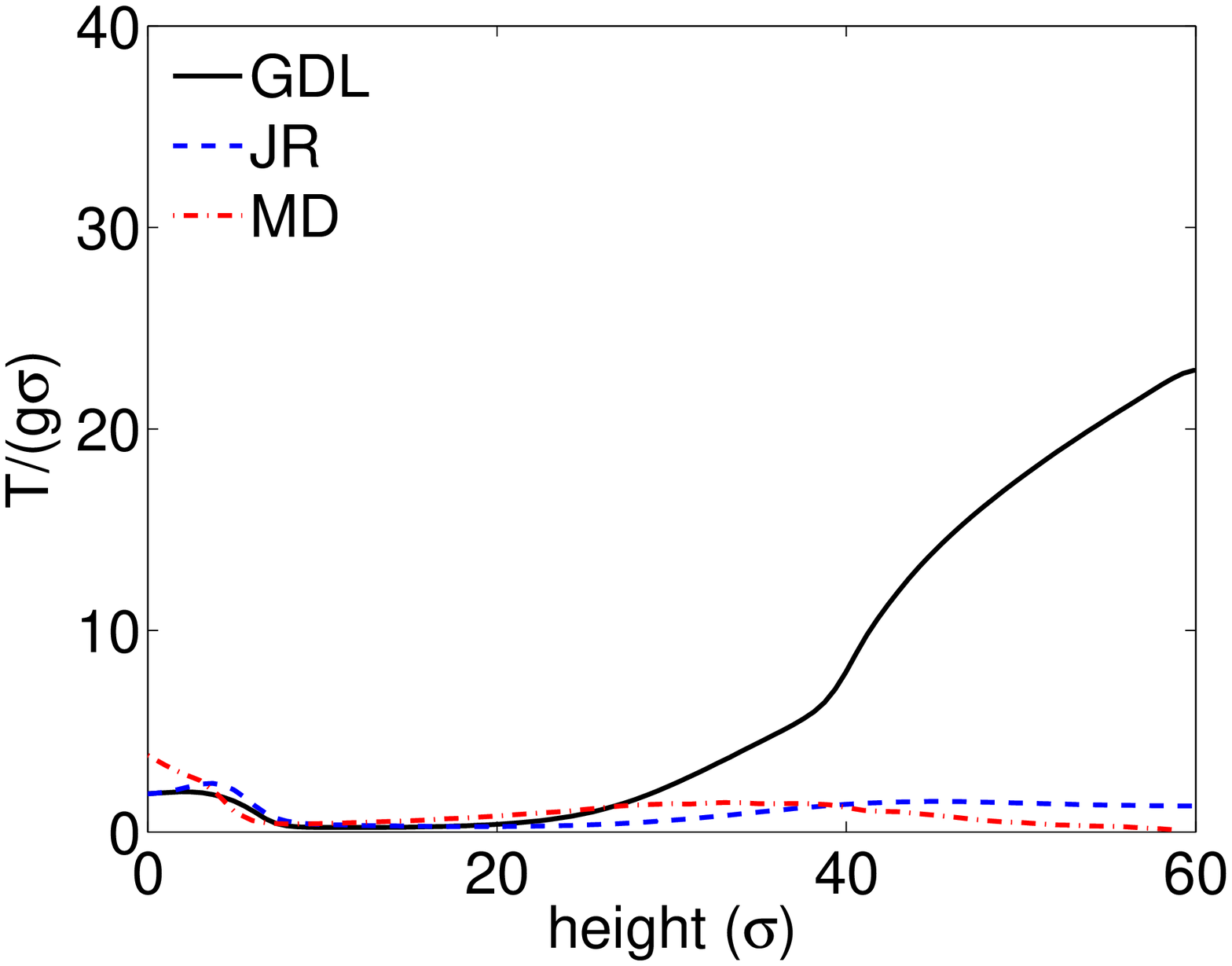}}

 \subfloat[$t=\nicefrac{2}{4} f^{-1}$]{\includegraphics[width=\halfcolumn]{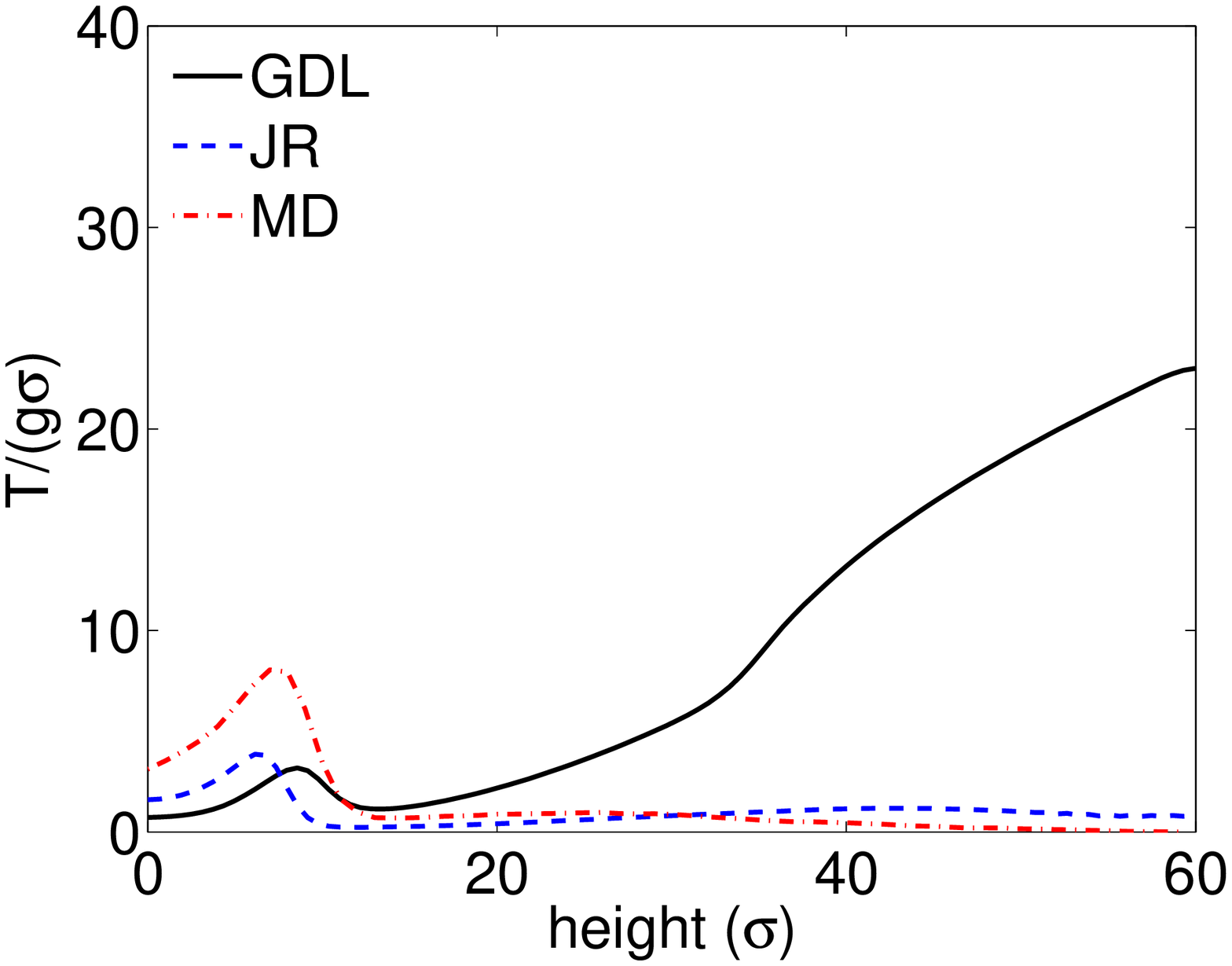}}
 \subfloat[$t=\nicefrac{3}{4} f^{-1}$]{\includegraphics[width=\halfcolumn]{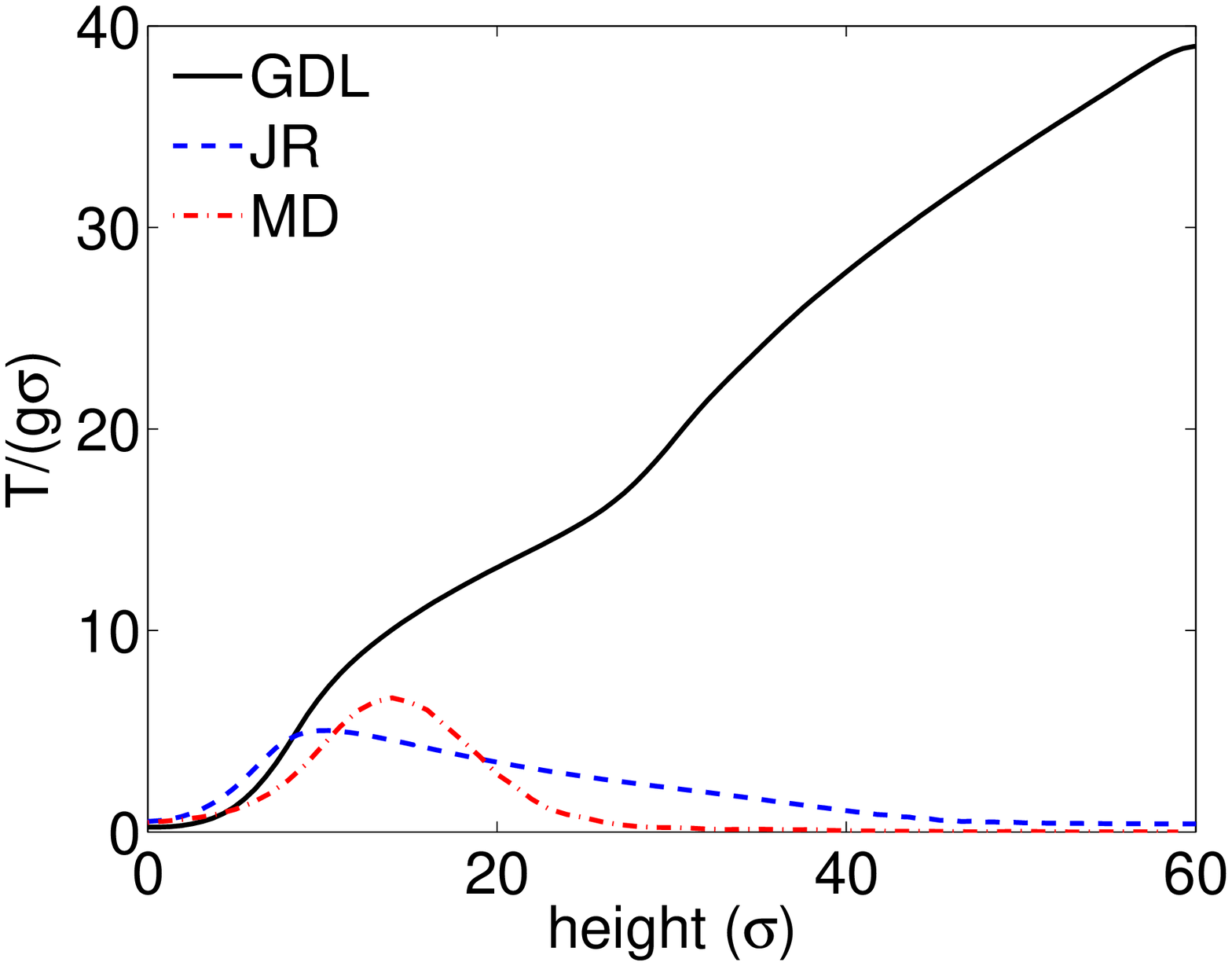}}

 \subfloat[$t=\nicefrac{4}{4} f^{-1}$]{\includegraphics[width=\halfcolumn]{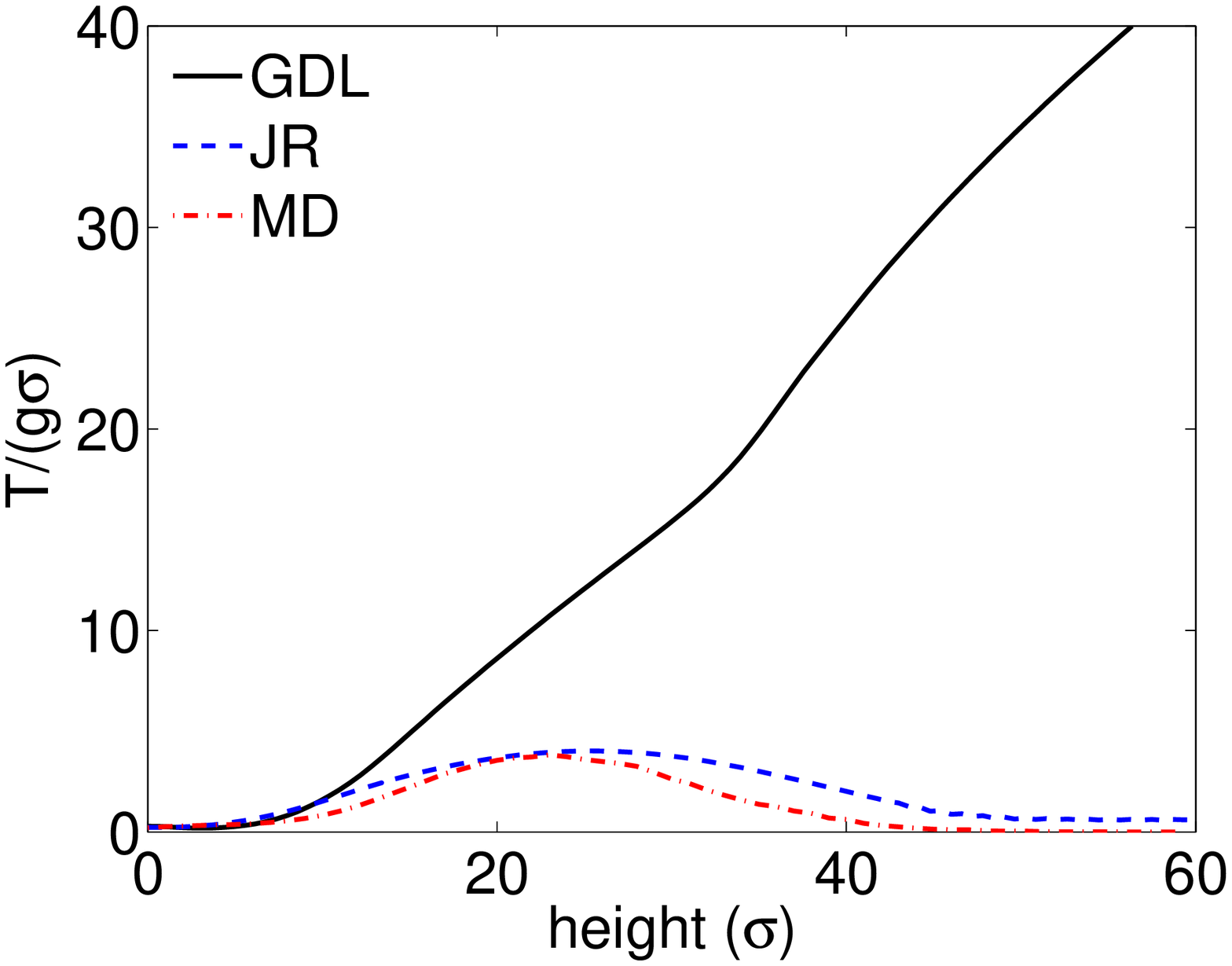}}
 \subfloat[$t=\nicefrac{5}{4} f^{-1}$]{\includegraphics[width=\halfcolumn]{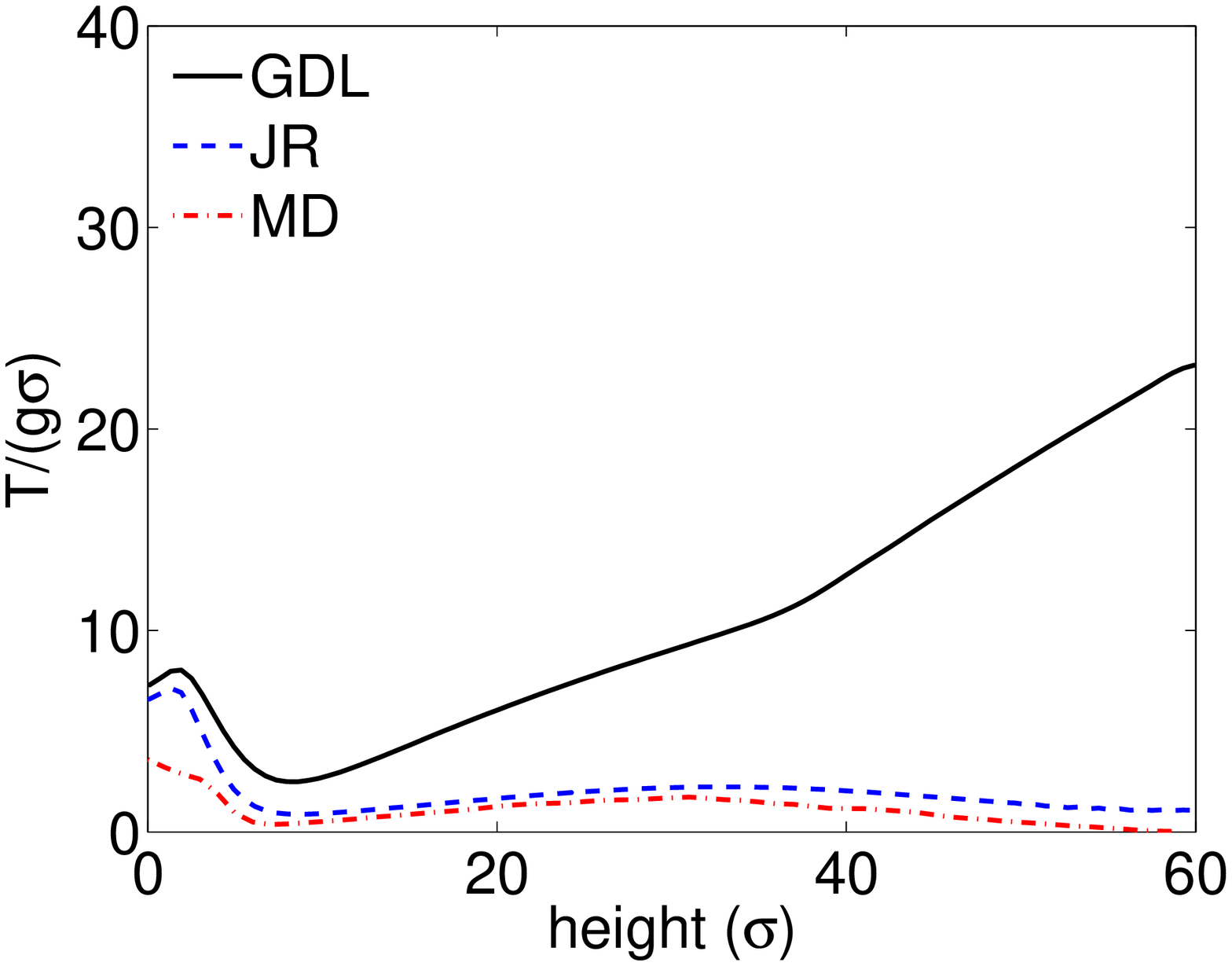}}

 \subfloat[$t=\nicefrac{6}{4} f^{-1}$]{\includegraphics[width=\halfcolumn]{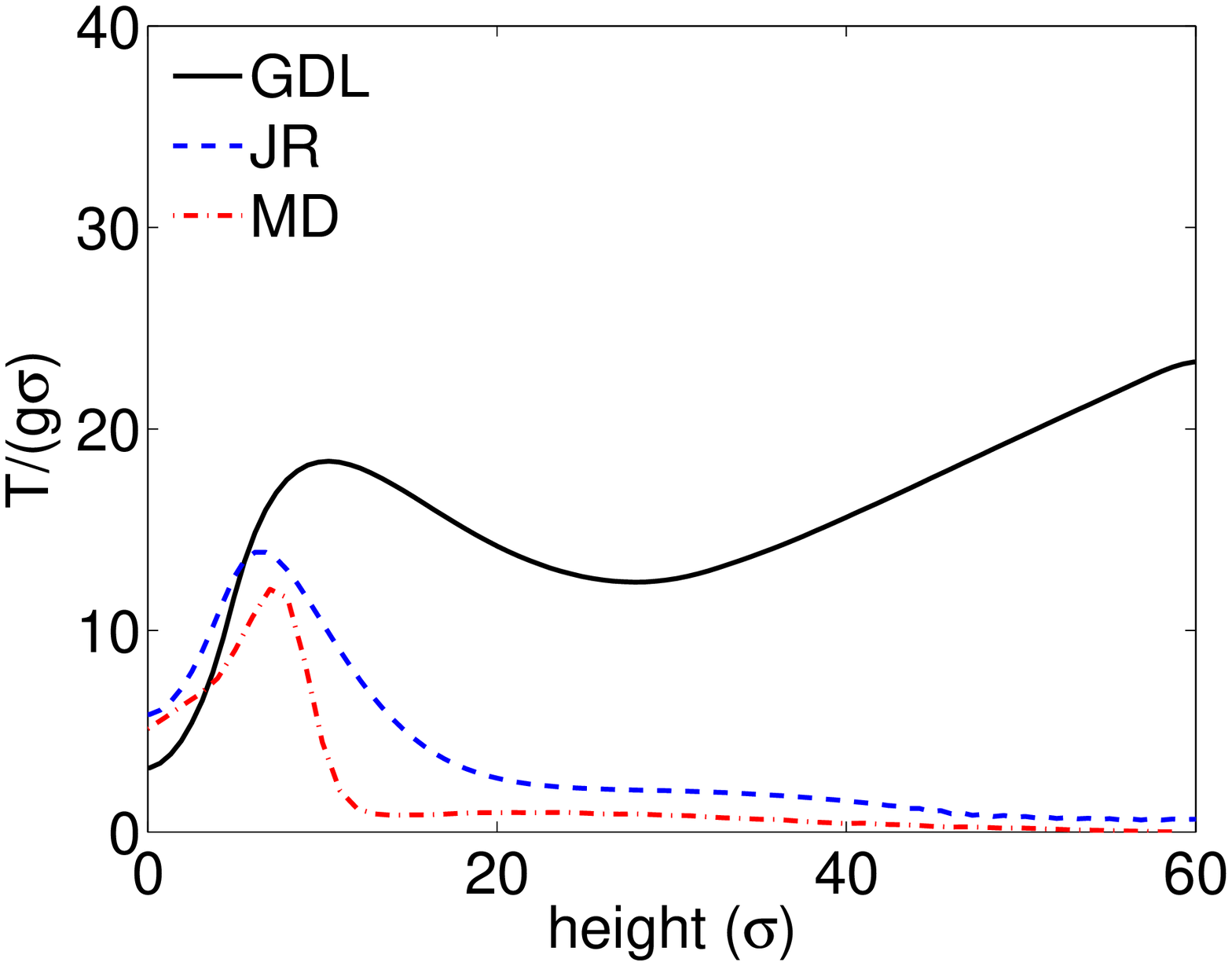}}
 \subfloat[$t=\nicefrac{7}{4} f^{-1}$]{\includegraphics[width=\halfcolumn]{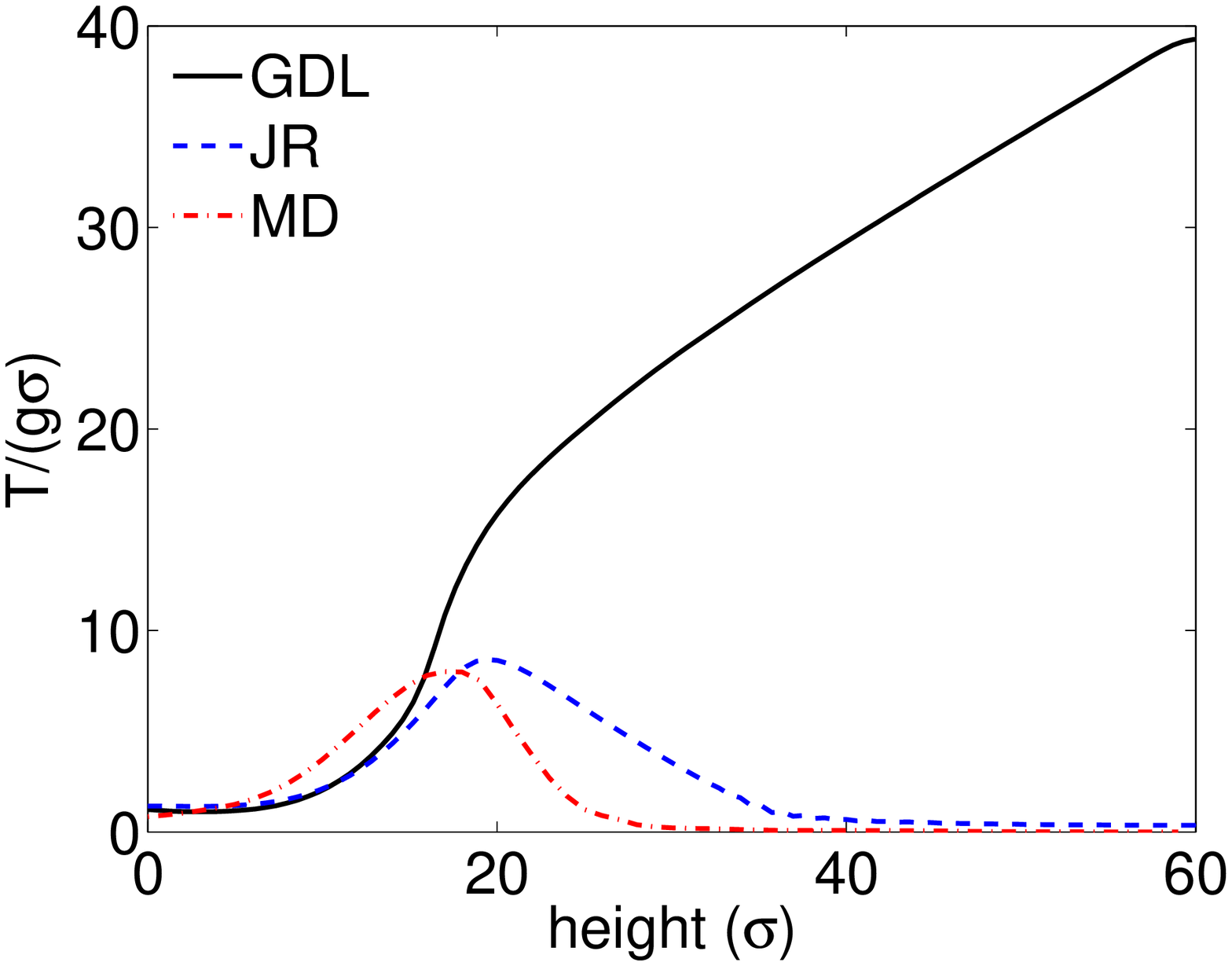}}
\end{center}
\caption{(color online) Profiles of the temperature ($T/(\sigma g)$) as a
function of height (in units of $\sigma$) at selected times over two oscillation
periods for the MD system and the JR and GDL solutions. For time evolution of
the profiles see \cite{online}.}
\label{fig:temperature}
\end{figure}

We disregard the transient originating from the initial condition until the
pattern of the Faraday instability has fully developed and no changes are
observed from period to period. After this transient time, which takes about 50
periods of forcing, the system reveals a subharmonic periodic dynamics where the
period is twice the period of the forcing $f^{-1}$. In this regime, we fix the
reference time, $t=0$ and consider the evolution of the profiles of packing
fraction, Fig. \ref{fig:xnu}, scaled thermal energy, Fig. \ref{fig:energy},
scaled granular temperature, Fig. \ref{fig:temperature} and scaled kinetic
energy, Fig. \ref{fig:Ec} using Eq. \eref{totalenergy}. The subfigures (a-h)
correspond to the times $t=0$; $\nicefrac{1}{4}f^{-1}$; $\nicefrac{2}{4}f^{-1}$;
$\dots$; $\nicefrac{7}{4}f^{-1}$. The corresponding position of the piston is $y
= -A \sin 2\pi f t$. Despite that the hydrodynamic fields are two
dimensional, we show 1D profiles for a more detailed quantitative analysis.
Thus, the profiles shown in Figs. \ref{fig:xnu}-\ref{fig:temperature} are taken
at a representative location along the abscissa, where the amplitude of the
Faraday pattern is developed. The evolution of these profiles over the period of
excitation is also presented as supplementary online material \cite{online},
showing the profiles at many more intermediate times.

\subsection{Density}

First of all we are going to discuss the behavior of the packing fraction,
Fig.~\ref{fig:xnu}. Since the packing fraction is proportional to the number
density $\phi=\pi\sigma^2n/4$, then we will use both terms indistinctly. As in
subsequent figures, the abscissa represents the height, in diameters. On the
ordinate we show here the packing fraction. The evolution is shown from left
to right, and then from top to bottom. Note that the integral of each curve is
not the same for the hydrodynamics and the MD simulations since it corresponds
just to a vertical cut at a position where the maximum height of the pattern is
achieved. Total conservation of mass is maintained in all simulations with
high accuracy, see \cite{CarrilloPoeschelSaluena:2008} for more details.

At time $t=0$, Fig. \ref{fig:xnu}(a), the piston is going down through the
equilibrium position. The height of the material at this location has already
grown to a maximum, formed at the end of the previous cycle (g, h). Shortly
after this time the granular layer experiences the impact against the bottom
wall and the propagation of a shock wave. Between (a) and (c), we see the
dissolution of the peak. We observe that the GDL {prediction} is denser than the
JR at a distance of 10 diameters from the plate. Just instants following frame
(c), the layer becomes flat --so does after frame (g), and the material floods
to neighboring positions to create peaks where valleys previously existed.
Shortly after (d), another impact with the plate takes place. From frame (d) to
frame (g), we see the evolution of the density at a valley.

The MD sequence reveals that the maximum density 0.69 in packing fraction is
smaller than in both hydrodynamic simulations, reaching the value 0.78. This can
be due to the irregularity of the MD pattern due to the elasticity of the system
at $\alpha=0.80$, which makes the location of any of the peaks of the MD
sequence somewhat uncertain. We recall that the granular Navier-Stokes solver
does not contain fluctuational --mesoscopic-- contributions, while the local
noise is enhanced by increasing the coefficient of restitution. That is why one
needs a factor of 20 times more cycles to obtain smooth fields, as compared with
the results at $\alpha=0.75$, obtained in our previous study
\cite{CarrilloPoeschelSaluena:2008}. There the regularity was much more
pronounced, and a much better agreement was achieved.

\textcolor{black}{The role of fluctuating hydrodynamics in
granular gases has been an object of study for the last decade, since Van
Noije and coworkers \cite{vanNoijeErnstBritoOrzaPRL1996}, or more recently,
Brey \cite{BreyMaynarGarciaPRE2009} and Costantini \cite{CostantiniPuglisiPRE2010}.
The essentially mesoscopic dynamics of the granular gas flow can not be
fully captured by means of macroscopic transport equations. This is easily emphasized,
for instance, by the need to apply mesoscoping averaging to MD
results, in order to compare particle and hydrodynamic simulations. Another
related effect is the diffusion found at the level of the bifurcation
threshold of the instability, as observed from MD simulations; the hydrodynamic
simulations show instead a sharp inception of the instability at about $\Gamma=2.0$,
when one represents the wavelength of the Faraday pattern as a function of the
reduced acceleration $\Gamma$ of the plate \cite{CarrilloPoeschelSaluena:2008}}.

\textcolor{black}{We expect (minor) differences between the two
models in the pattern wavelength, however we did not perform the
complete analysis of the bifurcation diagram in the case of the
GDL model for the following reason. As a result of what we have
explained above, at the threshold of the Faraday instability, the
uncertainty in the wavelength as determined by the Fourier
analysis of the density pattern is quite large
\cite{CarrilloPoeschelSaluena:2008}. The presence of noise turns
the transition into a continuous phenomenon, which the
hydrodynamic simulations without a source of fluctuations cannot
exactly reproduce. As a consequence, both the JR and the GDL
models will be providing somewhat different thresholds, none of
which will be accounting for the true effect. Beyond the
instability threshold, we do observe different wavelengths for the
case analyzed: 5.6 diameters of amplitude of vibration, and
reduced acceleration of 2.75. For these parameters, a system 500
wide shows 12 wavelengths in the JR simulation, whereas the GDL
shows 13. The MD also shows 13. This implies that the GDL approach
models better this feature as compared with JR, at least for the
values of the parameters chosen.}

While the GDL and JR profiles do not differ greatly, there are some differences:
the GDL density is higher at the core of peaks and valleys, as compared to the
JR prediction at equivalent times. Correspondingly, the packing fraction at
the bottom plate is smaller in the GDL simulation, and so is the minimum density
(0.054 vs. the value of 0.112 obtained in the JR simulation). However, the
minimum density in the averaged MD profile is still smaller: 0.004. Also, the
impact with the plate occurs later as compared with both hydrodynamic
simulations, the delay being about 0.16$f^{-1}$. Therefore we may argue that in
general the accurate expressions of the GDL approximation for the Navier-Stokes
transport coefficients does not greatly improve the density profile obtained
with the elastic forms of the JR approach to match the MD results in this
problem. A direct comparison of the time evolution of densities and velocity
fields in full spacial resolution can be found in the supplementary material
\cite{online}.

\begin{figure}
 \begin{center}
 \includegraphics[width=\columnwidth]{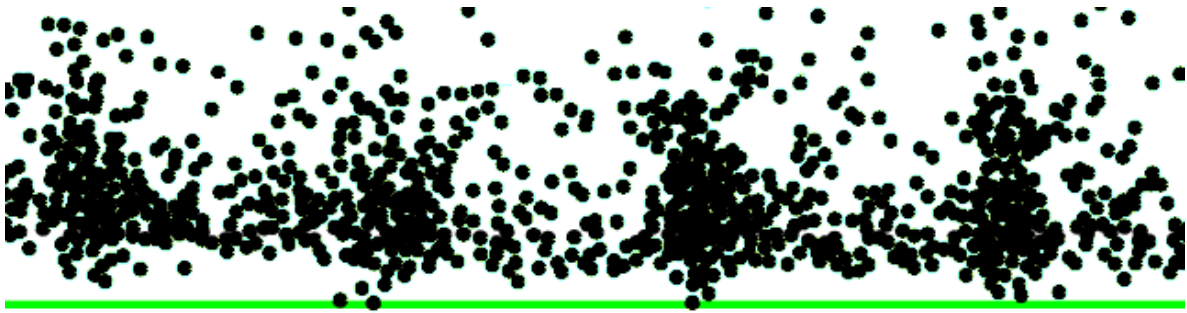}
\end{center}
\caption{A snapshot of the MD simulation at the maximum opening of the gap
($t\approx 0.12\,f^{-1}$), showing the material stuck at the bottom, between the
peaks and the plate, whereas there is a completely empty space below the
valleys.}
\label{fig:balls}
\end{figure}

A zoom of the region of the MD system close to the plate during the airborne
phase will show a few particles stuck to the base of the peaks and empty areas
with no particles at all below the valleys (Fig.~\ref{fig:balls}). As a
consequence, the impact of the wall against the material happens at
$t=0.16f^{-1}$ (instead of $t=0$). We want to remark that this piece of the
system is not in the hydrodynamic regime at this moment, but in the Knudsen
regime, and there is little hope that any hydrodynamic model can reproduce this
feature in full detail. However the GDL approach to the Navier-Stokes
equations improves the dynamics of the gap formed as compared with the JR
approach in the sense that the minimum density at the bottom plate is reduced.
On the other hand, the density gradients are higher in the GDL theory, a feature
which is not observed in the MD profiles, which are smoother. The differences
are basically due to the presence of the coefficient $\mu_\textrm{\small{GDL}}$ (Eq.
\eref{eq:mu} of the GDL model), which is absent ($\mu_\textrm{\small{JR}}=0$) in the
elastic case (the JR approach).

\subsection{Temperature and internal energy}

In Figure~\ref{fig:energy} we plot the scaled internal energy, $\phi T/(\sigma
g)$, where $g$ denotes the gravity acceleration. Here we see the evolution of
the shock wave travelling across the granular layer. We can observe that the
energy is smaller everywhere in the GDL system, except at intermediate and large
heights. Remarkably, the energy of the GDL shock wave is lower than the JR after
an impact with the wall, however the remnants persist for long at larger
heights. The MD profile indicates a higher energy at the bottom after an impact
(c), as compared with both GDL and JR results, but specially with the latter.
The GDL shock wave is very much damped. It also shows that the impact with the
bottom wall occurs effectively later, as pointed out when discussing the density
profiles. In addition, the MD profile shows that the energy vanishes quicker
than in the GDL solution. Let us examine then the temperature field.

The most striking difference between the GDL and JR solutions is the temperature
field, Fig.~\ref{fig:temperature}. At large heights, the GDL temperature is one
order of magnitude larger than the JR. Moreover, the GDL temperature gradient is
positive at middle heights (it starts to grow) whereas there the JR, like the MD
temperature gradient, is negative once the shock wave is dissipated. It is clear
that the term $\mu \nabla n$ helps to sustain large temperature gradients in the
system, transferring heat from the dense to the dilute regions at the top wall.
 {This term is the genuine contribution of the inelastic nature of the granular
gas to the transport coefficients}, although we find no hint in the obtained MD
profile that the temperature gradient should be positive instead of negative
when ascending from the dense to the dilute region. As mentioned in the
Introduction, the presence of the coefficient $\mu$ in the heat flux is an
\emph{exact} result of the inelastic Enskog equation and the JR approximation
fails in describing this  new effect. In addition, the existence of this term in
the heat flux has been already confirmed by computer simulation results
\cite{Soto}.

However one must recall that beyond 40 diameters in height the material gets
more and more rarefied (Fig.~\ref{fig:xnu}) and goes from densities of the order
of 1\% in packing fraction at 40 diameters to about 1\textperthousand ~at 60, as
obtained from MD results. Therefore one should find Knudsen layers when
approaching a virtual top wall --in our MD simulations there is none, making our
hydrodynamic simulations meaningless there. Note on the other hand that the
temperature field $T$ displayed in Fig.~\ref{fig:temperature}, when scaled with
the mean free path as the relevant unit length, will be proportional to the
quantity $\phi T$ displayed in Fig.~\ref{fig:energy}. In the latter  one can
appreciate that the mismatch between JR and GDL is reduced, although it still
persists. Also, by comparing the three figures (Figs.~\ref{fig:xnu},
\ref{fig:energy} and \ref{fig:temperature}) that the growth of the temperature
starts at intermediate heights, when the density is not specially small. For
this reason, one can conclude that the growth of the temperature is a true
result of the GDL approach and not a negligible product.

\begin{figure}
 \begin{center}
 \subfloat[$t=0$]{\includegraphics[width=\halfcolumn]{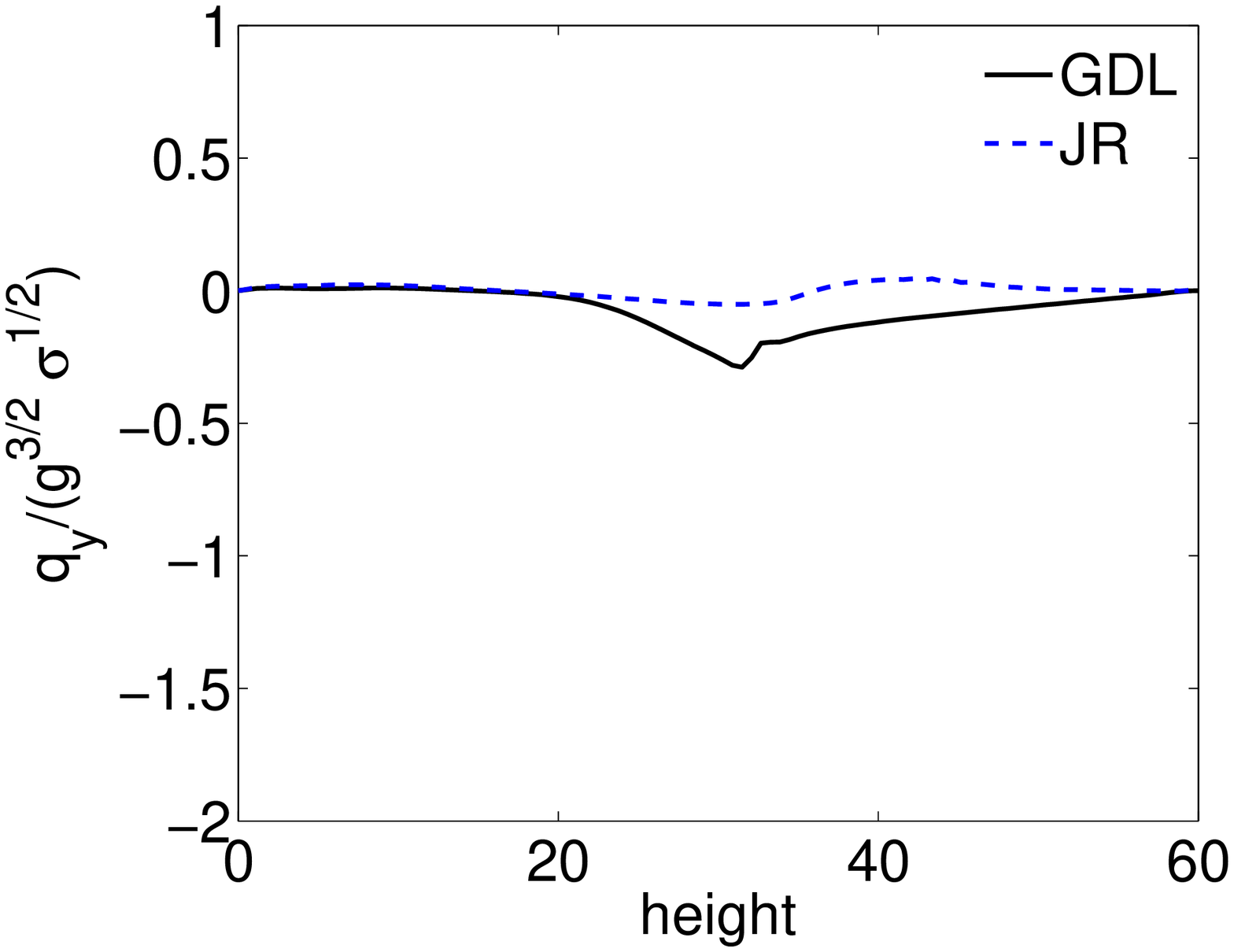}}
 \subfloat[$t=\nicefrac{1}{4} f^{-1}$]{\includegraphics[width=\halfcolumn]{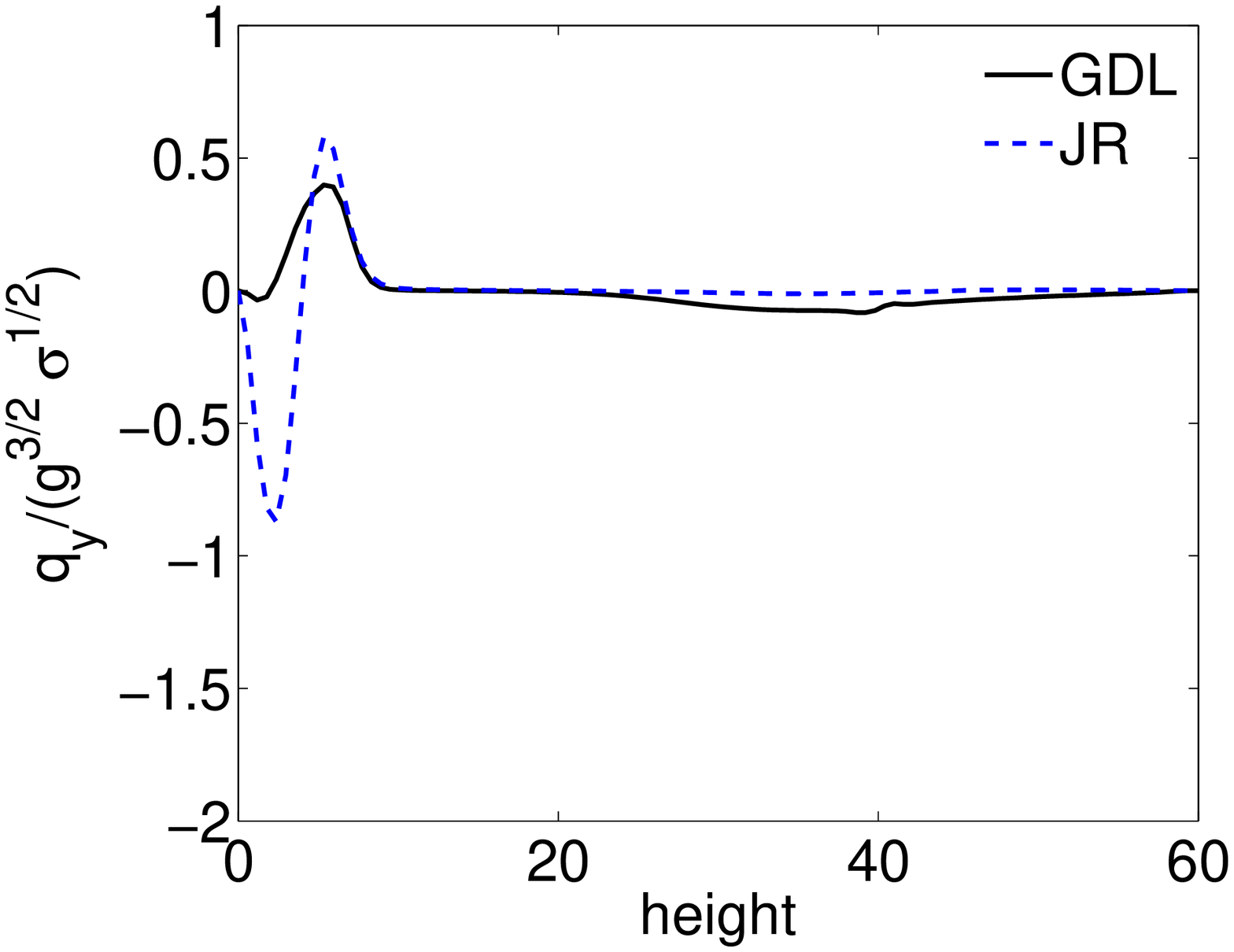}}

 \subfloat[$t=\nicefrac{2}{4} f^{-1}$]{\includegraphics[width=\halfcolumn]{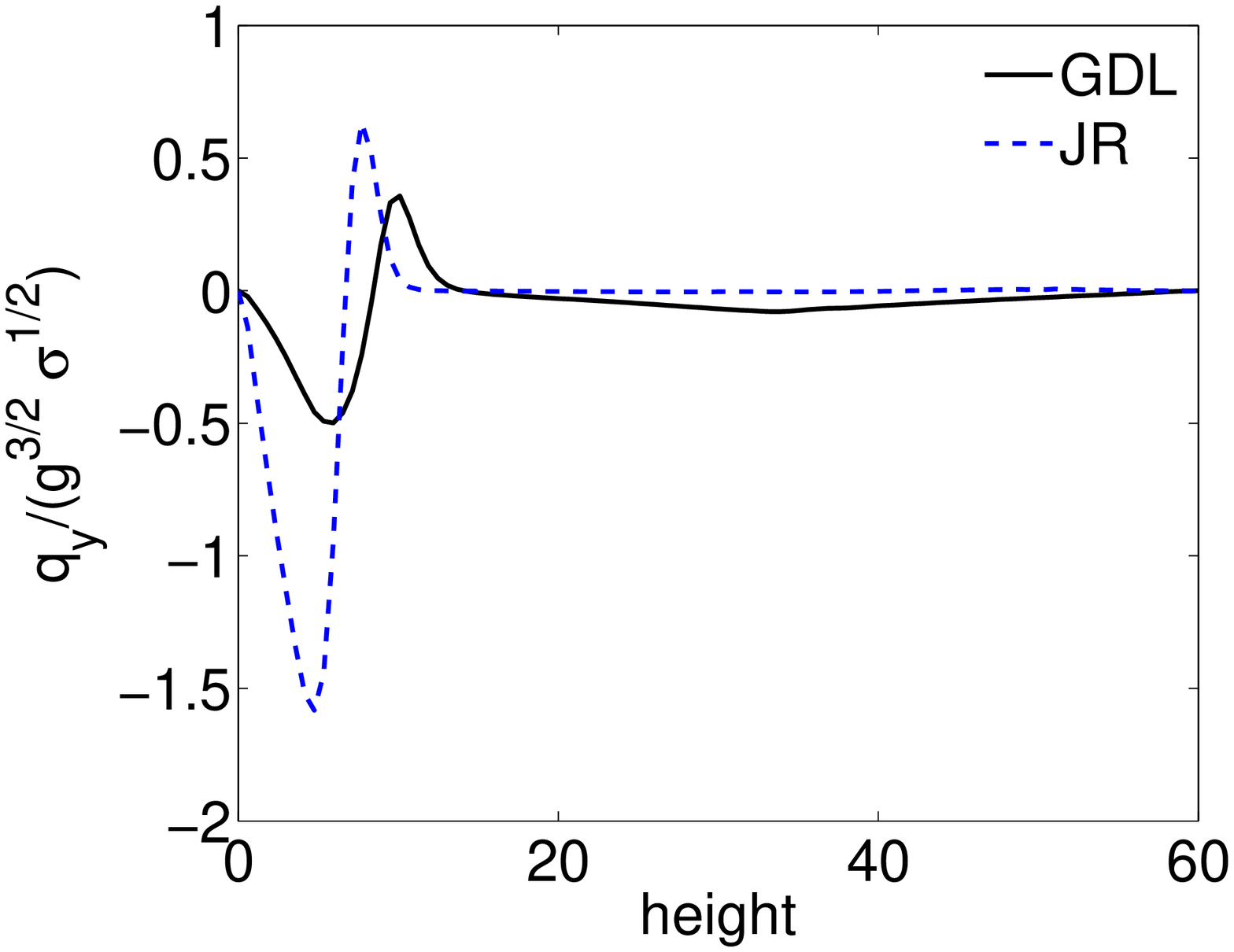}}
 \subfloat[$t=\nicefrac{3}{4} f^{-1}$]{\includegraphics[width=\halfcolumn]{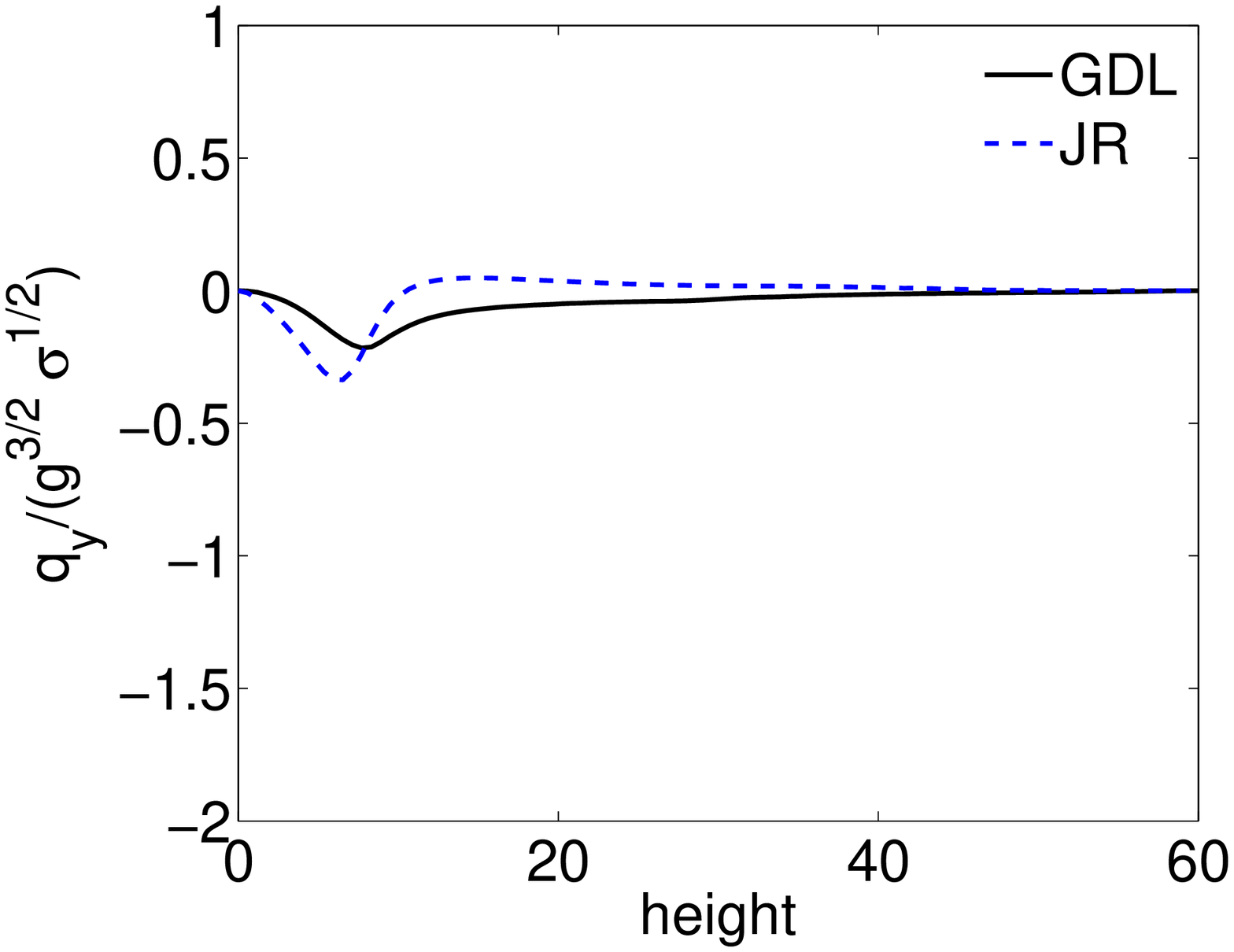}}

 \subfloat[$t=\nicefrac{4}{4} f^{-1}$]{\includegraphics[width=\halfcolumn]{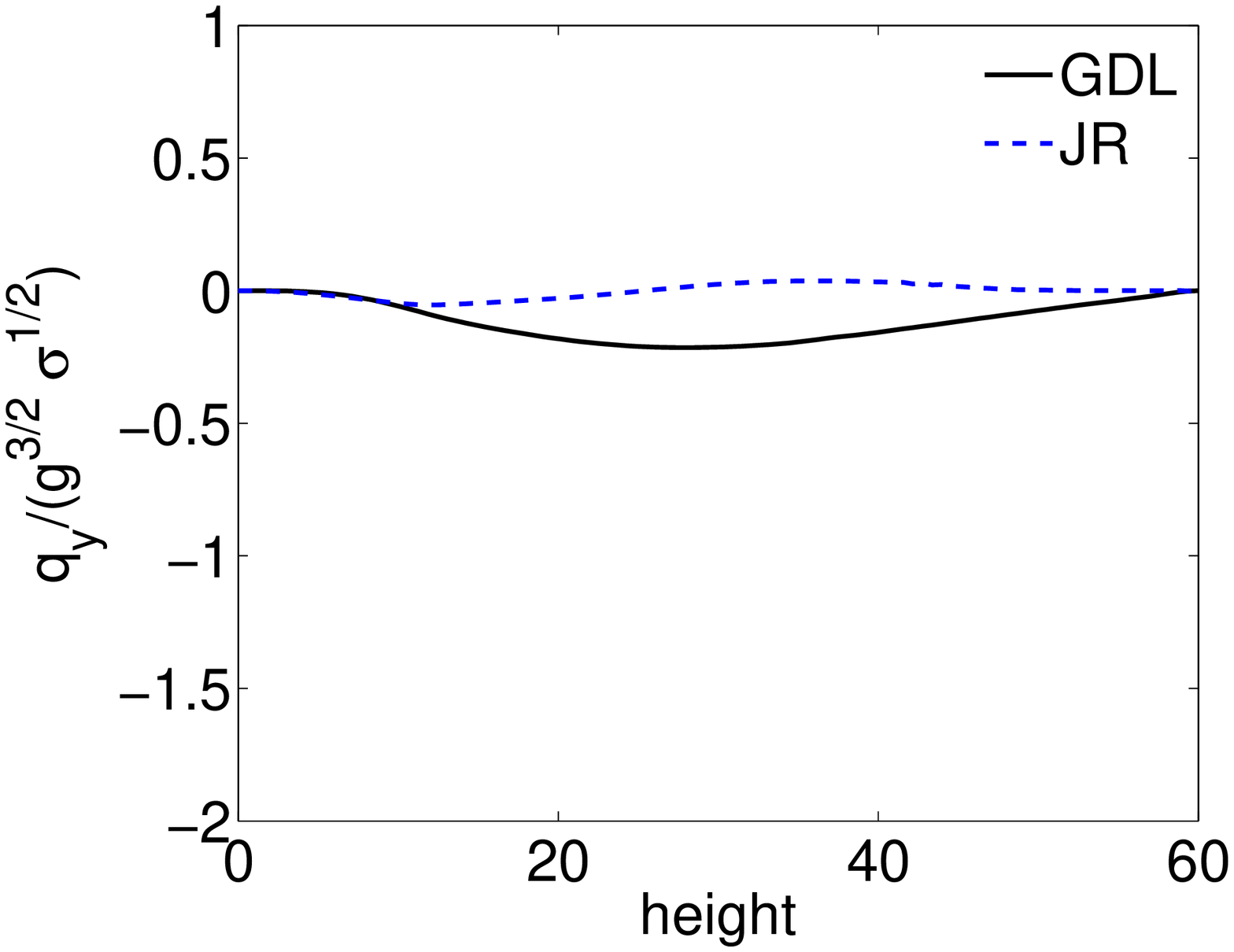}}
 \subfloat[$t=\nicefrac{5}{4} f^{-1}$]{\includegraphics[width=\halfcolumn]{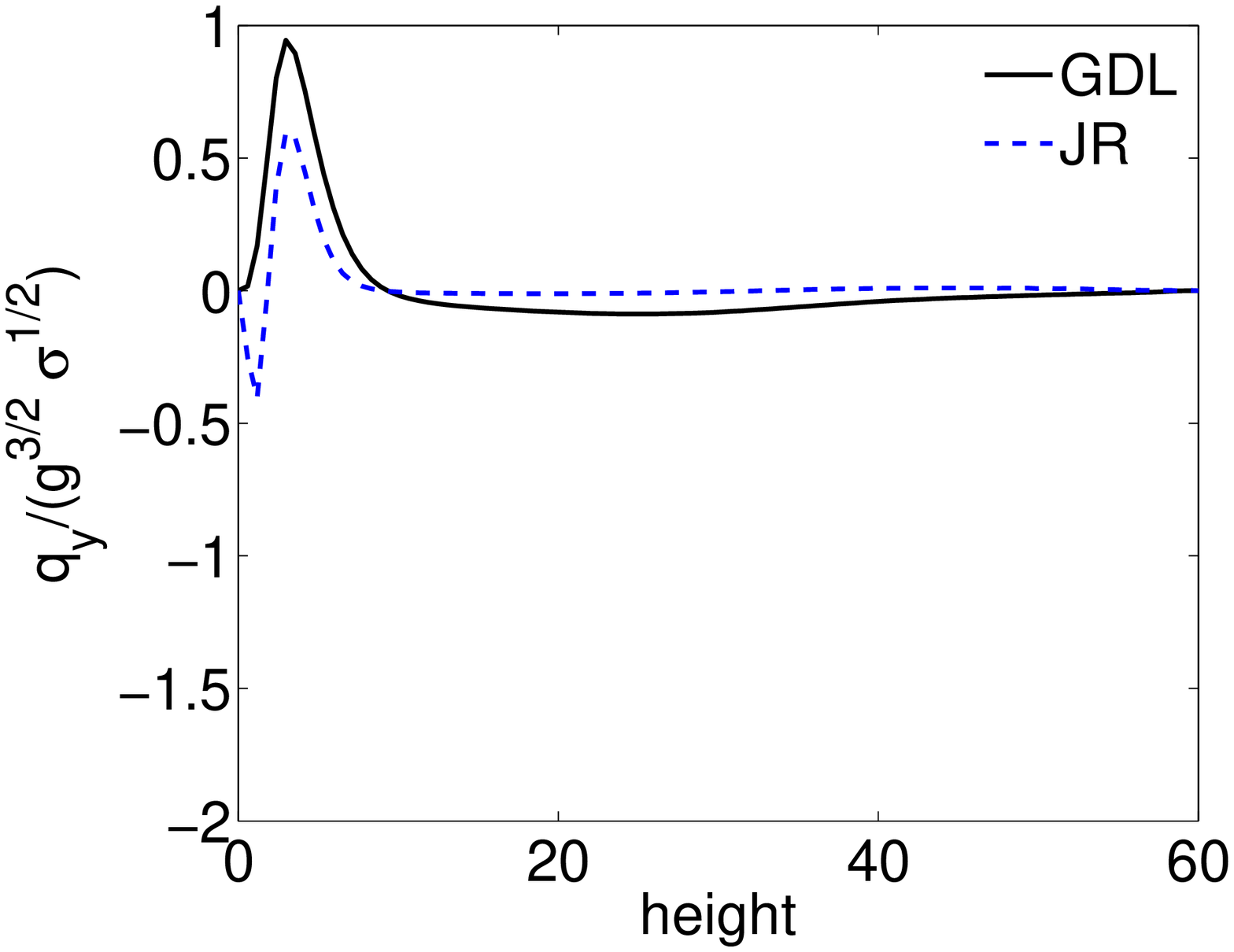}}

 \subfloat[$t=\nicefrac{6}{4} f^{-1}$]{\includegraphics[width=\halfcolumn]{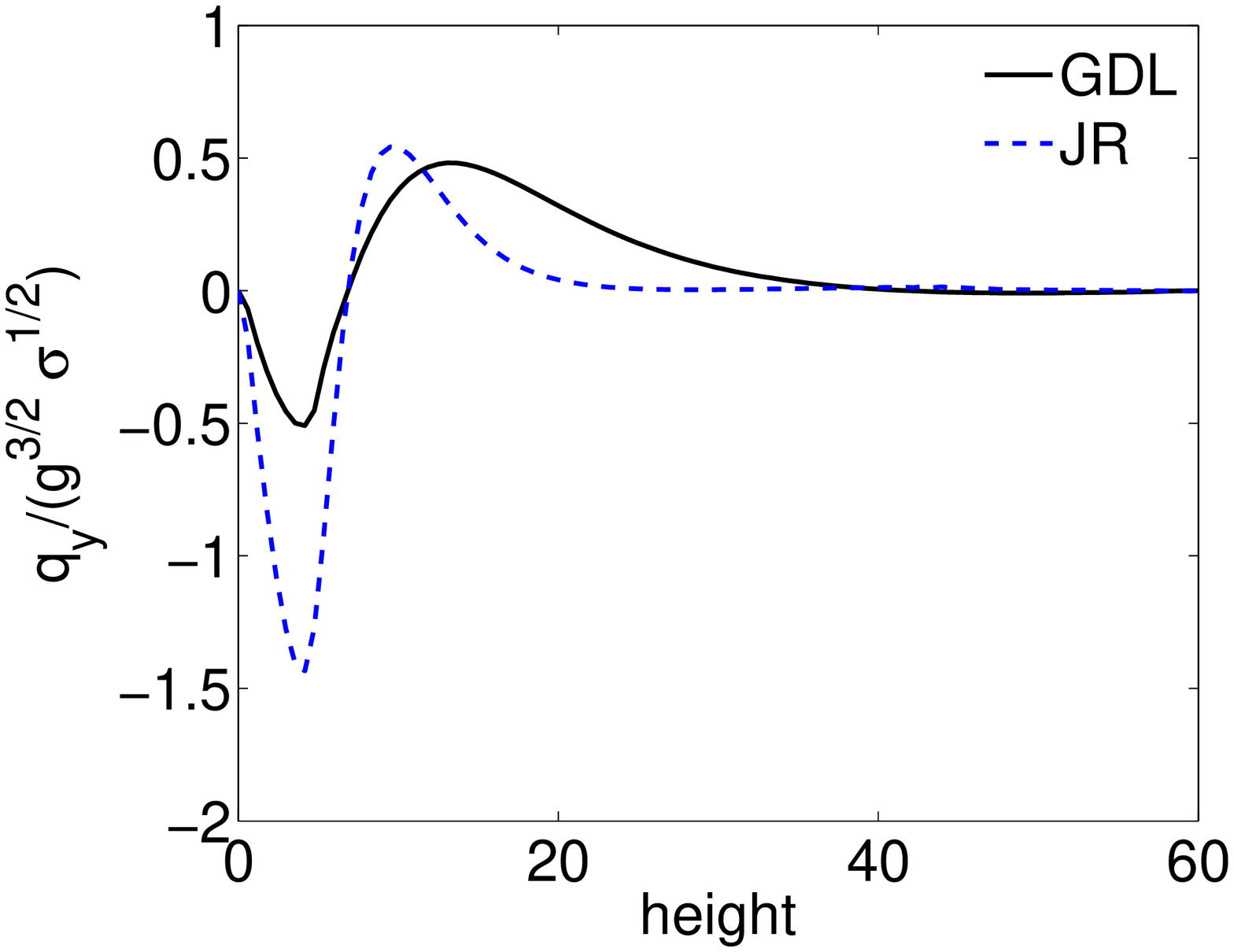}}
 \subfloat[$t=\nicefrac{7}{4} f^{-1}$]{\includegraphics[width=\halfcolumn]{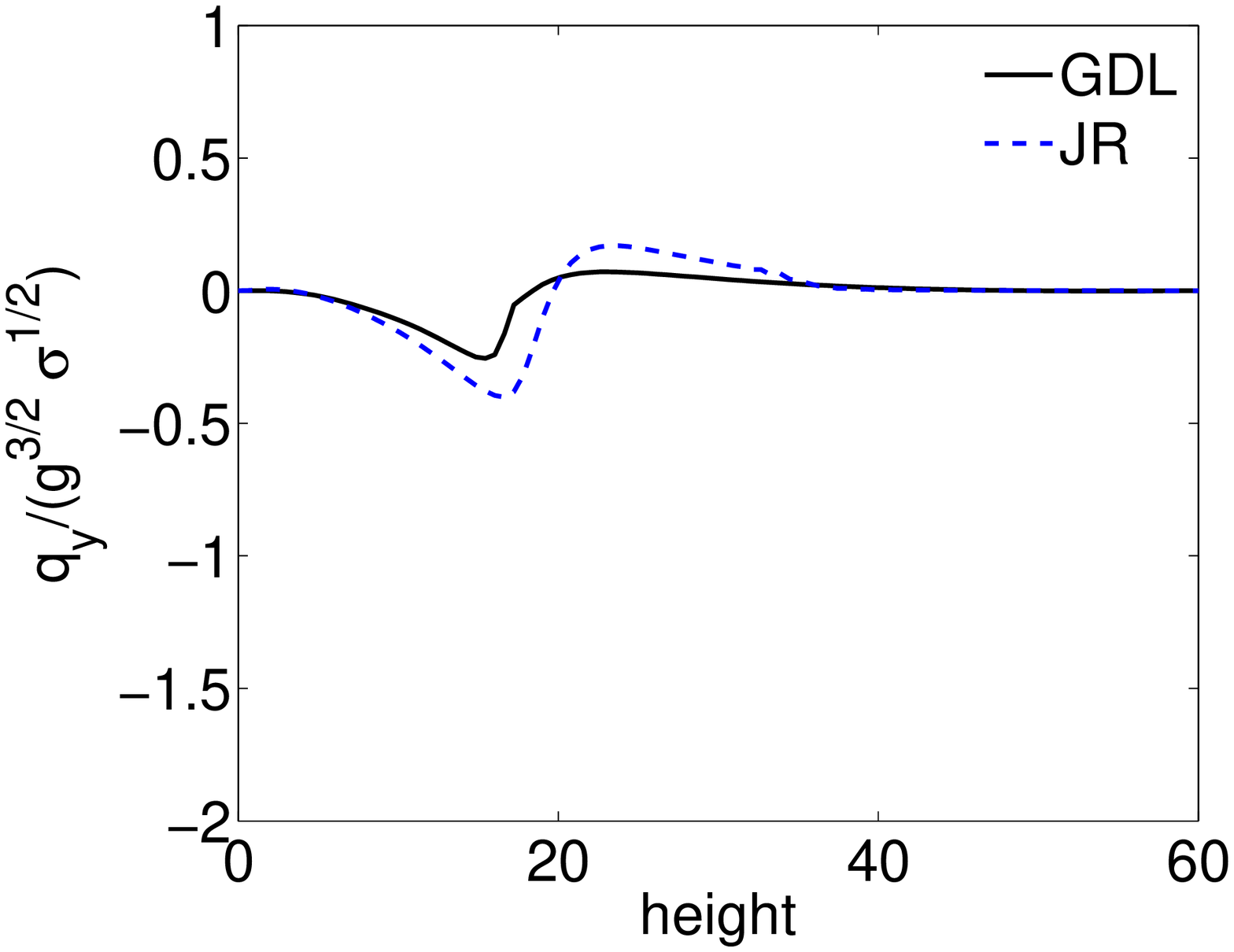}}
\end{center}
\caption{Vertical component of the (reduced) heat flux as a function of height
(in units of $\sigma$) at selected times over two oscillation periods, for the
JR and GDL simulations. For time evolution of the profiles see \cite{online}.}
\label{fig:q}
\end{figure}

As the GDL temperature is higher than the JR temperature at the top, the GDL
solution is more diffusive. Figure~\ref{fig:q}
shows the vertical component of the heat flux as a function of height, where
this effect is shown: note the enhanced heat transport at intermediate heights,
as compared with the JR solution. Unlike the JR case, the GDL heat flux consists
of two terms, the one coming from the temperature gradient, and the one
associated, through the coefficient $\mu$, to the density gradient. An analysis
of the data reveals that both terms have generally opposite signs. The role of
the latter contribution is to transfer heat from the dense towards the dilute
regions at the top, while the former brings energy into the granulate, from the
high temperature regions at the top. Both terms are relevant and contribute in
the same order of magnitude. So, the heat transfer dynamics is quite different
in the GDL and the JR approaches, not only at the top but also at the bottom
plate when the impacts occur, in such a way that gives rise to entirely
different solutions for the temperature field.

\begin{figure}
 \begin{center}
 \subfloat[$t=0$]{\includegraphics[width=\halfcolumn]{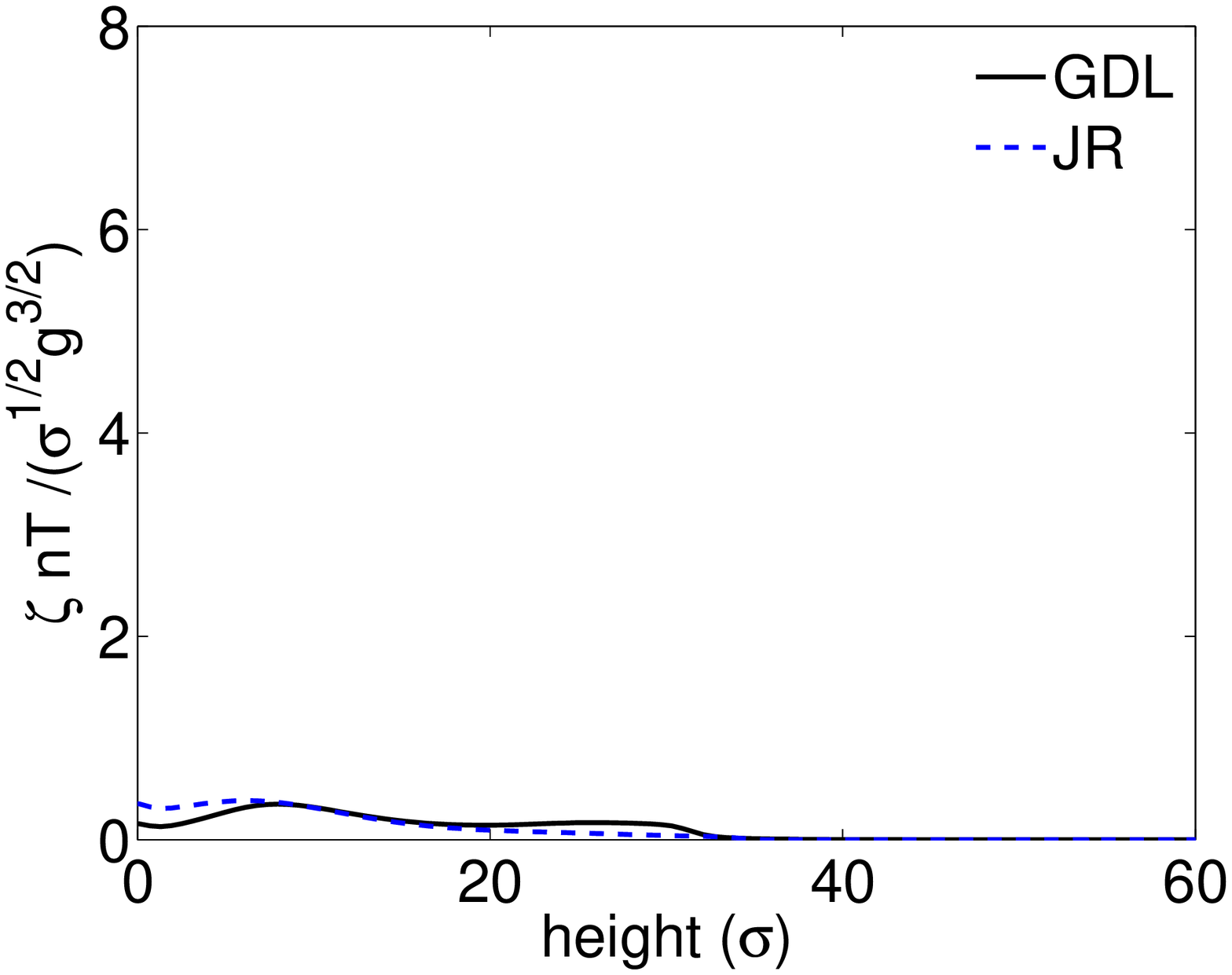}}
 \subfloat[$t=\nicefrac{1}{4} f^{-1}$]{\includegraphics[width=\halfcolumn]{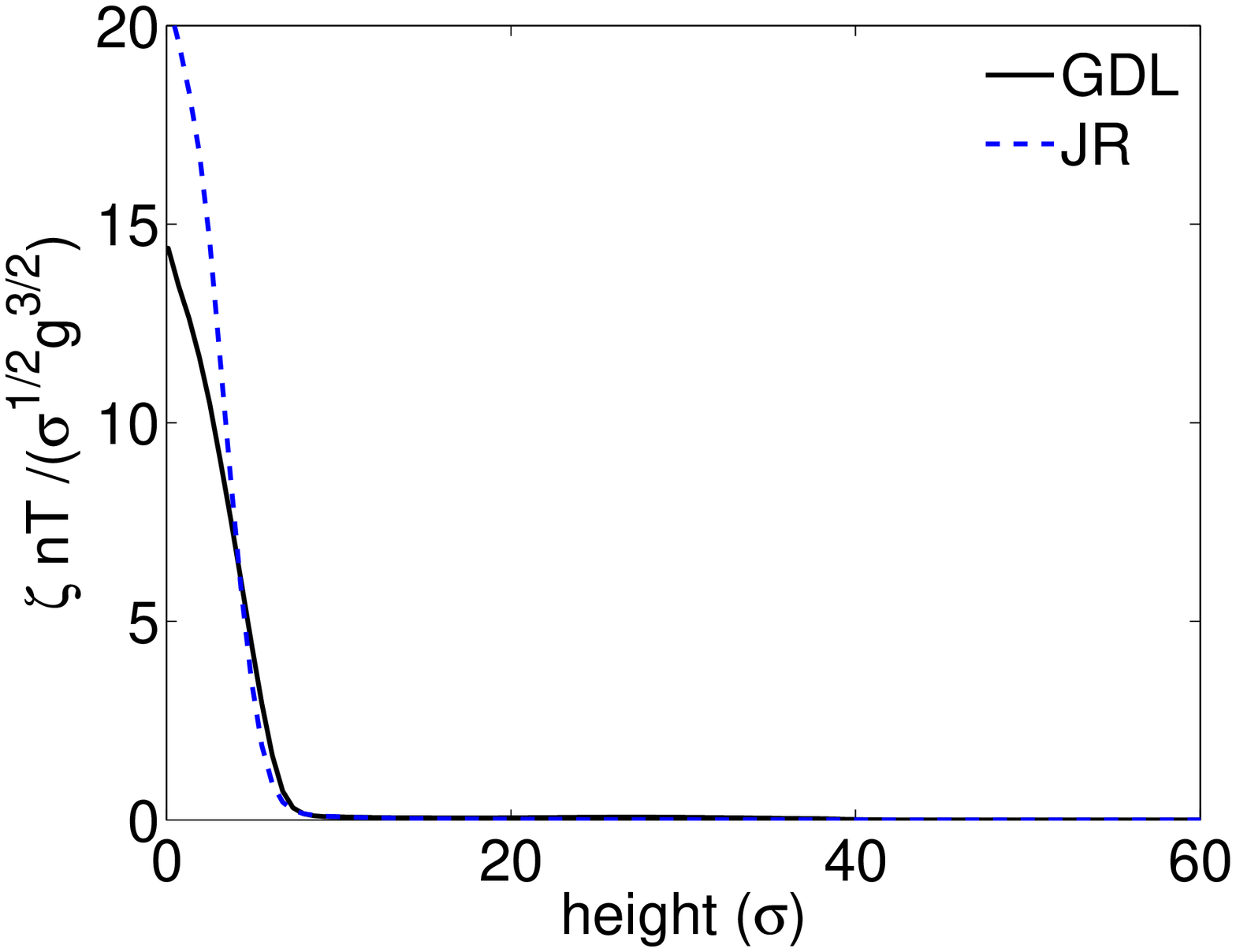}}

 \subfloat[$t=\nicefrac{2}{4} f^{-1}$]{\includegraphics[width=\halfcolumn]{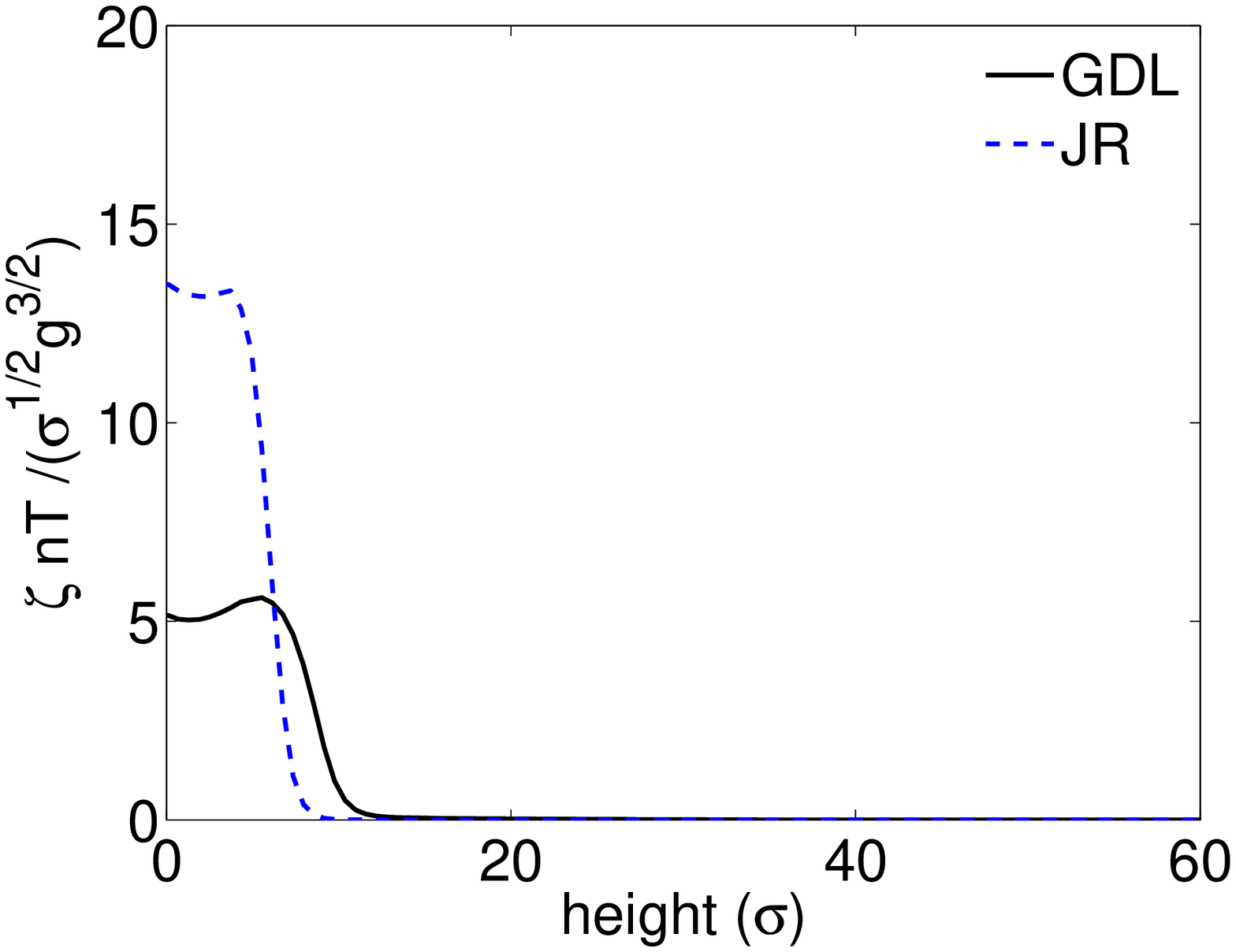}}
 \subfloat[$t=\nicefrac{3}{4} f^{-1}$]{\includegraphics[width=\halfcolumn]{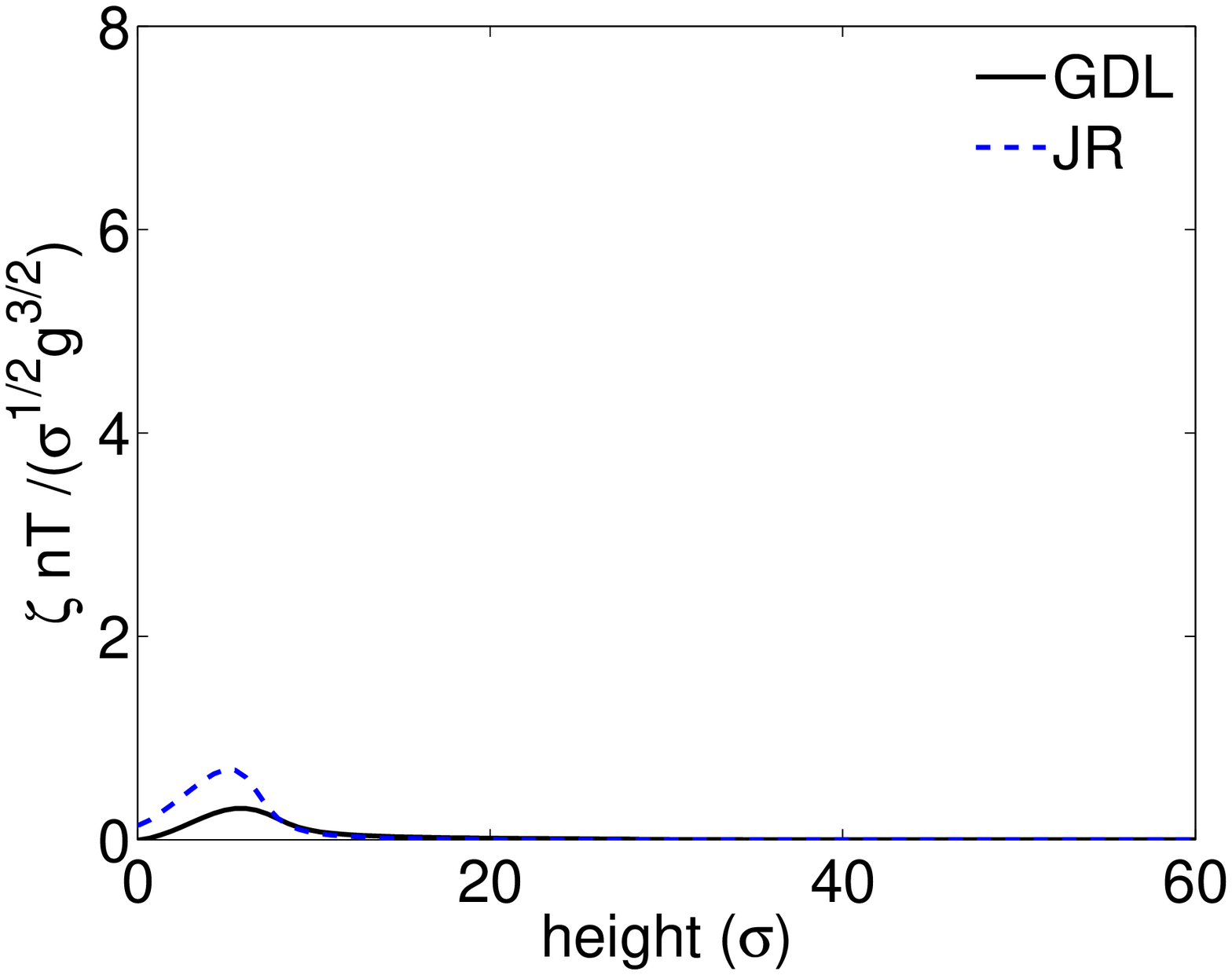}}

 \subfloat[$t=\nicefrac{4}{4} f^{-1}$]{\includegraphics[width=\halfcolumn]{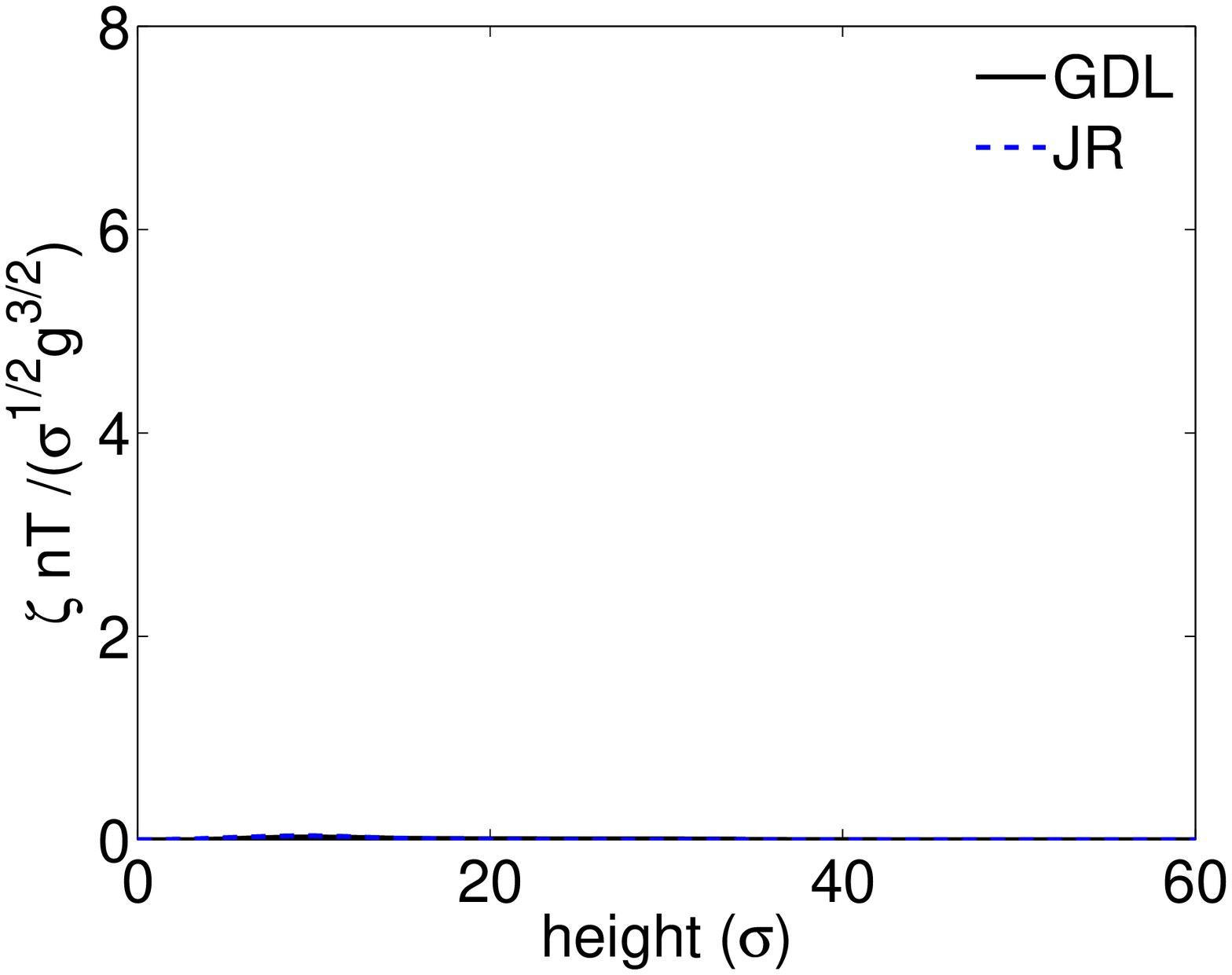}}
 \subfloat[$t=\nicefrac{5}{4} f^{-1}$]{\includegraphics[width=\halfcolumn]{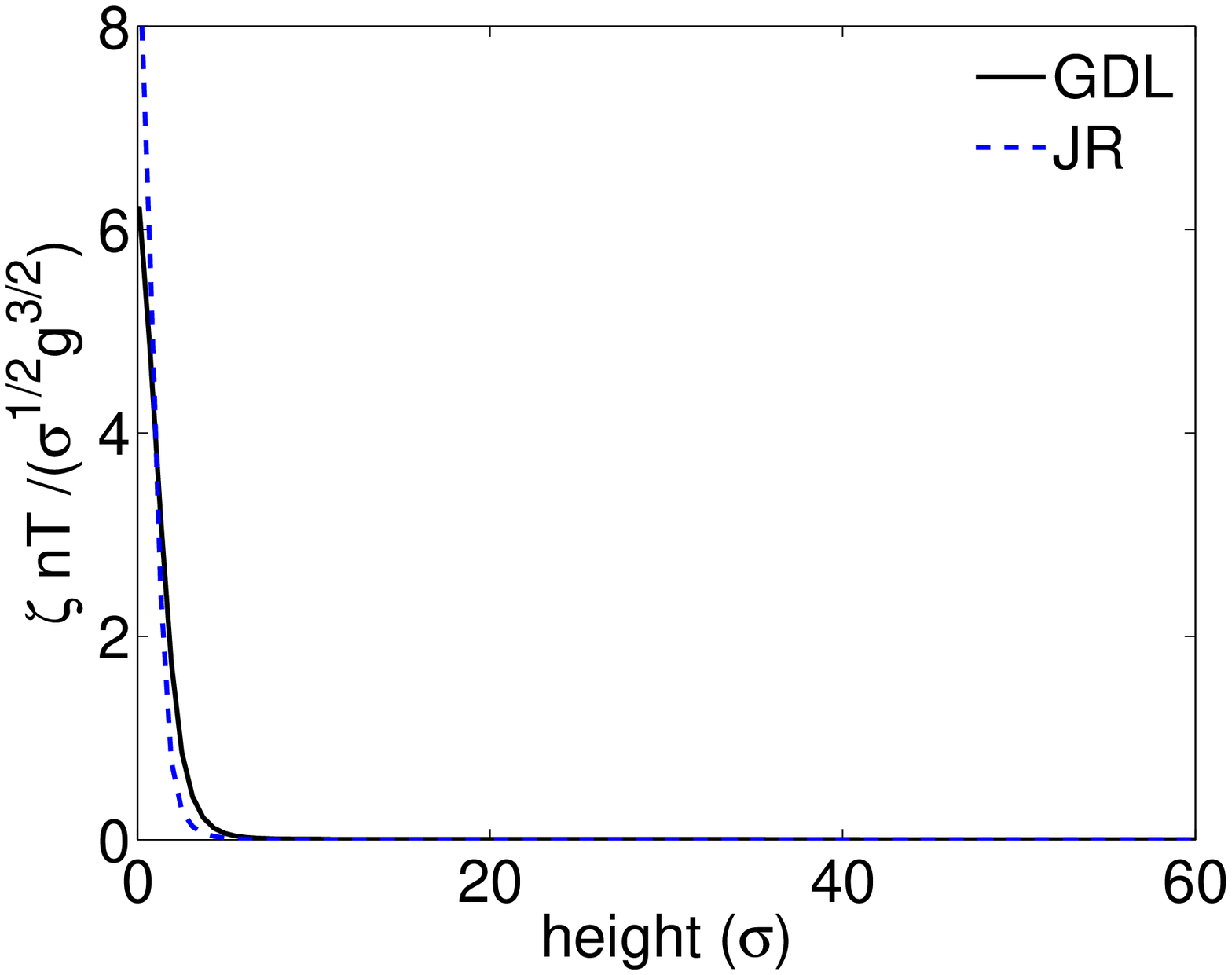}}

 \subfloat[$t=\nicefrac{6}{4} f^{-1}$]{\includegraphics[width=\halfcolumn]{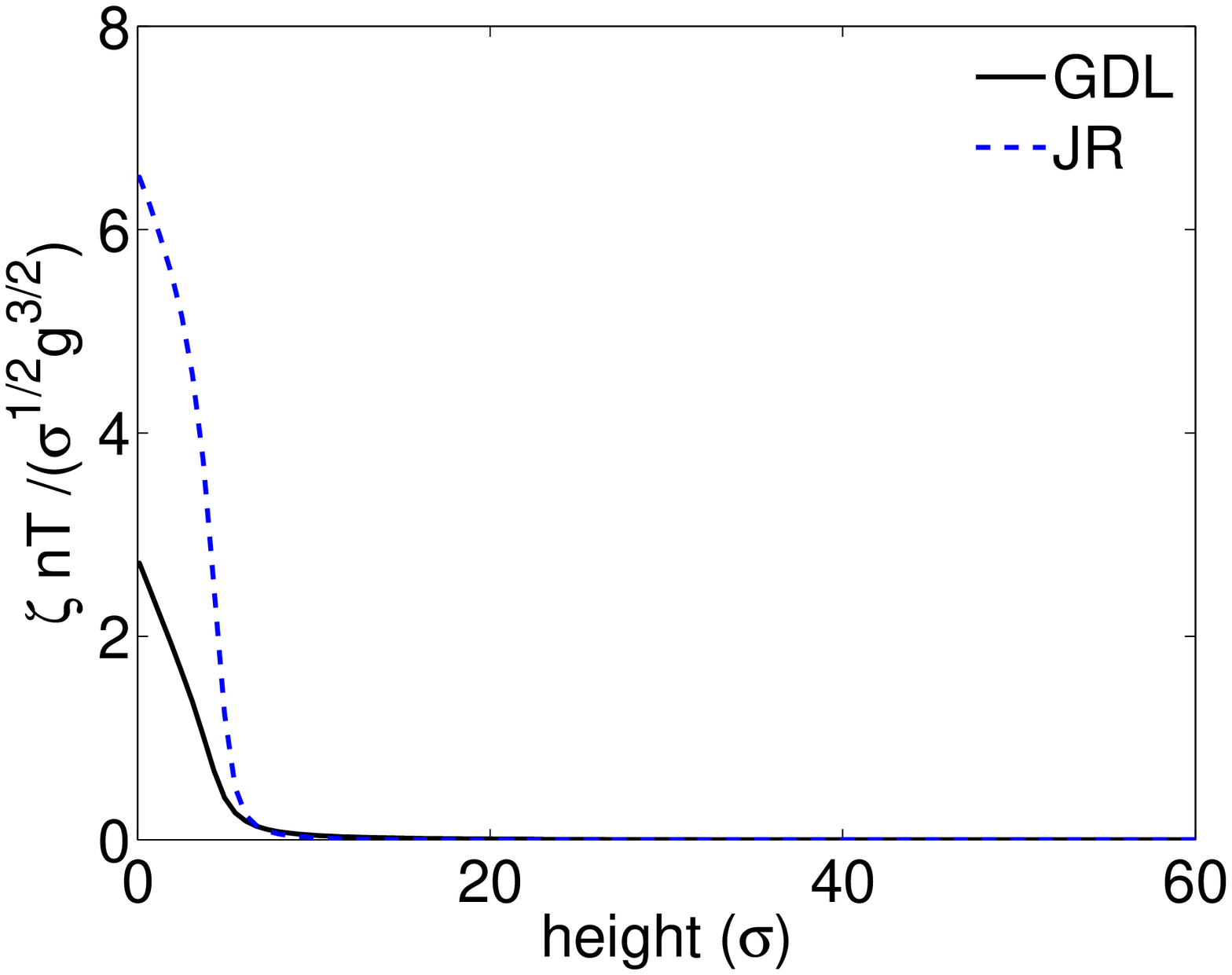}}
 \subfloat[$t=\nicefrac{7}{4} f^{-1}$]{\includegraphics[width=\halfcolumn]{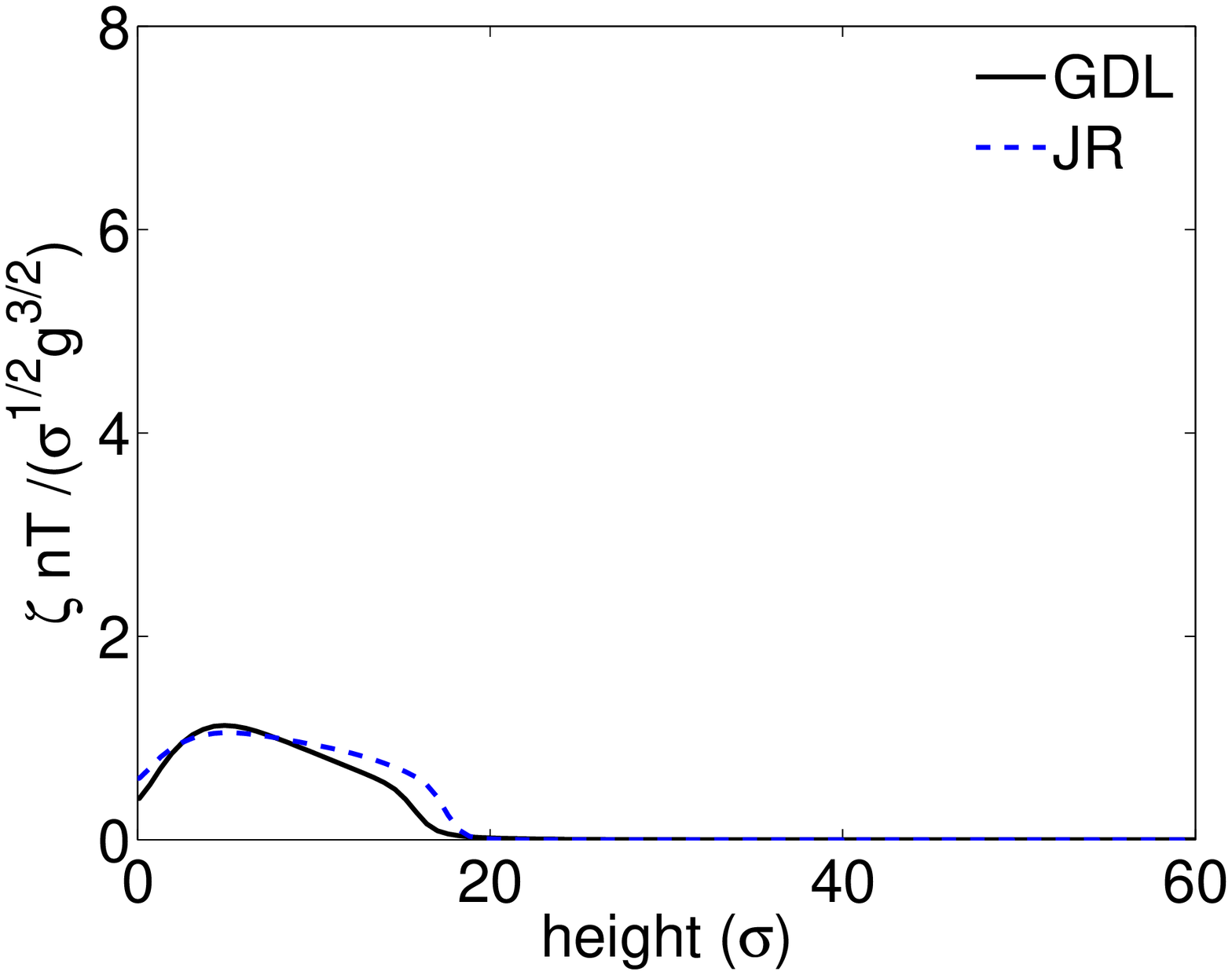}}
\end{center}
\caption{The profiles of the cooling term $\zeta n T$ as a function of height
(in units of $\sigma$), at selected times over two oscillation periods, for the
JR and GDL simulations. Note the change in the vertical scale in subfigures (b)
and (c). For  time evolution of the profiles see \cite{online}.}
\label{fig:coolingterm}
\end{figure}

In general, the GDL system is less diffusive very close to the plate and more at
intermediate heights and at the top, as compared with the JR system. The
viscosities and the cooling term (see Fig.~\ref{fig:coolingterm}) also follow
this pattern. The analysis of the results allows us to conclude that in the JR
system, most of the energy is dissipated very close to the plate, whereas much
less is diffused; in the GDL, comparatively, there is less dissipation at the
plate and more diffusion.

\subsection{Kinetic energy and Mach number}

\begin{figure}
\begin{center}
 \subfloat[$t=0$]{\includegraphics[width=\halfcolumn]{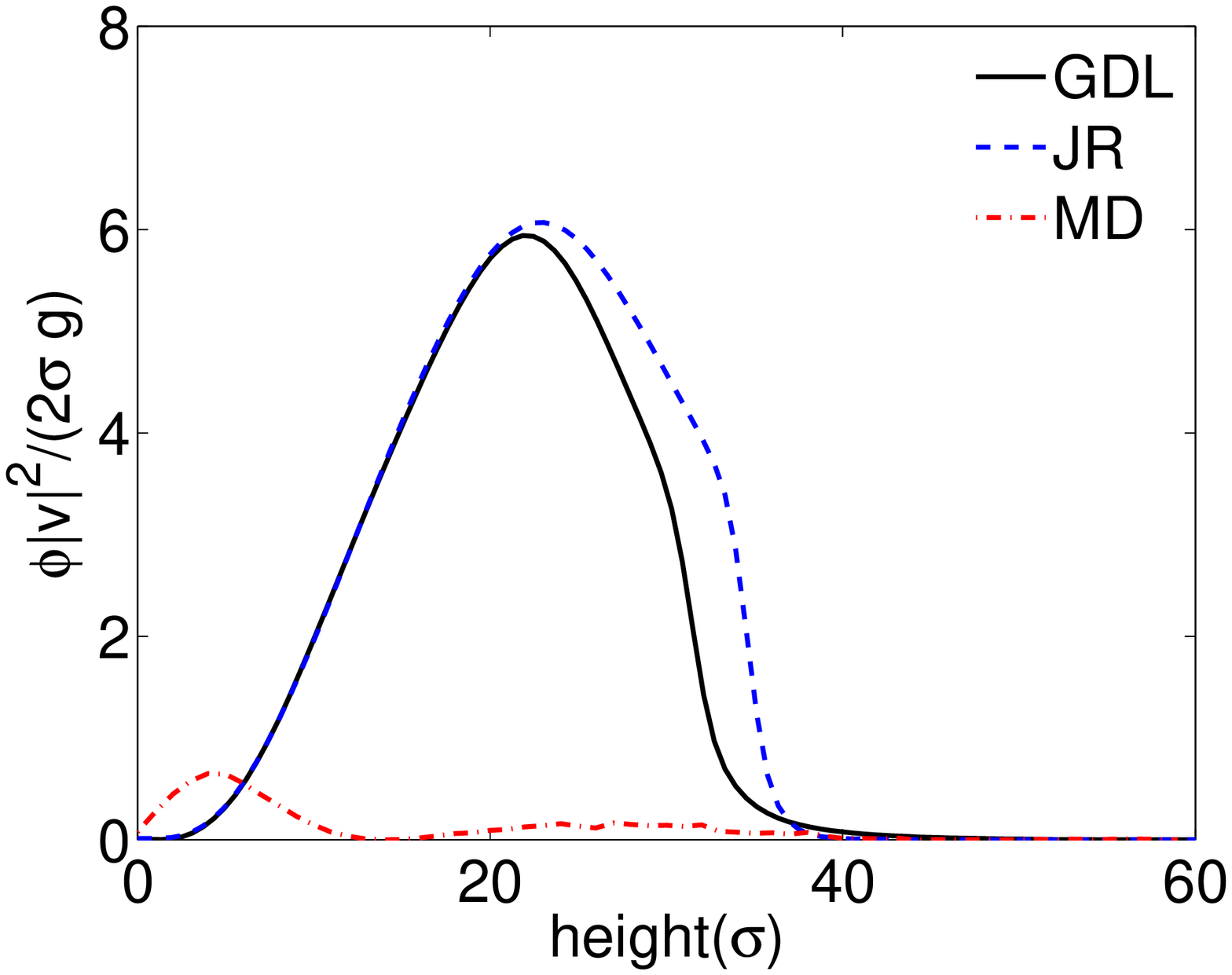}}
 \subfloat[$t=\nicefrac{1}{4} f^{-1}$]{\includegraphics[width=\halfcolumn]{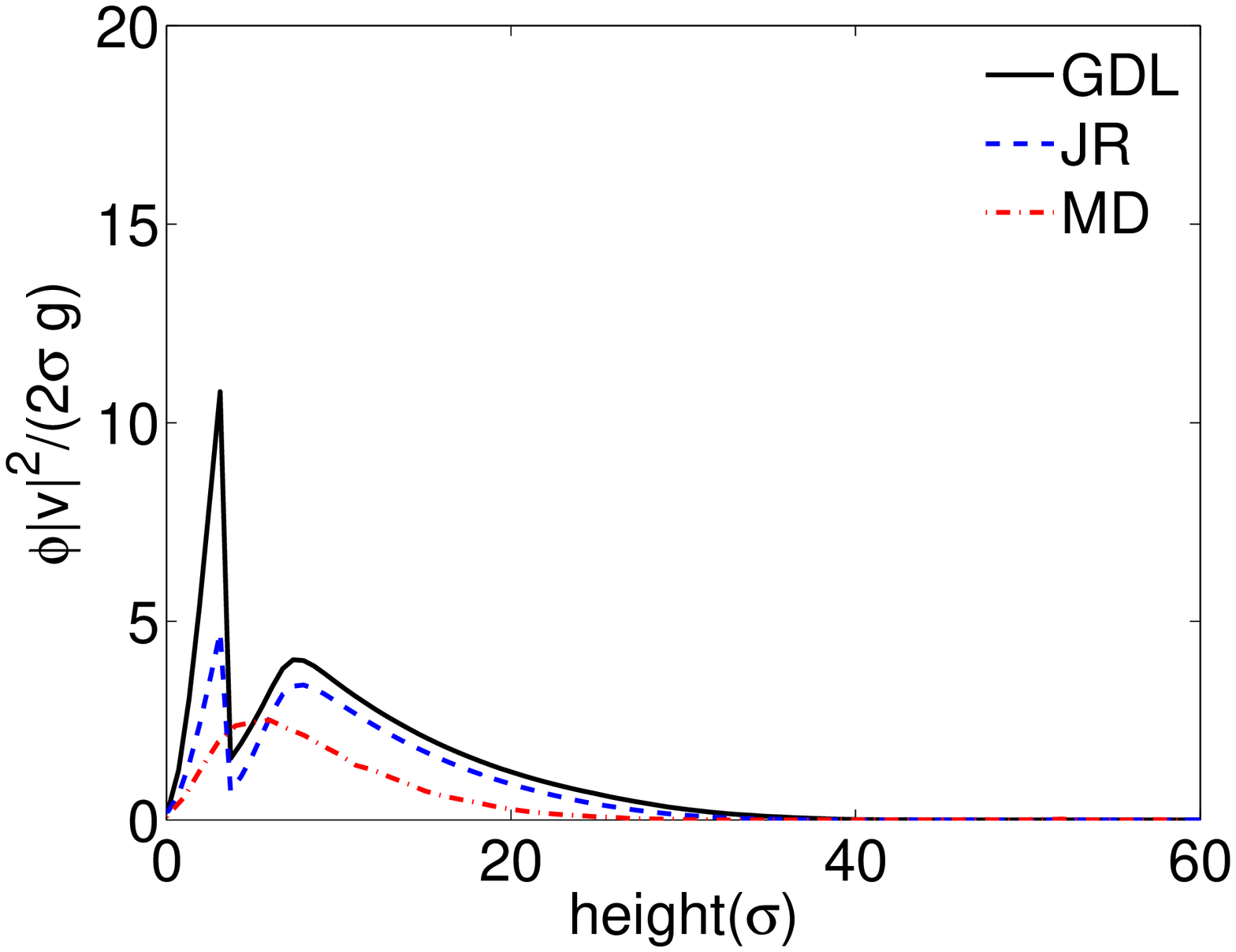}}

 \subfloat[$t=\nicefrac{2}{4} f^{-1}$]{\includegraphics[width=\halfcolumn]{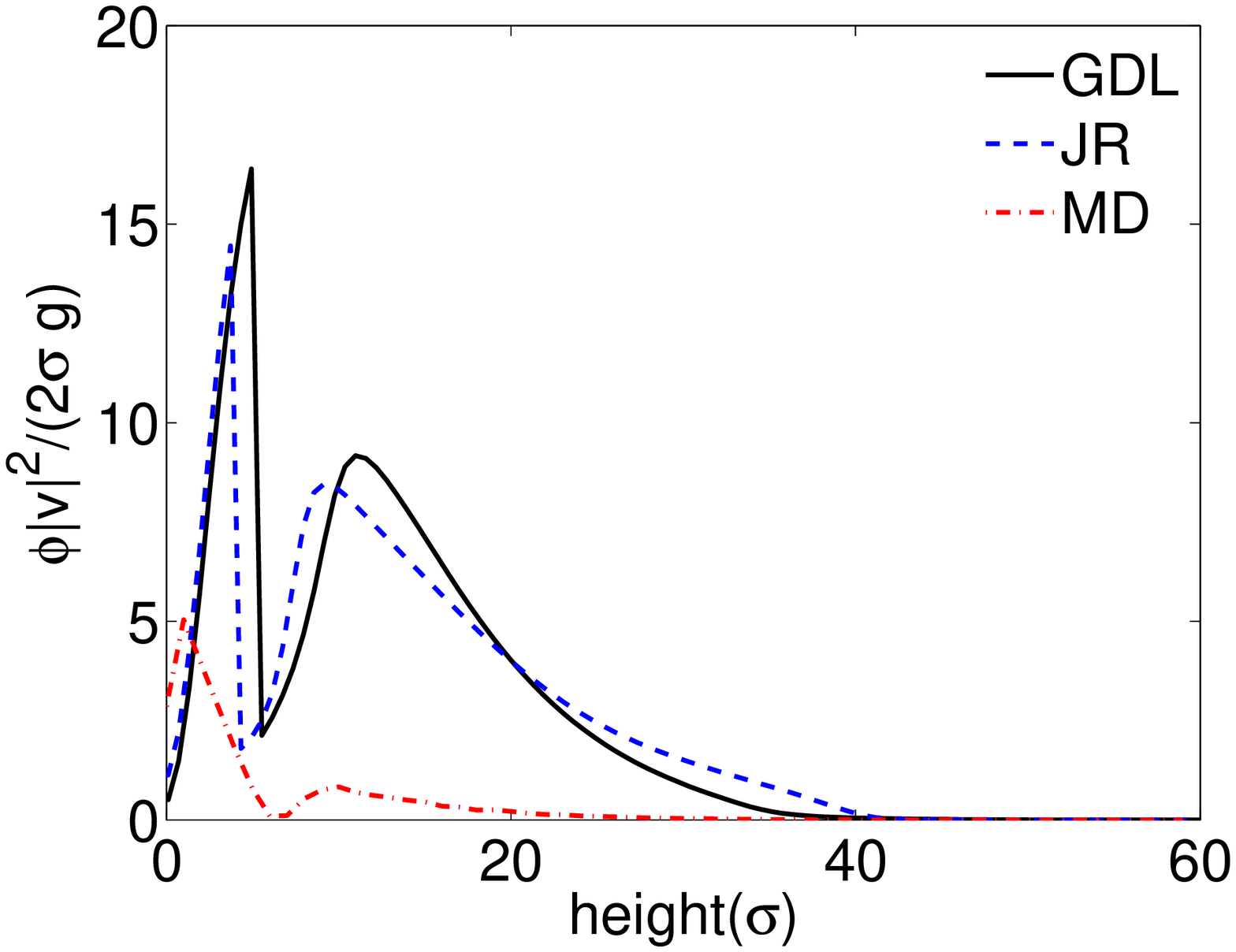}}
 \subfloat[$t=\nicefrac{3}{4} f^{-1}$]{\includegraphics[width=\halfcolumn]{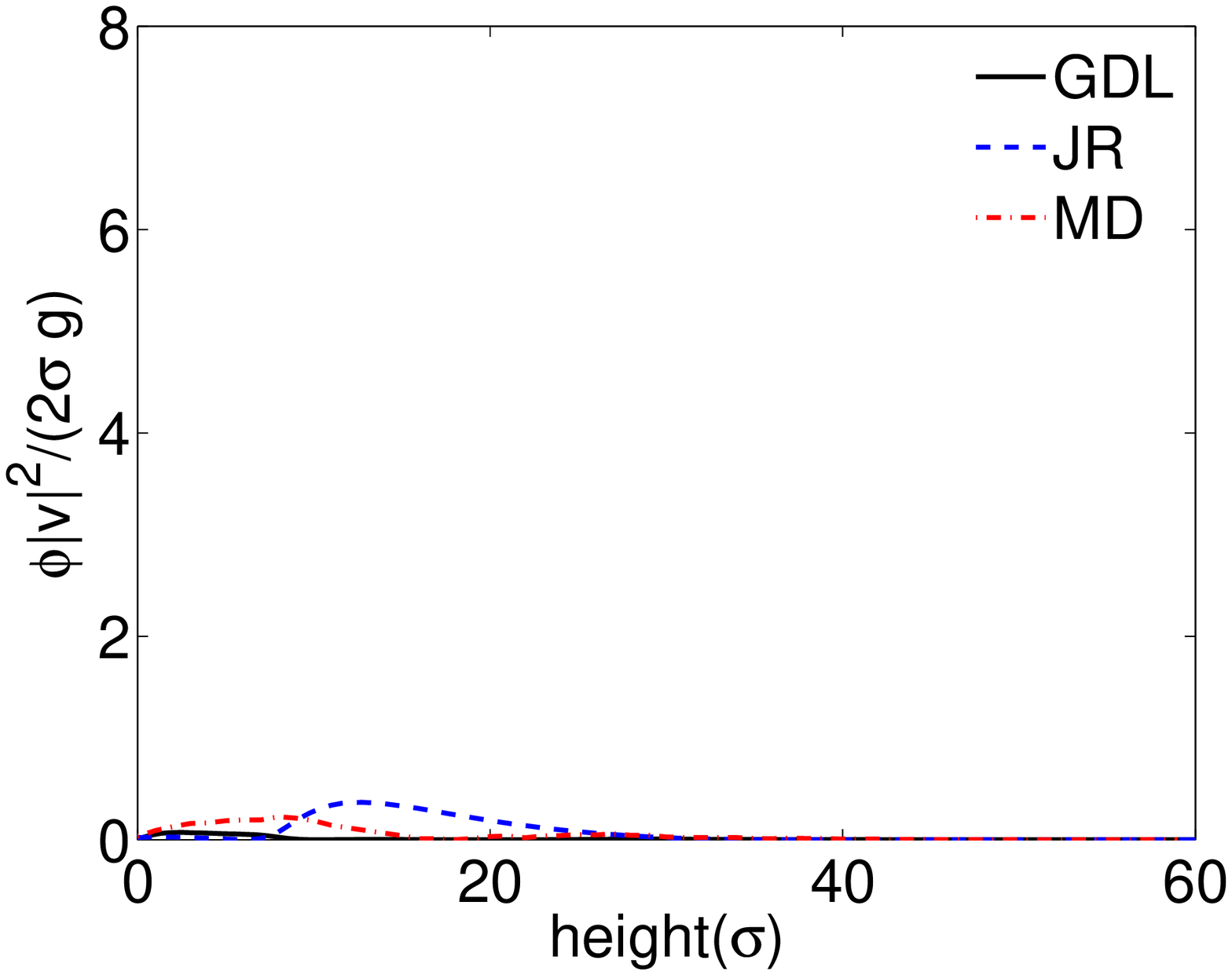}}

 \subfloat[$t=\nicefrac{4}{4} f^{-1}$]{\includegraphics[width=\halfcolumn]{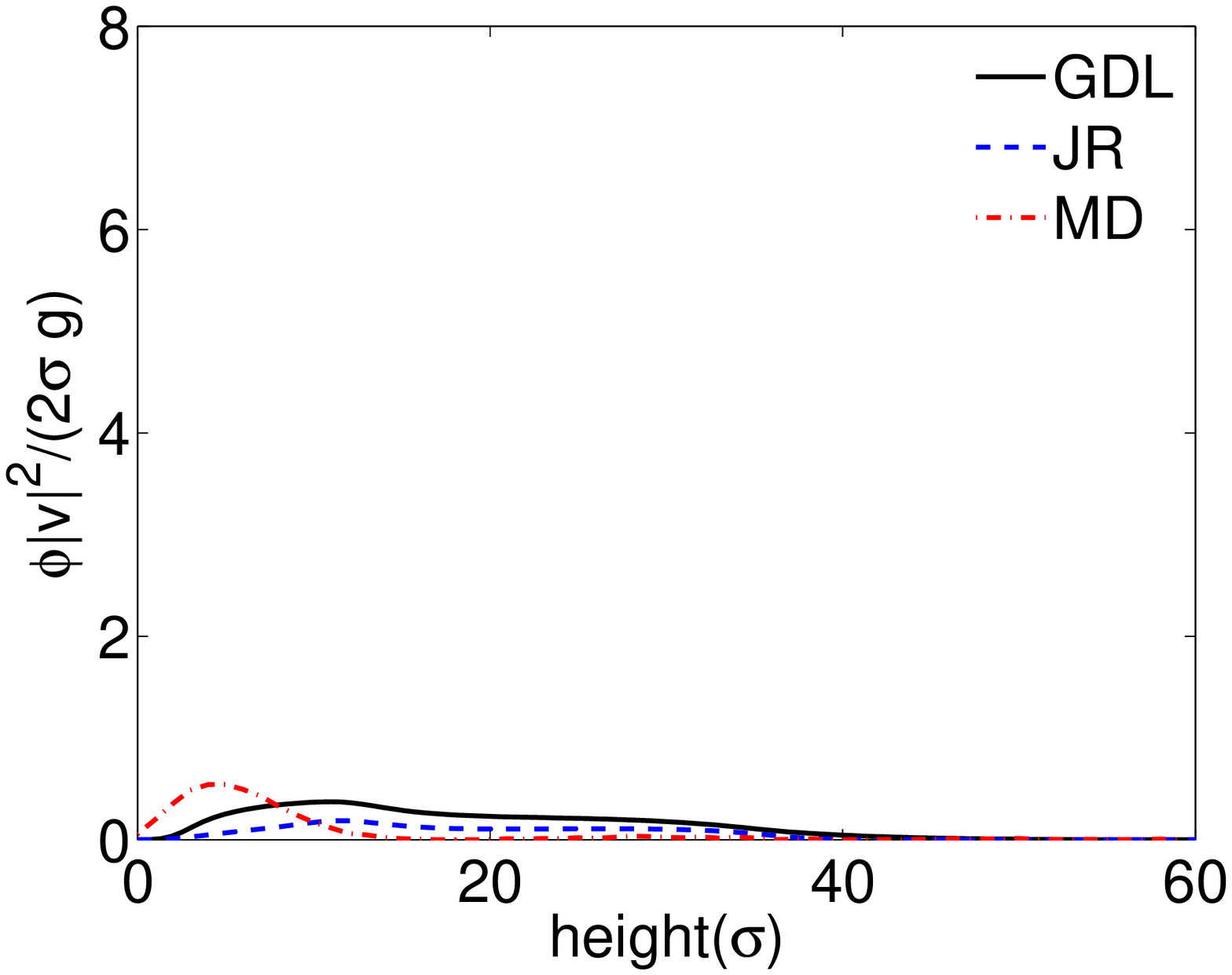}}
 \subfloat[$t=\nicefrac{5}{4} f^{-1}$]{\includegraphics[width=\halfcolumn]{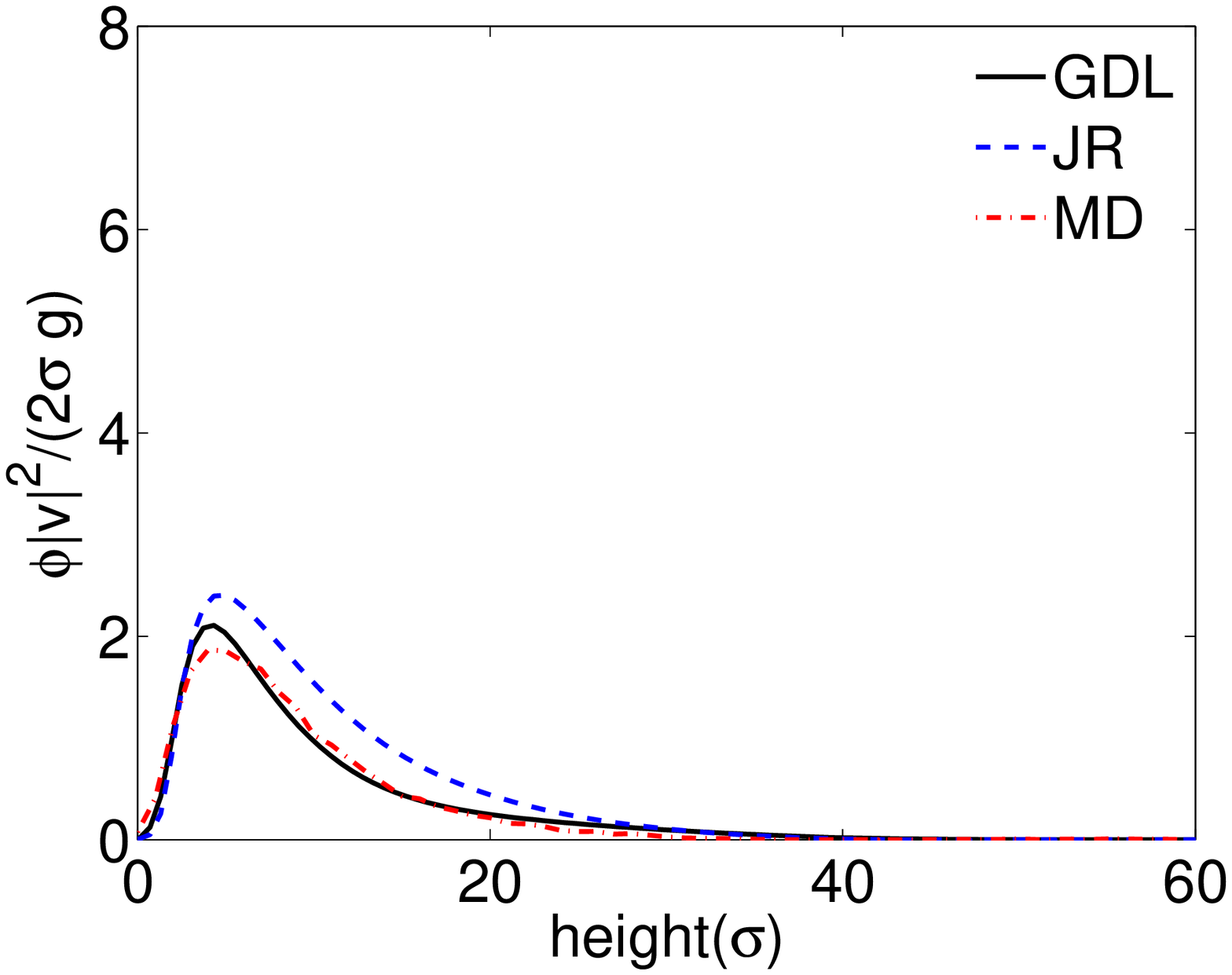}}

 \subfloat[$t=\nicefrac{6}{4} f^{-1}$]{\includegraphics[width=\halfcolumn]{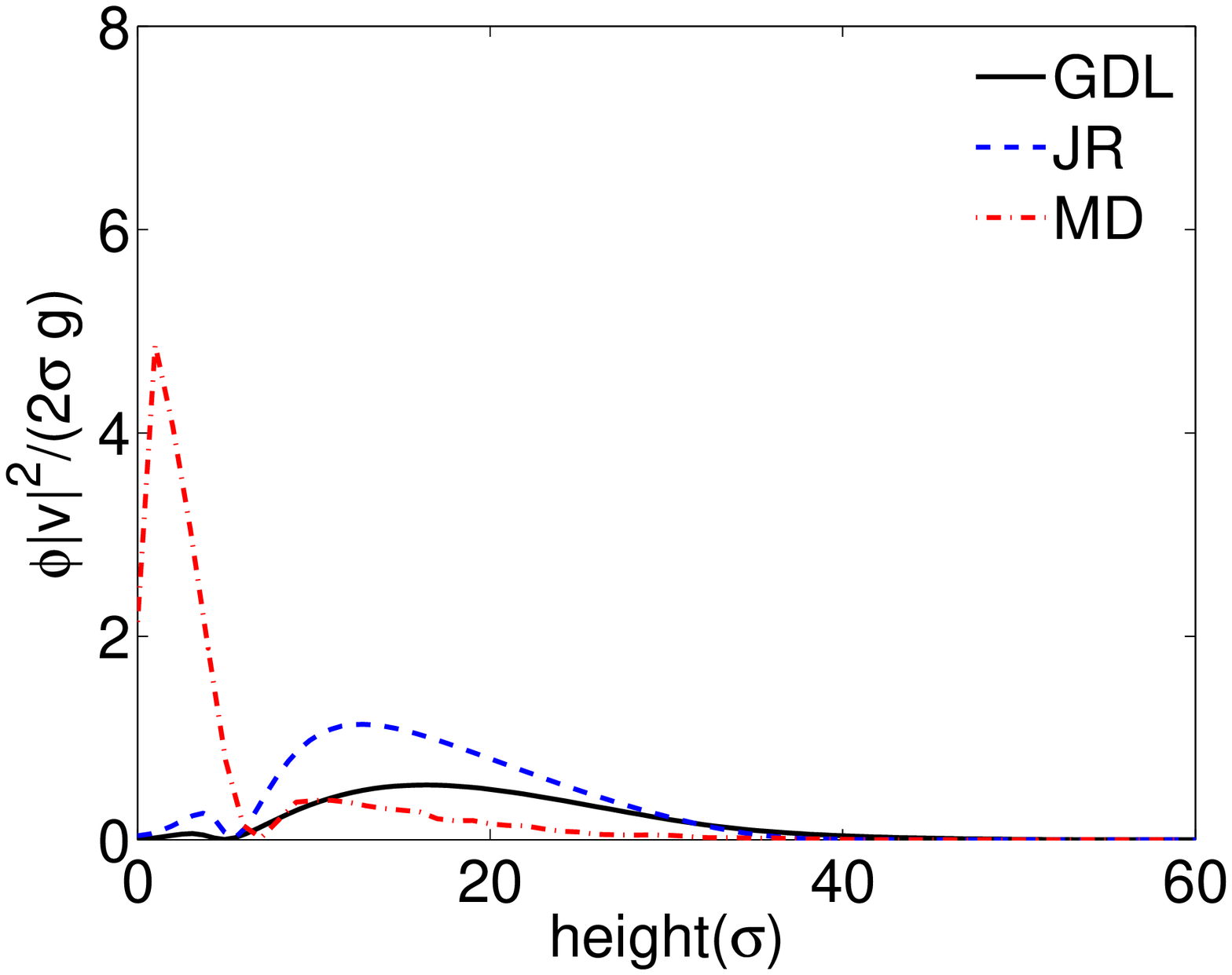}}
 \subfloat[$t=\nicefrac{7}{4} f^{-1}$]{\includegraphics[width=\halfcolumn]{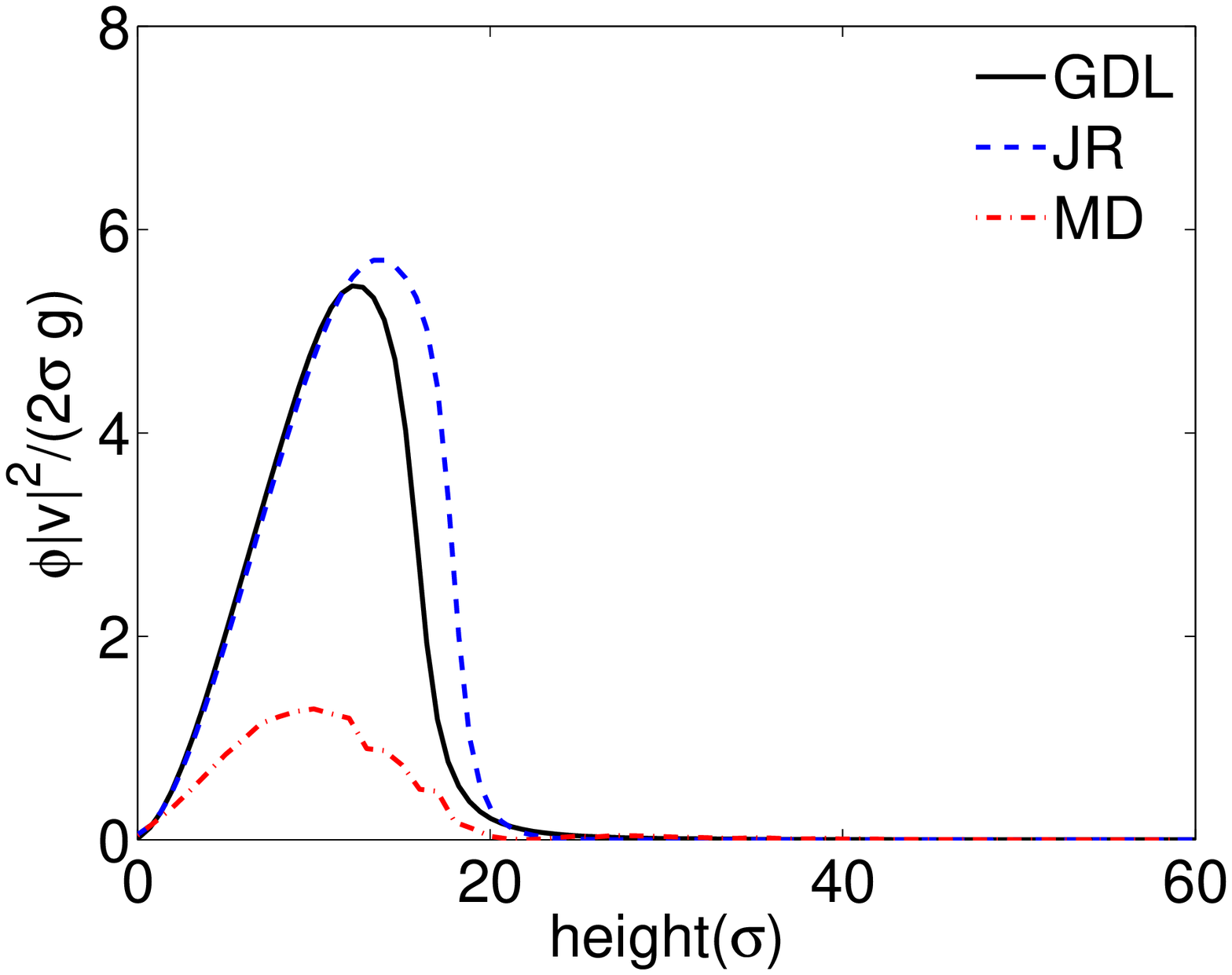}}
\end{center}
\caption{The scaled kinetic energy profiles as a function of
height (in units of $\sigma$), at selected times over two
oscillation periods for the MD, JR and GDL systems. For time
evolution of the profiles see \cite{online}.} \label{fig:Ec}
\end{figure}

Figure~\ref{fig:Ec} shows the scaled kinetic energy profiles. An examination of
the entire sequence shows that the maximum of the kinetic energy is achieved at
$t=0.38 f^{-1}$ in the GDL simulation, at $t=0.42f^{-1}$ in the JR and at
$t=0.54f^{-1}$ in the MD. The GDL peak is the highest, more than 4 times bigger
than the MD, and about 50\% bigger than the JR. This shows that the GDL solution
for the velocity field is also quantitatively different from the JR, a
consequence of the inelasticity contributing to the viscosities. Leaving aside
the mismatch at the maximum, the JR and GDL solutions go close to each other,
and differently from the MD profile, due to the delayed landing of the granular
layer in the MD simulation. In any case the comparison of the kinetic energy
profiles reinforces the quite unexpected result that the GDL solution is not
closer to the MD, but even further away, than the JR.

\begin{figure}
 \begin{center}
\center{\includegraphics[width=0.7\columnwidth]{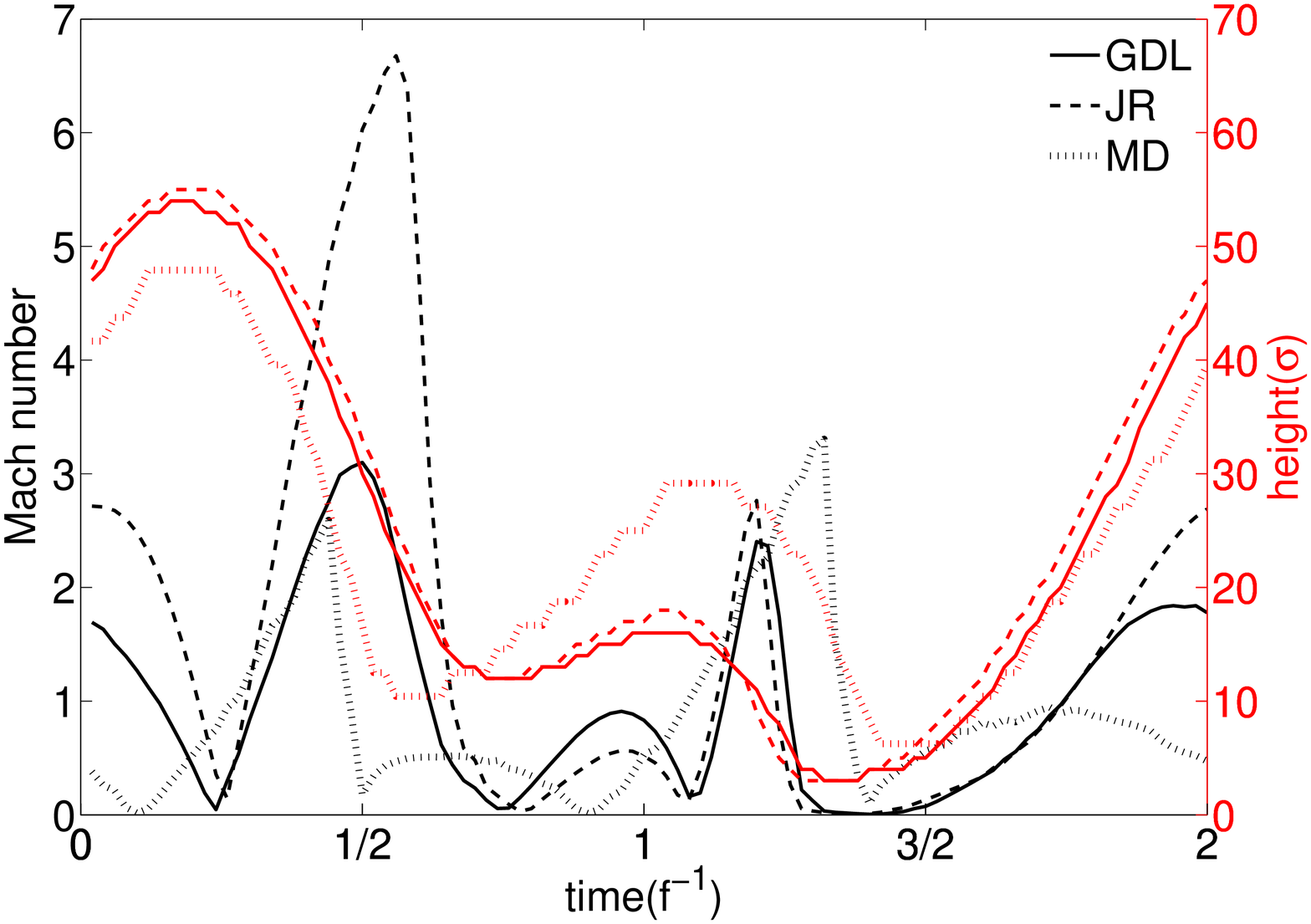}}
\end{center}
\caption{(color online) Mach number for MD system and JR and GDL theories as a
function of time, along two periods ($f^{-1}$) of oscillation of the plate. The
red curves indicate the variable height (in diameters) which corresponds to the
Mach number shown. These heights are found as the first point in vertical
direction from the plate where the packing fraction is 0.1.}
\label{fig:Mach}
\end{figure}

Since the GDL temperature is about one order of magnitude higher than JR in the
dilute region, the Mach number is also smaller. In Fig.~\ref{fig:Mach} we can
see how the differences are very relevant during the stages (c)-(d), when the
layer has achieved its maximal extension, and where the JR Mach number is about
twice that of the GDL. This is another fact showing that the GDL system is
more diffusive than the JR.

The MD curve for the Mach number has been produced using the averaged density
and temperature fields into Eq.\ \eref{soundspeed} for the sound speed,
supplied with the equation of state \eref{hydrostaticLG}.

Unlike JR and GDL approximations, the second MD peak in the Mach number is
higher than the first one. Anyway GDL predicts better the behavior of the Mach
number than JR. The values of the Mach number have been computed at the heights
shown by the red curves  in Fig.~\ref{fig:Mach}. They correspond to the first
point, going from the dense to the dilute phase, where the packing fraction is
0.1. There we also find discrepancies when comparing the MD results with those
of JR and GDL simulations. This is a consequence of the discrepancies in the
density field discussed above.

\section{Conclusions}

In this paper, we have compared the predictions of the Navier-Stokes
hydrodynamic equations of two dimensional granular gases with MD simulations in
a highly nonlinear, far-from-equilibrium problem such as the periodic impact of
a horizontal piston which gives rise to the characteristic pattern formation of
the Faraday instability. Given that the corresponding Navier-Stokes transport
coefficients are not \emph{exactly} known, two different approaches to them have
been considered: the JR and GDL approximations. While the first approach applies
for nearly elastic particles (in fact their forms for the transport coefficients
are the same as the elastic ones), the latter approximation is much more
accurate for granular gases (as verified for instance in Ref.\ \cite{GSM07} by
comparing the GDL theory with computer simulations at quite extreme values of
dissipation) since it incorporates the effect of inelasticity on the transport
coefficients. In particular, while the JR theory neglects the term $-\mu \vec
\nabla n$ in the heat flux, the new transport coefficient $\mu$ is clearly
different from zero in the GDL theory (see fourth panel of Fig.\
\ref{fig:ratiotransportcoef}). After comparing both approaches with
coarse-grained MD results, we can conclude on the following relevant aspects.

First, we conclude that the granular Navier-Stokes hydrodynamics with the
proper GDL forms for the transport coefficients $\eta$, $\gamma$, $\kappa$ and
$\mu$ is not capable of reducing the discrepancy between discrete particle
simulations and hydrodynamic simulations of moderately dense, inelastic gases.
This quantitative disagreement can be due to the fact that while the
Navier-Stokes constitutive equations \eref{vic1} and \eref{vic2} for the
pressure tensor and the heat flux, respectively, apply to first order in the
spatial gradients, the problem analyzed here might be outside the strict
validity of the Navier-Stokes approximation as the comparison with MD
simulations shows. Surprisingly, the discrepancies between theory and
simulations decrease if one considers the elastic forms of the transport
coefficients. We think that there are no physical reasons behind this
improvement. A similar conclusion has been found in the simple shear flow
problem for dilute gases since the non-Newtonian shear viscosity
$\eta_\textrm{s}(\alpha)$ to be plugged into the Navier-Stokes hydrodynamic
equations is better modeled by the elastic shear viscosity than its
corresponding inelastic version $\eta_\textrm{\small{GDL}}(\alpha)$ (see Fig. 1 of Ref.\
\cite{SGD04} where while $\eta_\textrm{s}$ decreases as decreasing $\alpha$ the
opposite happens for $\eta_\textrm{\small{GDL}}$).

Apart from the different $\alpha$-dependence of $\eta$, $\gamma$
and $\kappa$, the main difference between the JR and GDL
approaches lies in the presence of the coefficient $\mu$ in the
heat flux. This new transport coefficient, characteristic of
inelastic gases and thus vanishing in the JR theory, constitutes
the significant contribution to an enhanced heat transfer
mechanism which leads to a high temperature solution in the dilute
region, which is not supported by the particle simulations. Even
in the dilute region at the top of the system
(Fig.~\ref{fig:ratiotransportcoef}), where shock waves should be
completely damped, the GDL model produces an unrealistic energy
excess. There, at densities of the order of 0.001 in packing
fraction, the coefficient $\mu$ is very different from zero.

Thus, we can conclude that the transport coefficient $\mu$ is
clearly overestimated by the Navier-Stokes approximation and
consequently, the influence of the diffusion term $-\mu \vec
\nabla n$ on the heat flux is larger than the one observed in the
simulations. As mentioned before, these discrepancies do not imply
that the GDL approach is deficient in any respect. Rather
differently, they show the limits of the Navier-Stokes description
applied to certain regimes of complex granular flow.

As a matter of fact, in the theory of granular gases it is well
accepted that in certain cases the Navier-Stokes description is
insufficient. It was clearly evidenced on the mathematical level,
e.g., in \cite{G08}, that Burnett-order terms are important for
the kinetics of granular gases. These terms in the constitutive
relations are of second-order in the gradients and therefore
beyond the Navier-Stokes level of description. In \cite{G08} it
was shown that these terms are even necessary for a consistent
description due to the lack of scale separation in granular gases.
The presence of large gradients is quite usual in granular flow,
where physical variables may change several orders of magnitude
within a distance of a few grains, due to inelastic interactions.
That typically manifests into strong shock waves propagating into
the system from the boundaries, where the energy source is
located. In this way, granular flow is very often supersonic or
even hypersonic, and in this regime of extreme gradients the
Navier-Stokes description may reveal inadequate.

The inclusion of higher-order terms (beyond the Navier-Stokes domain) in the
constitutive equations for the momentum and heat fluxes might prove a better
approximation to problems like this one, where the first order in the gradients
expansion looks insufficient. However, the determination of these nonlinear
contributions to the fluxes becomes a very hard task if one starts from the
revised Enskog equation. In these cases it is useful to consider kinetic models
with the same qualitative features as the Enskog equation but with a
mathematically simpler structure \cite{BDS99}. The use of these models allows to
derive explicit forms for generalized constitutive equations in complex states
driven far from equilibrium, such as the simple shear flow state \cite{MGSB99}.

In spite of the discrepancies found here, the Navier-Stokes approximation with
the GDL forms for the transport coefficients is still appropriate and accurate
for a wide class of flows. Some examples include applications of Navier-Stokes
hydrodynamics to symmetry breaking and density/temperature profiles in vibrated
gases \cite{BRM01,BRMG02}, binary mixtures \cite{DHGD02} and supersonic flow
past a wedge in real experiments \cite{RBSS02,YHCMW02,HYCMW04}. Another group
refers to spatial perturbations of the homogeneous cooling state for an isolated
system where MD results of the critical length for the onsets of vortices and clusters 
\cite{MDCPHPhysFluids2011,Peter12} are successfully compared with the
predictions from linear stability analysis \cite{G05} performed on the basis of
the GDL transport coefficients.

As a summary, the Navier-Stokes theory has shown limitations when exploring the
highly nonlinear problem of the granular Faraday instability. In particular,
the presence of rarefied regions where strong transient shock wave fronts
propagate seem to justify the inclusion of higher order gradients in the
transport equations, going beyond the Navier-Stokes approximation \cite{G08}.
In spite of that, both GDL and JR models work quite well, although here the main
discrepancy is attributed to the term in the heat flux coupled to the density
gradient, which is the missing contribution in the JR approach that the
inelastic theory comes to fix. More work has to be done in this respect to
clarify the conditions under which the Navier-Stokes approximation fails to
describe appropriately the granular heat transport. Finally, as a complementary
route to the Navier-Stokes approximation, one could numerically solve the Enskog
equation via the direct simulation Monte Carlo method \cite{Bird,Chema1}.
Presumably, the numerical solution would give better quantitative agreement with
MD simulations than the Navier-Stokes results reported here. This is quite an
interesting problem to be addressed in the future for the Faraday and other
different problems such as the simple shear flow or homogeneous states.


\ack{
LA and JAC were partially supported by the project
MTM2011-27739-C04-02 DGI (Spain) and 2009-SGR-345 from
AGAUR-Generalitat de Catalunya. LA, JAC, and CS acknowledge
support of the project Ingenio Mathematica FUT-C4-0175. CS and VG
appreciate funding from the projects DPI2010-17212 (CS) and
FIS2010-16587 (VG) of the Spanish Ministry of Science and
Innovation. The latter project (FIS2010-16587) is partially
financed by FEDER funds and by the Junta de Extremadura (Spain)
through Grant No. GR10158. LA and TP were supported by Deutsche
Forschungsgemeinschaft through the Cluster of Excellence {\it
Engineering of Advanced Materials}. JAC acknowledges support from
the Royal Society through a Wolfson Research Merit Award. This
work was supported by Engineering and Physical Sciences Research
Council grant number EP/K008404/1.}

\section*{References}
\bibliographystyle{iopart-num}

\bibliography{biblio}

\end{document}